\documentclass[11pt]{article}
\pdfoutput=1
\usepackage{booktabs} % To thicken table lines

\usepackage[normalem]{ulem}

\usepackage{epsfig}
\usepackage{cite}
\usepackage{amsmath}
\usepackage{amssymb}
\usepackage[footnotesize]{caption}
\usepackage{slashed}
\usepackage{xcolor}
\usepackage[utf8]{inputenc}
\usepackage{tikz}
\usetikzlibrary{shapes.geometric}
\usetikzlibrary{arrows.meta,arrows}
\usetikzlibrary{quotes,angles}

\usepackage{multicol}
\usepackage{multirow}
\usepackage{array}
\numberwithin{equation}{section}

\newcommand{\be}{\begin{equation}}
\newcommand{\ee}{\end{equation}}
\newcommand{\beq}{\begin{eqnarray}}
\newcommand{\eeq}{\end{eqnarray}}

\begin{document}

\title{
%\vspace*{-3.7cm}
%\phantom{h} \hfill\mbox{\small KA-TP-10-2017}\\[-1.1cm]
%\vspace*{0.7cm}
%\\[1cm]
%\vspace{13mm}
\textbf{Light Higgs searches in $t \bar t \phi$ production at the LHC \\[4mm]}}

\date{}
\author{
Duarte Azevedo$^{1\,}$\footnote{E-mail:
  \texttt{drpazevedo@fc.ul.pt}} ,
Rodrigo Capucha$^{1\,}$\footnote{E-mail:
\texttt{rodrigocapucha@hotmail.com}} ,
Emanuel Gouveia$^{2\,}$\footnote{E-mail:
\texttt{e.gouveia@cern.ch}} ,\\
Ant\'{o}nio Onofre$^{3\,}$\footnote{E-mail:
\texttt{antonio.onofre@cern.ch}} ,
Rui Santos$^{1,4\,}$\footnote{E-mail:
  \texttt{rasantos@fc.ul.pt}} 
\\[5mm]
{\small\it $^1$Centro de F\'{\i}sica Te\'{o}rica e Computacional,
    Faculdade de Ci\^{e}ncias,} \\
{\small \it    Universidade de Lisboa, Campo Grande, Edif\'{\i}cio C8
  1749-016 Lisboa, Portugal} \\[3mm]
{\small\it
$^2$ LIP, Departamento de F\'{\i}sica , Universidade do Minho, 4710-057 Braga, Portugal} \\[3mm]
{\small\it
$^3$ Departamento de F\'{\i}sica , Universidade do Minho, 4710-057 Braga, Portugal} \\[3mm]
{\small\it
$^4$ISEL -
 Instituto Superior de Engenharia de Lisboa,} \\
{\small \it   Instituto Polit\'ecnico de Lisboa
 1959-007 Lisboa, Portugal} \\[3mm]
}

\maketitle

\begin{abstract}
\noindent
In this paper we propose a new reconstruction method to explore the low mass region in the associated production of top-quark pairs ($t\bar{t}$) with a generic scalar boson ($\phi$) at the LHC. 
The new method of mass reconstruction shows an improved resolution of at least a factor of two in the low mass region when compared to previous methods, without the loss of sensitivity of previous analyses. It turns out that it also leads to an improvement of the mass reconstruction of the 125 GeV Higgs for the same production process. 
We use an effective Lagrangian to describe a scalar with a generic Yukawa coupling to the top quarks. A full phenomenological analysis was performed, using Standard Model background and signal events generated with \texttt{MadGraph5\_aMC@NLO} and reconstructed using a kinematic fit. The use of CP-sensitive variables allows then to maximize the distinction between CP-even and CP-odd components of the Yukawa couplings. 
Confidence Levels (CLs) for the exclusion of $\phi$ bosons with mixed CP (both CP-even and CP-odd components) were determined as a function of the top Yukawa couplings to the $\phi$ boson. 
The mass range analysed starts slightly above the $\Upsilon$ mass up to 40 GeV, although the analysis can be used for an arbitrary mass. %We focus on dileptonic final states of the $t\bar{t}\phi$ system, with $\phi\rightarrow b\bar{b}$. 
If no new light scalar is found, exclusion limits at 95\% CL for the absolute value of the CP-even and CP-odd Yukawa are derived.
%couplings are expected to be, approximately, as low as 0.10 and 0.50, respectively, at the end of the High Luminosity phase of the LHC (HL-LHC). 
Finally, we analyse how these limits constrain the parameter space of the complex two-Higgs doublet model (C2HDM).

\end{abstract}

\thispagestyle{empty}
\vfill
\newpage
\setcounter{page}{1}

\section{Introduction}
\hspace{\parindent} %forca identacao

The Large Hadron Collider (LHC) will soon restart operation. It is now time to prepare the searches for both lighter and heavier scalars than the already discovered Higgs with a mass of 125 GeV. These scalars are predicted by most of the extensions of the Standard Model (SM) with an enlarged scalar sector. We have recently concluded a study~\cite{Azevedo:2020vfw} on the searches for a scalar with indefinite CP in the associated production of top-quark pairs ($t\bar{t}$) at the LHC. The study was performed for a scalar in the mass region between 40 GeV and 200 GeV. The reason to stop at 40 GeV was mainly due to the fact that in the low mass regime the jets resulting from the $\phi$ boson decay may overlap in the detector and appear as one single jet. This in turn causes a potential loss of sensitivity of the analysis. Therefore, in order to correctly identify the jet(s) coming from the $\phi$ boson, a new approach to the kinematic reconstruction was used, extending the one considered in~\cite{Azevedo:2020vfw}. With the new approach we are now able to probe the low mass region down to the mass of the $\Upsilon$ meson with a mass of 9.46 GeV. We therefore limit our analysis to the mass range between 12 GeV and 40 GeV. Furthermore, the new method leads to a resolution improvement by roughly a factor of two for a scalar with a 40 GeV mass. As we will see it turns out that this resolution improvement also happens for the 125 GeV Higgs.
 
The current measurements of the properties of the Higgs boson at the LHC revealed that it is consistent with the SM prediction. 
Nevertheless, the LHC experiments cannot, currently, exclude the possibility of Physics Beyond the SM (BSM) in the Higgs sector. Despite the fact that ATLAS and CMS established that the discovered 125 GeV Higgs~\cite{Aad:2012tfa, Chatrchyan:2012ufa} is not a pure pseudoscalar state with a 99\% confidence level (CL), mixed states with significant contributions from CP-odd components are still possible, even for the discovered Higgs boson. 
As additional sources of CP-violation, as discussed by Sakharov~\cite{Sakharov:1967dj}, are required to explain the matter anti-matter asymmetry observed in the Universe, the study of the CP nature of the discovered Higgs boson couplings to bosons and fermions is of utmost importance at the LHC. 
Moreover, the fact that additional Higgs bosons may exist, with masses that are allowed to range from few GeV up to the TeV scale, implies that dedicated searches must be improved in order to increase the sensitivity to detect such Higgs bosons, in particular in the very challenging low mass region. 
One simple extension of the SM with a CP-violating scalar sector is the CP-violating version of the 2-Higgs doublet model (2HDM) known as C2HDM.
The model has an extra scalar doublet and has been the subject of many studies~\cite{Lee:1973iz, Ginzburg:2002wt, Khater:2003wq, ElKaffas:2007rq, Grzadkowski:2009iz, Arhrib:2010ju, Barroso:2012wz, 
Inoue:2014nva,Cheung:2014oaa, Fontes:2014xva, Fontes:2015mea, Chen:2015gaa, Muhlleitner:2017dkd, Fontes:2017zfn}.
The C2HDM is an excellent benchmark model to test the scalar's CP quantum numbers at the LHC. It contains three neutral scalars which have a mixture of CP-even and CP-odd components with no restrictions on the values of the masses other than the ones from experimental and theoretical constraints. Any of the three scalars can be the discovered 125 GeV Higgs boson. 
The search for BSM physics and in particular the measurement of the Yukawa couplings has become a primary target of the next LHC run. 
The relation between the CP-even and the CP-odd Yukawa couplings can be directly probed both in the production or in the decays of these scalars. 
There are many proposals in the literature for production, in the case of the top quark~\cite{Gunion:1996xu, Boudjema:2015nda, Santos:2015dja, AmorDosSantos:2017ayi, Goncalves:2018agy, Faroughy:2019ird}, 
in the decays of the tau leptons~\cite{Berge:2008wi, Berge:2008dr, Berge:2011ij, Berge:2014sra, Berge:2015nua, Antusch:2020ngh} and more recently also for bottom quarks~\cite{Ghosh:2019dmv, Grojean:2020ech} .

ATLAS and CMS have so far studied the CP nature of the 125~GeV Higgs boson couplings to the top quarks and to the $\tau$ leptons. 
The CP nature of the couplings is more accessible experimentally with fermions because it is a tree-level coupling, as opposed to gauge bosons CP-odd contributions that appear via higher-order corrections to the Higgs vertices~\cite{Huang:2020zde} and are suppressed by powers of the energy scale associated to possible new physics. Using the Higgs boson two photons decay channel $H\rightarrow \gamma\gamma$ in associated production of top quarks and Higgs bosons $pp\rightarrow t\bar{t}H$, both ATLAS and CMS~\cite{Sirunyan:2020sum, Aad_2020} were able to exclude the purely CP-odd hypothesis at best with 3.9 standard deviations and to establish a 95\% CL observed (expected) exclusion upper limit for the mixing angle of 43$^{\circ}$~(63$^{\circ}$).  
Recently  CMS~\cite{CMS-PAS-HIG-20-006} has performed the first measurement of the CP mixing angle of the tau lepton Yukawa coupling, using data collected at $\sqrt{s} =$ 13~TeV, corresponding to an integrated luminosity of 137~fb$^{-1}$. 
The CP mixing angle was found to be 4$^{\circ}$ $\pm$ 17$^{\circ}$, allowing to set an observed (expected) exclusion upper limit for the mixing angle of 36$^{\circ}$~(55$^{\circ}$).

In this paper, using the new reconstruction method we will also study the possibility of probing the CP nature of the couplings of low mass Higgs bosons ($\phi$) to top quarks in the associated production process $t\bar{t}\phi$, still considering the main decay channel of the Higgs boson i.e., $\phi\rightarrow b\bar{b}$.  
We now cover the low mass $\phi$ region, with masses in the range 12~GeV $\leq m_\phi \leq 40$~GeV, without the loss of sensitivity observed previously.

This paper is organised as follows. Following the Introduction, the theoretical Higgs boson phenomenological framework is presented in Section~\ref{sec:TH}, as well as the relevant parameters of the model. The event generation and kinematic reconstruction are described in Section~\ref{sec:generation}. The full event selection is discussed in Section~\ref{sec:selection}. The main results are presented in Section~\ref{sec:results} and their impact in the framework of the C2HDM is analysed in Section~\ref{sec:C2HDM}. Finally, our main conclusions are drawn in Section~\ref{sec:conclusions}.

%%%%%%%%%%%%%%%. Lagrangian
\section{The Lagrangian \label{sec:TH}}
\hspace{\parindent} %forca identacao

In the SM, the Yukawa coupling of the top quark to the Higgs boson is CP-even with a strength given by $y_t=\sqrt{2}m_t/v$, where $m_t$ is the top quark mass and $v$ is the electroweak vacuum expectation value. If a CP-odd component would contribute to the Yukawa interaction, the Higgs boson  ($\phi$)  would no longer have a well defined CP number. A Lagrangian that describes this generalised interaction can be written as 
\begin{equation}
{\cal L} = \kappa_t y_t  \bar t (\cos \alpha + i \gamma_5 \sin \alpha) t \phi \, =   y_t  \bar t (\kappa + i \tilde \kappa \gamma_5) t \phi \,,
\label{eq:higgscharacter}
\end{equation}
where  $\kappa_t$ parametrises the total coupling strength relative to the SM and the angle $\alpha$ parametrises the CP-phase, which is related to the parameters in the Higgs potential. 
We will refer to $\phi = H$ for the CP-even scenario and $\phi=A$ for the CP-odd case. The CP-even case is recovered by setting $\cos \alpha = \pm 1$ while the CP-odd case is obtained by fixing $\cos \alpha = 0$.

Several angular observables have been proposed~\cite{Gunion:1996xu, Boudjema:2015nda, Santos:2015dja, AmorDosSantos:2017ayi} to probe the CP nature of a scalar boson in the top quark Yukawa coupling using $t\bar t\phi$ production at colliders. These observables are sensitive not only to the nature of the scalar but also allow for the discrimination 
of Higgs boson signals from irreducible backgrounds at the LHC. Moreover, the results obtained with a phenomenological analysis where $t\bar{t}\phi$ signals (assuming  $m_\phi$ = 125~GeV) and dominant backgrounds were generated at the LHC, including simulated detector effects (resolutions and acceptances)~\cite{Santos:2015dja, AmorDosSantos:2017ayi}, showed that these observables can be classified in two major categories. 
The first category are observables that can discriminate signals from dominant backgrounds. In these observables the differential distribution is similar
between the signals, which makes these observables particularly suited for cross section measurements comparison regardless of signal type. 
The second category are distributions that have a significant discriminating power between signals i.e., are sensitive to the CP-phase. Recently~\cite{Azevedo:2020vfw}, we have extended the use of these angular observables to a wider mass range, from 40~GeV to 500~GeV. The low mass boundary was imposed by the analysis which became inefficient due to the $t\bar{t}\phi$ reconstruction methods applied. In this paper we consider an even more challenging lower mass range, between 12~GeV and 40~GeV (the low mass regime), where a new reconstruction algorithm was used, with significantly improved performance. 
For the studies presented in this paper, and in order to compare with previous ones published, we will use the variables $b_2$ and $b_4$ 
as defined in~\cite{Gunion:1996xu, Ferroglia:2019qjy} in the laboratory (LAB) and $t\bar{t}\phi$ centre-of-mass frames ($b_2^{t\bar{t}\phi}$ and $b_4^{t\bar{t}\phi}$, respectively),
\begin{equation}
b_2  = ( \vec{p}_{t} \times \hat{k}_z ).( \vec{p}_{\bar{t}} \times \hat{k}_z )/ (|\vec{p}_{t}| .  |\vec{p}_{\bar{t}}|), \ \ \ \  b_4 = (p^z_t . p^z_{\bar{t}}) / (|\vec{p}_{t}| . |\vec{p}_{\bar{t}}| ),
\end{equation}
where the $z$-direction corresponds to the beam line. 
It is worth noting that $b_2$ and $b_4$ have a natural physics interpretation. They depend on the $t$ and $\bar{t}$ polar angles, $\theta_t$ and $\theta_{\bar{t}}$ respectively, with respect to the $z$-direction, and on the azimuthal angle difference between the top quarks $\Delta\phi_{t\bar t}$, and can be expressed as $b_2=\cos{\Delta\phi_{t\bar t}} \times \sin{\theta_t} \times \sin{\theta_{\bar{t}}}$ and $b_4=\cos{\theta_t} \times \cos{\theta_{\bar{t}}}$.

%%%%%%%%%%%%%%%. Event Generation and Kinematic Reconstruction
\section{Event generation and kinematic reconstruction  \label{sec:generation}}
\hspace{\parindent} %forca identacao

Signal events from double $pp\to t\bar t\phi$ and single $pp\to t\phi + jets$ top quark associated production at the LHC with $\phi=\{H,A\}$, were generated at next-to-leading order (NLO) with the Higgs Characterisation model \texttt{HC\_NLO\_X0}~\cite{Artoisenet:2013puc}, using \texttt{MadGraph5\_aMC@NLO}~\cite{Alwall:2011uj}. The pure CP-even and the pure CP-odd samples were generated by setting $\cos \alpha=1$ and  $\cos \alpha=0$, respectively, following Equation~\ref{eq:higgscharacter}, with $\kappa_t=1$. Four samples, for both scalar and pseudoscalar signals, were generated with masses $m_\phi$ equal to 12, 20, 30 and 40~GeV. While the CP-even and CP-odd bosons were only allowed to decay to a pair of $b$-quarks ($\phi\to b \bar b$), the $t\bar{t}$ system was assumed to decay to a pair of $b$-quarks and two intermediate $W^\pm$ gauge bosons which, in turn, decay to two charged leptons and two neutrinos $t (\bar t)\to b W^+ (\bar b W^-)\to b \ell^+ \nu_\ell (\bar b \ell^- \bar \nu_\ell )$. Only $W$ boson decays to electrons ($e$) and muons ($\mu$) were considered as signal. 
This configuration defines the dileptonic channel. In addition to signal samples, backgrounds from SM $t\bar{t}H$, $t\bar{t}$ + jets, with up to 3 jets, $t\bar{t}V$ + jets, single top quark production ($t$-, $s$- and $Wt$-channels), $W$($Z$) + 4~jets, $W$($Z$)$b\bar{b}$ + 2~jets and $WW, ZZ, WZ$ diboson processes were also generated using \texttt{MadGraph5\_aMC@NLO}.  
As the details of signal and backgrounds Monte Carlo generation, hadronization and DELPHES detector simulation are the ones in~\cite{Azevedo:2020vfw}, they will not be repeated here.  
The event analysis is performed using the \texttt{MadAnalysis5}~\cite{Conte:2012fm} framework.

As the main decay mode of the Higgs boson searched for in this paper is the $\phi\to b \bar b$ channel, only events with at least two opposite charge leptons and four or more jets are selected for kinematic reconstruction. Both leptons and jets were required to have transverse momentum $p_T\geq 20$ GeV and pseudo-rapidity $|\eta| \leq 2.5$. 
These criteria lead to signal selection efficiencies that vary from 5\% (9\%) to 9\% (12\%) for masses of the scalar (pseudoscalar) from 12~GeV to 40~GeV, respectively. 
This set of cuts constitute what we call the {\it pre-selection}.

\begin{figure}[h!]
	\begin{center}
		\begin{tabular}{ccc}
		\hspace*{-5mm}\includegraphics[height = 5.7cm]{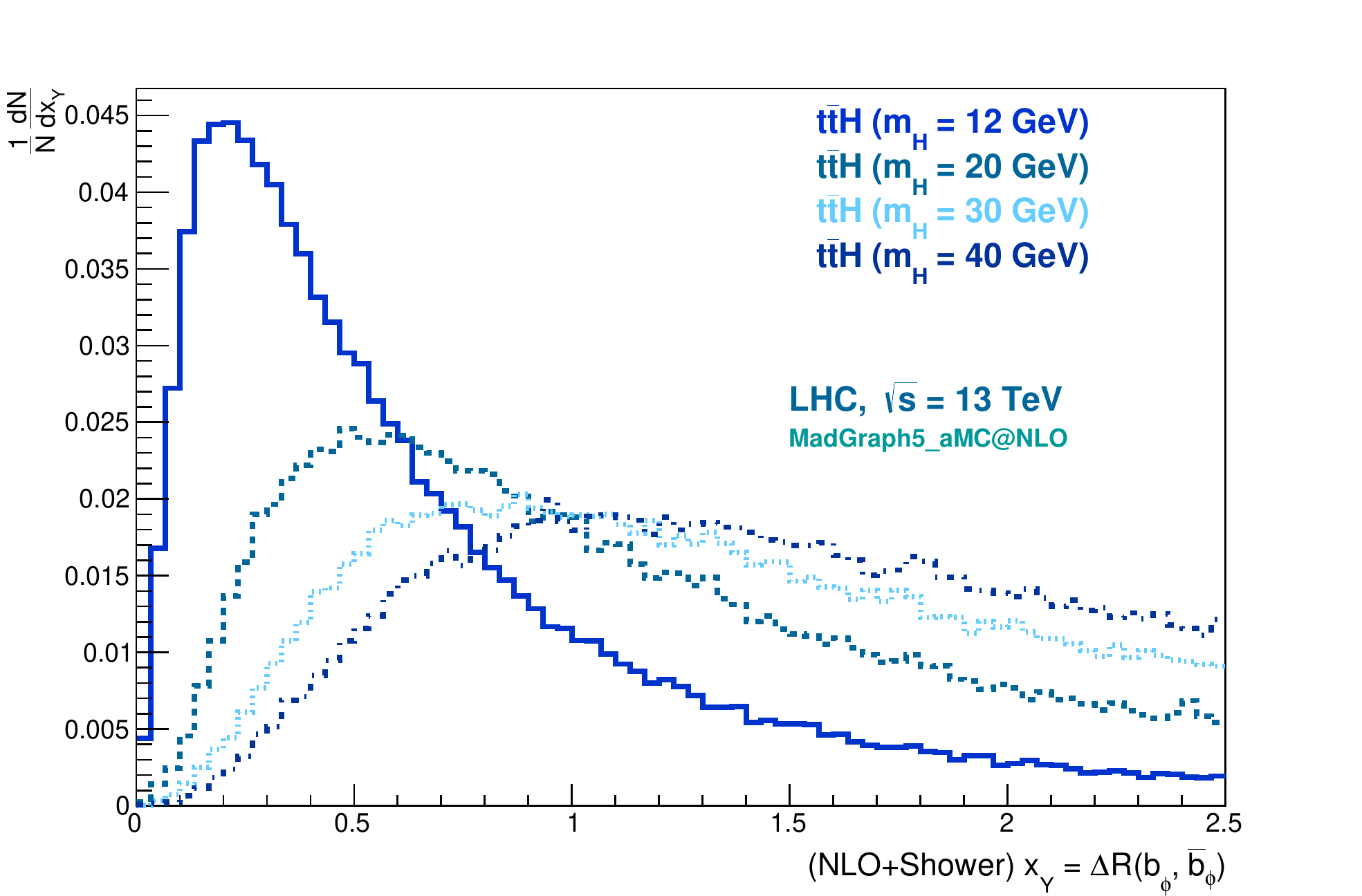}
		\hspace*{-7mm}\includegraphics[height = 5.7cm]{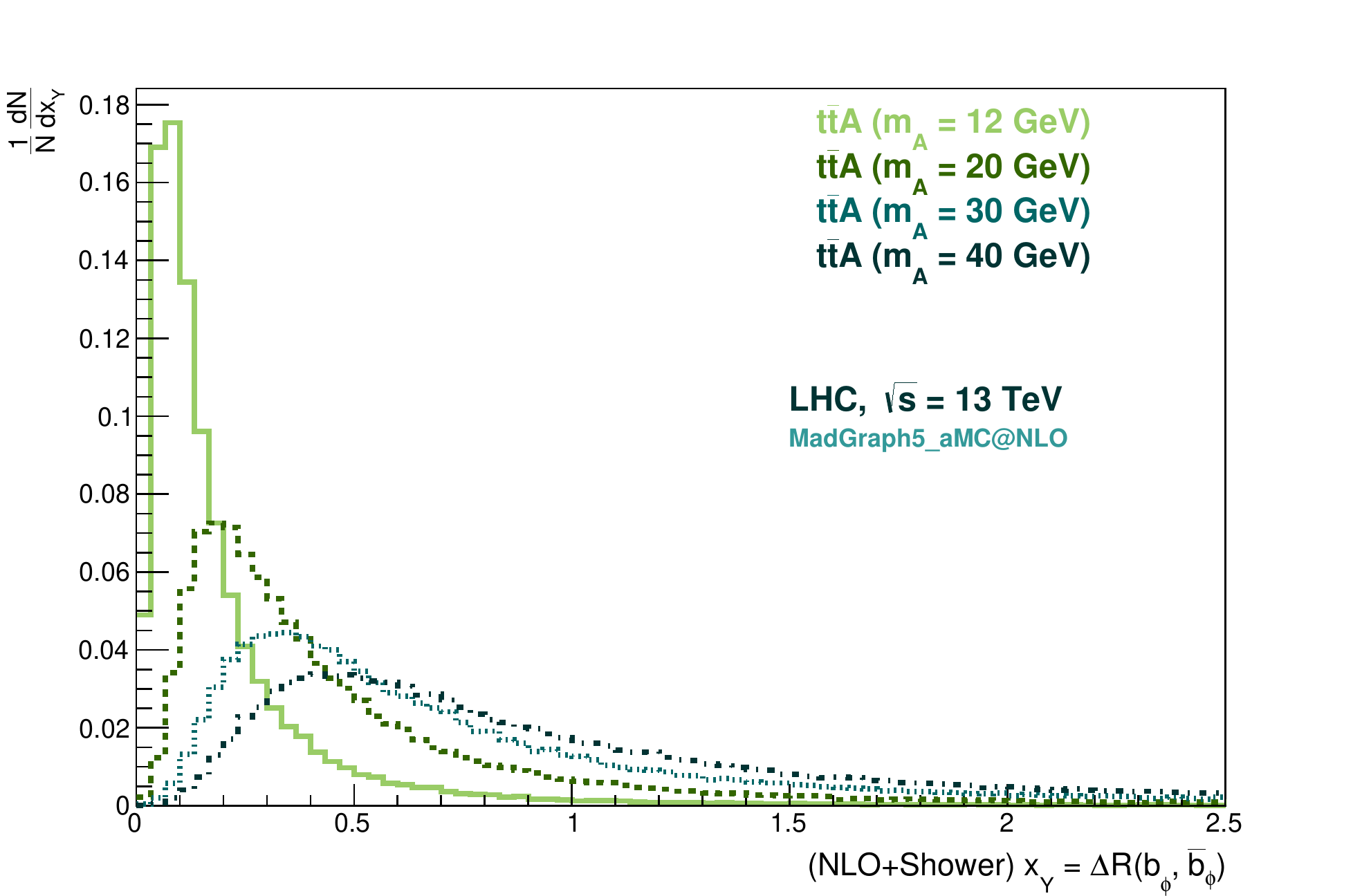} 
		\end{tabular}
		\caption[]{Parton level $\Delta R(b_\phi, \bar{b}_\phi)$ distributions with NLO corrections and shower effects (NLO+Shower), for $m_\phi= 12, 20, 30$ and 40~GeV. The CP-even case is shown on the left, and the CP-odd one on the right.}
		\label{fig:deltaRs}
	\end{center}
\end{figure}

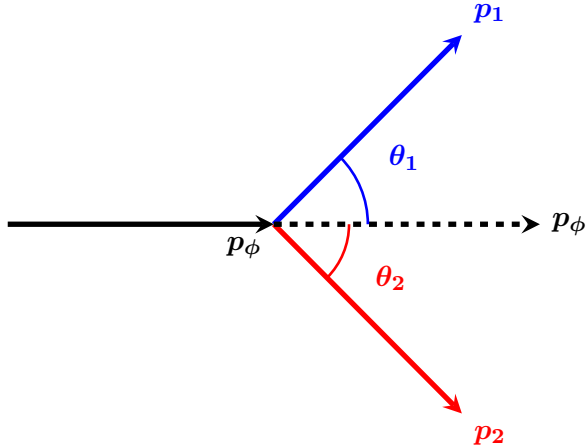
\begin{figure}[!h]
	\centering
	\begin{tikzpicture}[scale=1.25]
	\draw[line width=2pt,blue,-stealth] (2.83,0)--(4.83,2) node[anchor=south west]{$\boldsymbol{p_1}$};
	\draw[line width=2pt,red,-stealth](2.83,0)--(4.83,-2) node[anchor=north west]{$\boldsymbol{p_2}$};
	\draw[line width=2pt,black,-stealth](0,0)--(2.83,0) node[anchor=north east]{$\boldsymbol{p_\phi}$};
	\draw[line width=2pt,black,-stealth, dashed](2.83,0)--(5.66,0) node[anchor=west]{$\boldsymbol{p_\phi}$};
	\draw[line width=1pt,blue] (3.83,0) arc (0:45:1) node[right=5mm]{$\boldsymbol{\theta_1}$};
	\draw[line width=1pt,red] (3.63,0) arc (0:-45:0.8) node[right=5mm]{$\boldsymbol{\theta_2}$};
	\end{tikzpicture}
	\caption{Schematic representation of the $\phi$ boson decay and angles between the Higgs and its decay products.}
	\label{fig:scheme}
\end{figure}

One of the main challenges of the kinematic reconstruction in the low mass regime is that the jets resulting from the $\phi$ boson decay may overlap in the detector and appear as one single jet. 
This effect causes a significant loss of sensitivity of the analysis as can be inferred from Figure~\ref{fig:deltaRs}, which shows the 
$\Delta R$~\footnote{$\Delta R\equiv\sqrt{\Delta \phi^2+\Delta \eta^2}$, where $\Delta \phi \, (\Delta \eta)$ correspond to the difference in the azimuthal angle (pseudo-rapidity) of two objects.} 
between the decay products of the $\phi$ boson. It is clear that the $b$ and $\bar{b}$ quarks from the Higgs decay, labelled $b_\phi$ and $\bar{b}_\phi$ respectively, get progressively close to each other as the $\phi$ boson mass decreases. This is particularly true for the CP-odd signals (Figure~\ref{fig:deltaRs} right), at least in this mass regime. To understand the implications of this overlap, across the different signal mass samples, we set the cone size in the jet reconstruction algorithm to $\Delta R=0.7$, which is slightly larger than the value usually used by 
ATLAS and CMS i.e., 0.4 and 0.5, respectively. This will increase the number of events with a single jet topology formed from the decay products of the Higgs boson, hence will allow us to better understand this single jet population of events, even for the signal samples with higher masses ($\sim$40~GeV). 
Thus, in order to correctly identify the jet(s) coming from the $\phi$ boson, a new approach to the kinematic reconstruction was used, different from the one considered in~\cite{Azevedo:2020vfw}. 
We start by computing the mass of each individual jet reconstructed in the event and also the mass of each pair of jets. For each two jet ($j_1$, $j_2$) combination, besides the invariant mass calculation ($m_{j_1j_2}$), an additional mass value is computed, $m^{(1)}_\phi$, using the following equation~\footnote{It is assumed that $p_1$ ($p_2$) $\gg$ $m_{j_1}$ ($m_{j_2}$), where $m_{j_1}$ ($m_{j_2}$) corresponds to the mass of $j_1$ ($j_2$).}

\begin{equation}
	\begin{aligned}
		m^{(1)}_\phi = p_1 \, \sqrt{2 \, \frac{\sin \theta_1}{\sin \theta_2} \bigg(1 - \cos (\theta_1 + \theta_2)\bigg)} \, \, .  \\ 
	\end{aligned}
	\label{eq:mass1}
\end{equation}
Here, $p_1$ corresponds to the magnitude of the 3-momentum of $j_1$, and $\theta_1$ ($\theta_2$) is the angle between the 3-momentum vectors of $j_1$ ($j_2$), with respect to the total 3 momentum ($p_\phi$) of the $j_1 + j_2$ system (see Figure~\ref{fig:scheme}; an identical mass can be obtained by interchanging the indices 1 and 2). It should be stressed that if the $t\bar{t}$ momentum would be assumed as recoiling against the momentum of the Higgs boson, this would allow an additional mass value to be available for the reconstruction. In the end, out of the three methods described, the one that gives the closest reconstructed mass to the input value (labelled as the best of all methods or best method) is chosen, in each event. The jets (or jet) used by the best method are the ones associated by the kinematic reconstruction to the Higgs boson decay partons. 

%~\footnotemark
%\footnotetext{$\Delta R\equiv\sqrt{\Delta \phi^2+\Delta \eta^2}$, where $\Delta \phi \, (\Delta \eta)$ correspond to the difference in the azimuthal angle (pseudo-rapidity) of two objects.}  

%\begin{equation}
%\begin{aligned}
%m^{(2)}_\phi = p_2 \, \sqrt{2 \, \frac{\sin \theta_2}{\sin \theta_1} \bigg(1 - \cos (\theta_1 + \theta_2)\bigg)} \, \,  \\
%\end{aligned}
%\label{eq:mass2}
%\end{equation}

\begin{table}[hbt!] 
	\renewcommand{\arraystretch}{1.35}
	\begin{center}
		\begin{tabular}{|c|cc|cc|cc|}
			\hline 
			\multirow{2}{2.cm}{\centering $m_\phi$ (GeV)} &
			\multicolumn{2}{c|}{$m^{inv}_\phi$ (1 jet)}     & \multicolumn{2}{c|}{$m^{inv}_\phi$ (2 jets)} 	& \multicolumn{2}{c|}{$m^{(1)}_\phi$} \\
			& $\phi = H$  & $\phi = A$ & $\phi = H$ & $\phi = A$ & $\phi = H$ & $\phi = A$ \\ \hline
			$ 12$ & 99.96 & 99.98 & 0.00  & 0.00   & 0.04  & 0.02  \\
			$ 20$ & 94.05 & 96.94 & 0.58  & 0.21   & 5.37  & 2.85  \\
			$ 30$ & 63.52 & 76.37 & 10.64 & 7.72   & 25.84 & 15.91 \\
			$ 40$ & 27.76 & 44.27 & 33.74 & 27.20  & 38.50 & 28.53 \\ \hline
		\end{tabular}
		\caption{Efficiencies (in \%), rounded to two decimal places, of the three methods used to reconstruct the $\phi$ boson. $m^{inv}_\phi$ (1 jet) is the invariant	mass from 1 jet only, $m^{inv}_\phi$ (2 jets) is the invariant mass from 2 jets, and $m^{(1)}_\phi$ is the mass from Equation~\ref{eq:mass1}.}
		\label{table:eff_massRec}
	\end{center}
\end{table}

\begin{figure}[h!]
	\begin{center}
			\hspace*{-1mm}\includegraphics[height = 6.6cm]{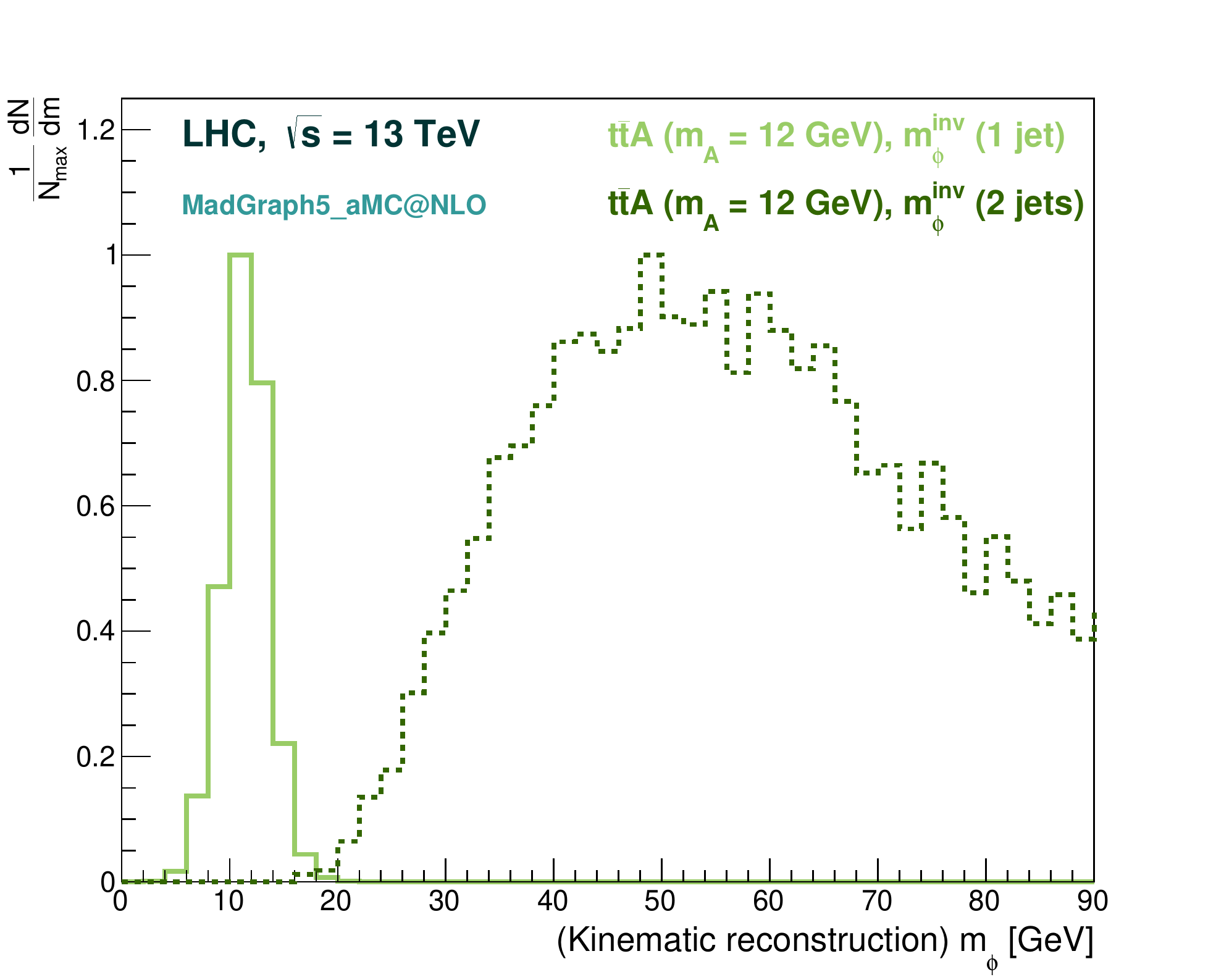}
			\hspace*{-4mm}\includegraphics[height = 6.6cm]{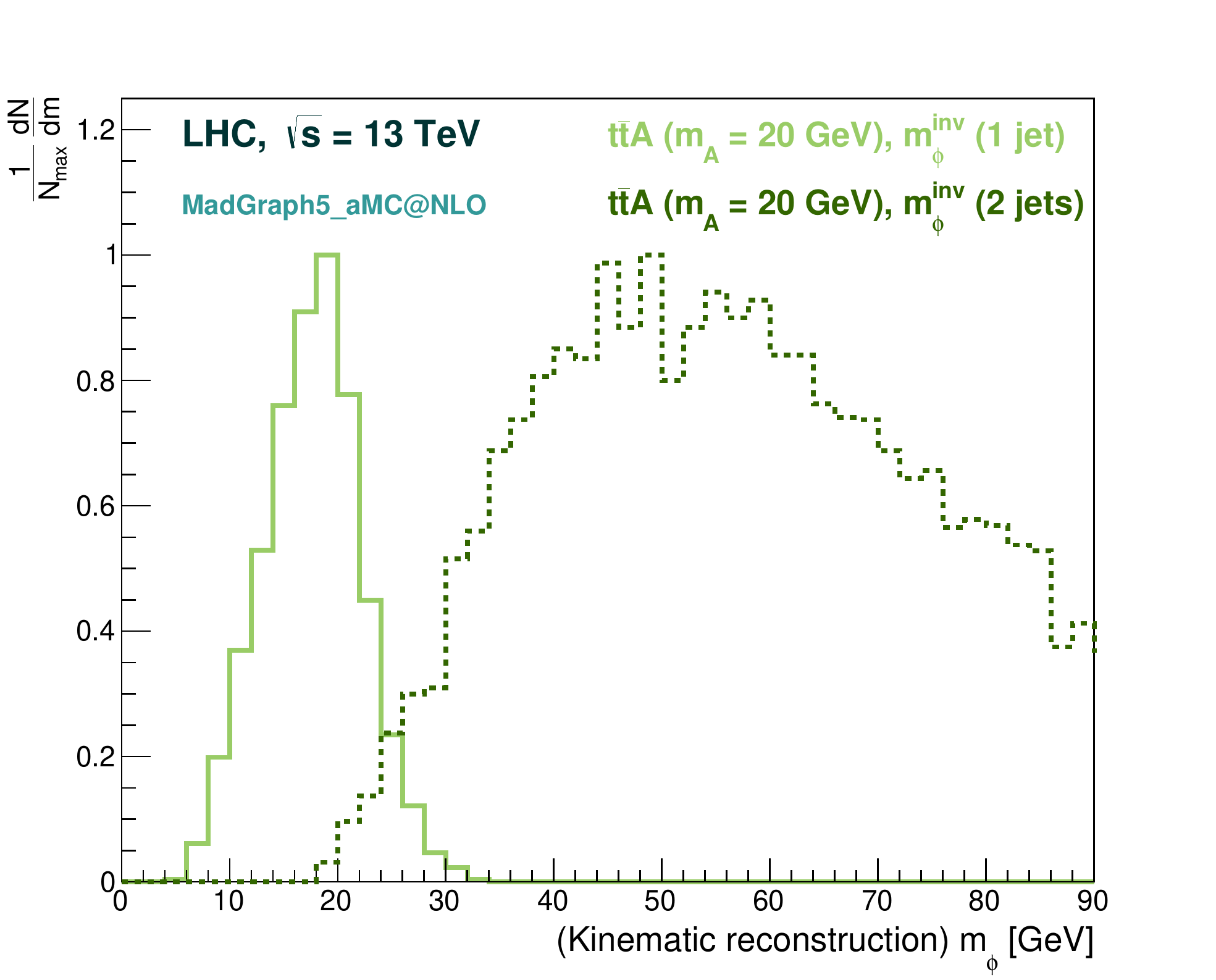}
			\\ 
			\hspace*{-1mm}\includegraphics[height = 6.6cm]{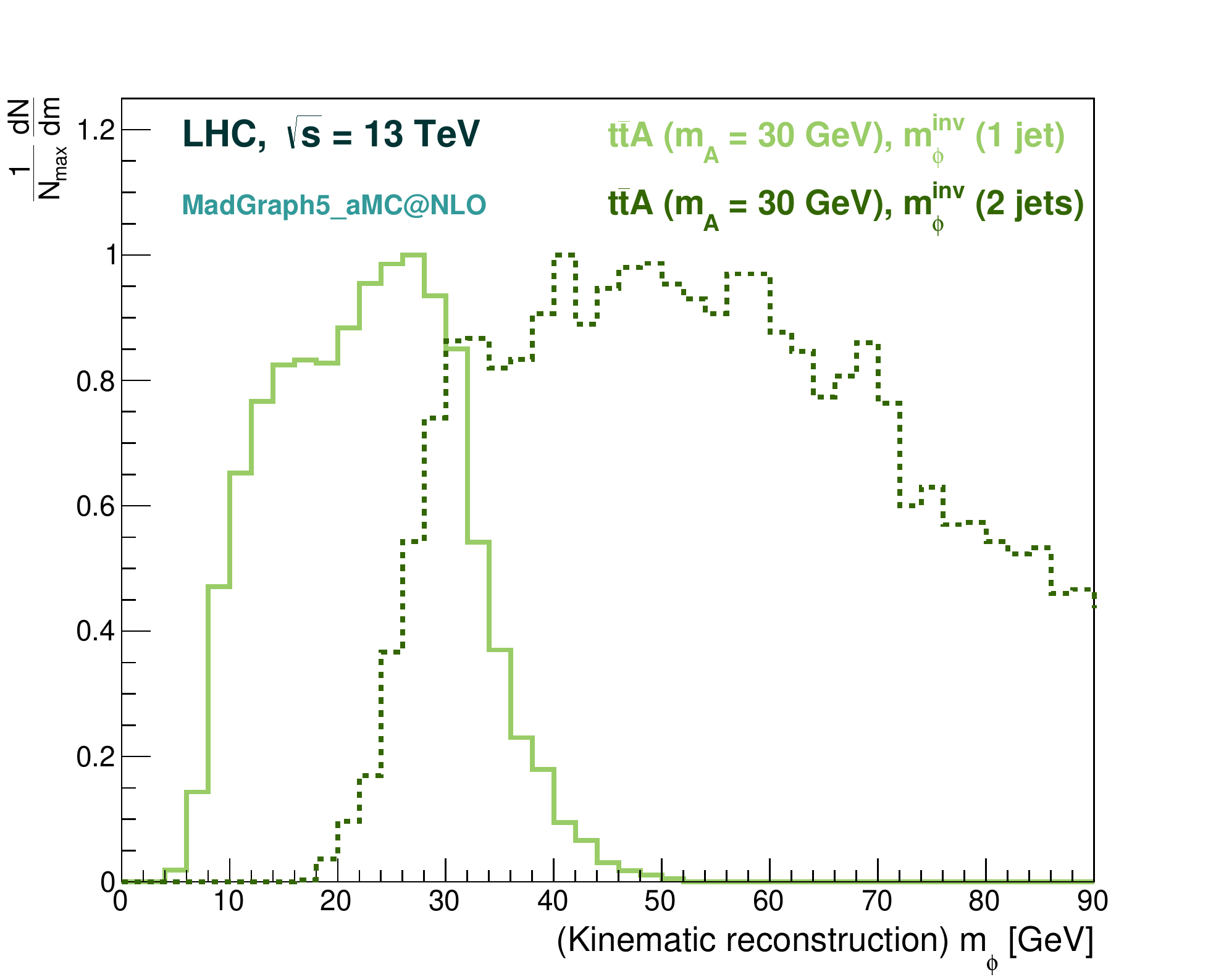}
			\hspace*{-4mm}\includegraphics[height = 6.6cm]{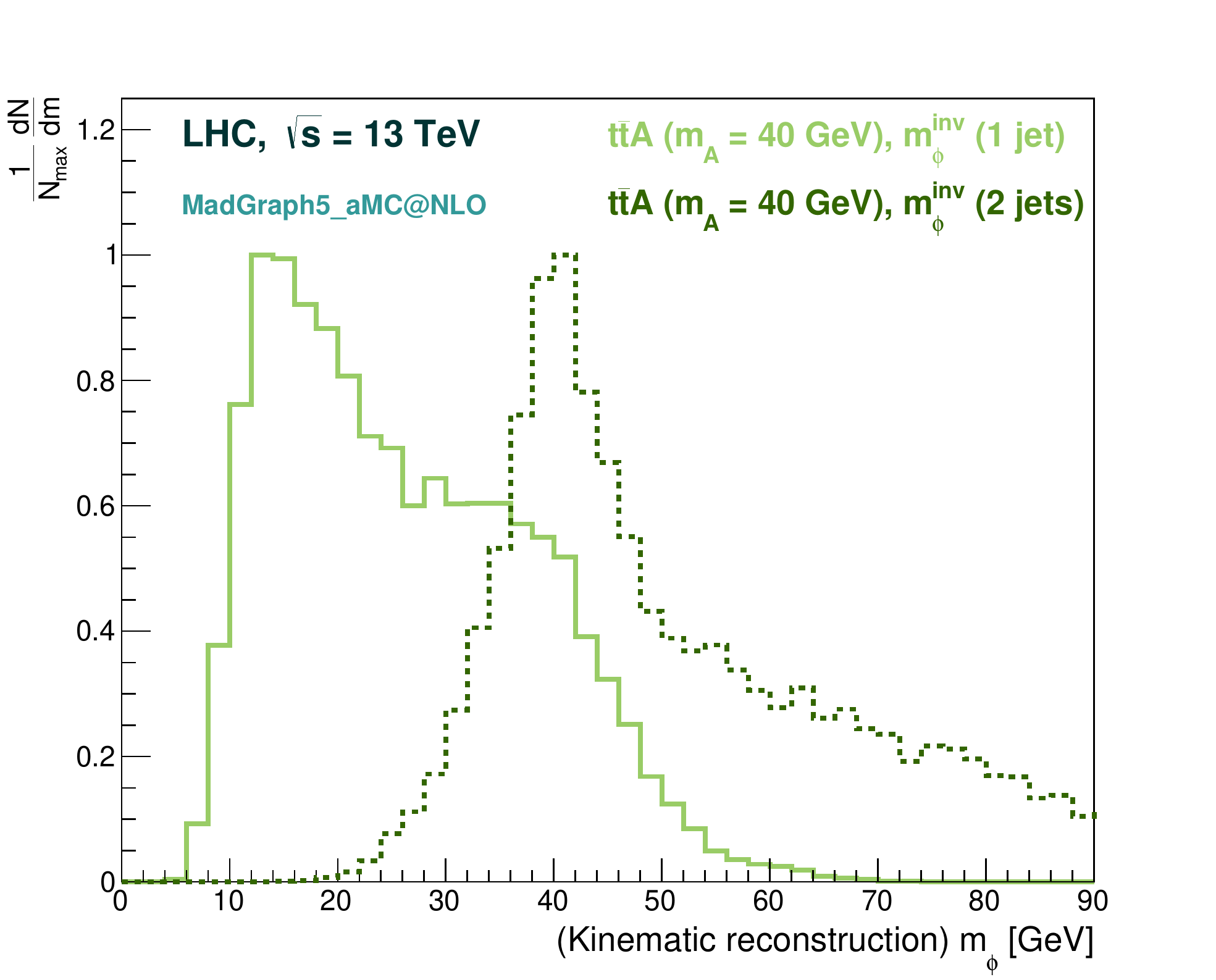}
		\caption{Higgs mass distributions after kinematic reconstruction, for $m_\phi= 12, 20, 30$ and 40~GeV, for the pseudoscalar case. The solid lines show the invariant Higgs mass from one jet, and the dashed lines the invariant mass from 2 jets. The distributions are normalised to the maximum number of events in a given bin, $N_{max}$.}
		\label{fig:1vs2jets}
	\end{center}
\end{figure}

\begin{figure}[h!]
	\begin{center}
		\hspace*{-1mm}\includegraphics[height = 6.6cm]{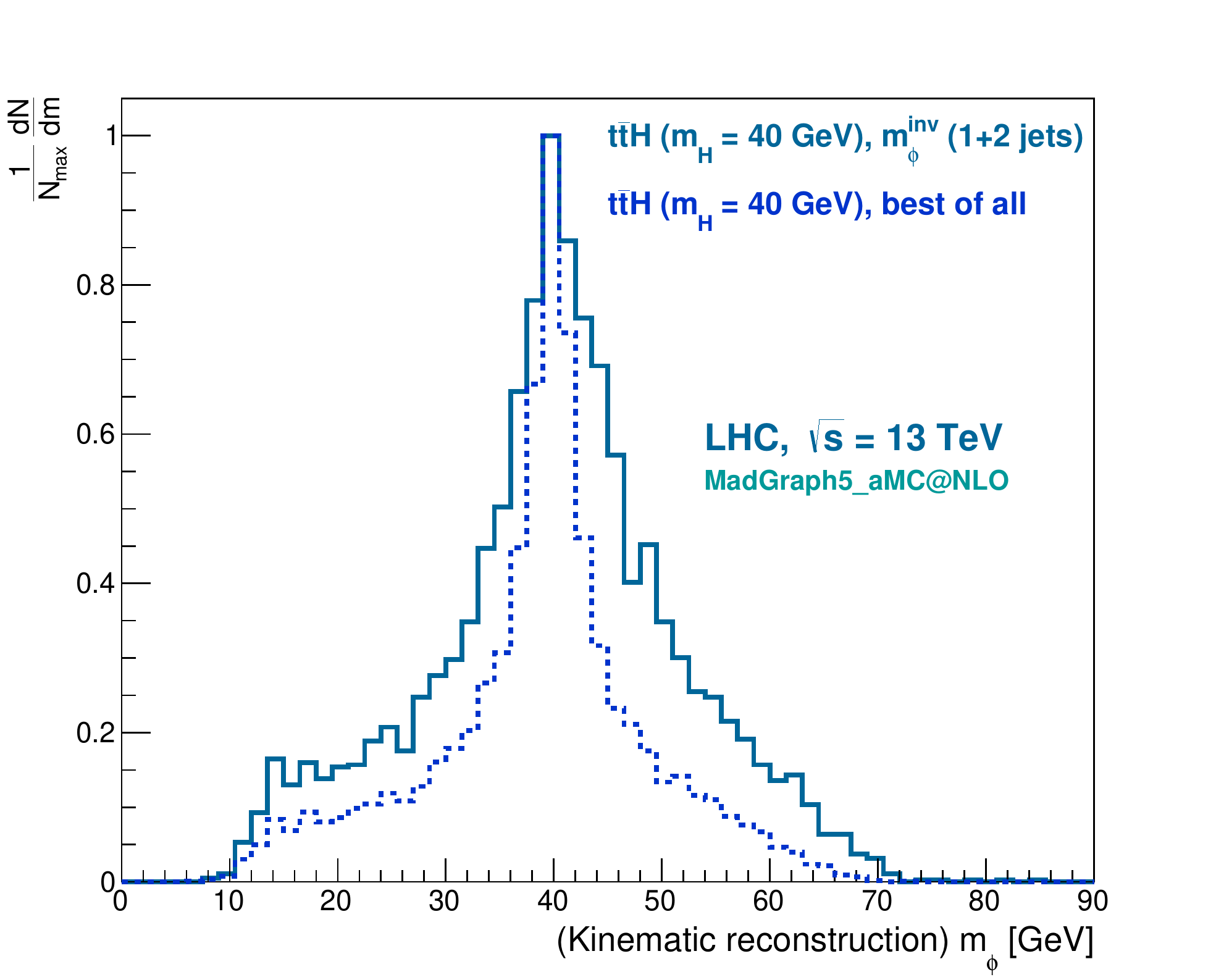}
		\hspace*{-4mm}\includegraphics[height = 6.6cm]{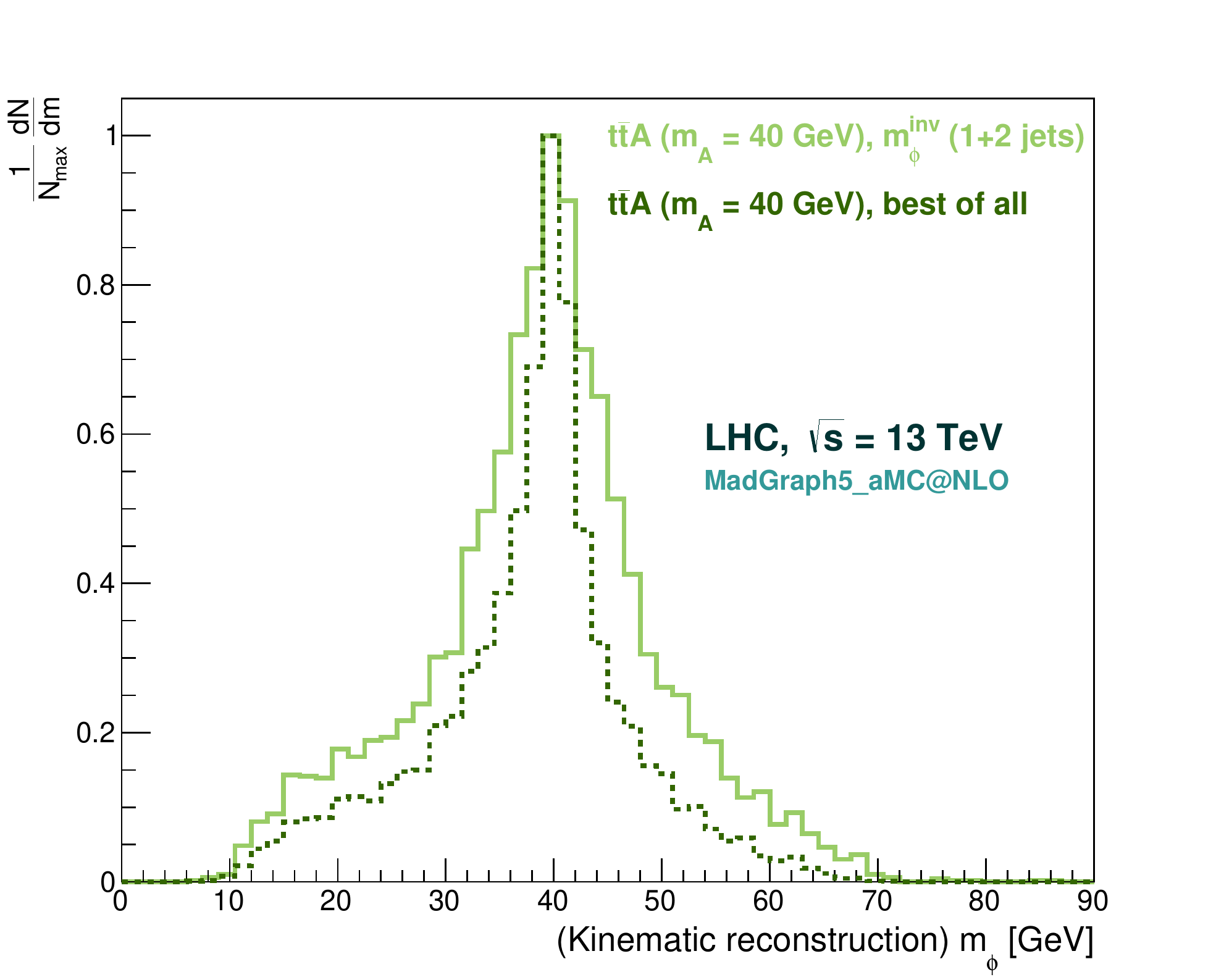}
		\caption{Higgs mass distributions after kinematic reconstruction, for $m_\phi= 40$~GeV, for the scalar (left) and pseudoscalar (right) cases. The solid line shows the best invariant Higgs mass from one or two jets i.e., $m^{inv}_\phi$ (1+2 jets) = $m^{inv}_\phi$ (1 jet) or $m^{inv}_\phi$ (2 jets), and the dashed line represents the best of all methods (best of all).}
		\label{fig:2jets_2methods}
	\end{center}
\end{figure}

\begin{figure}[h!]
	\begin{center}
		\hspace*{-5mm}\includegraphics[height = 7.5cm]{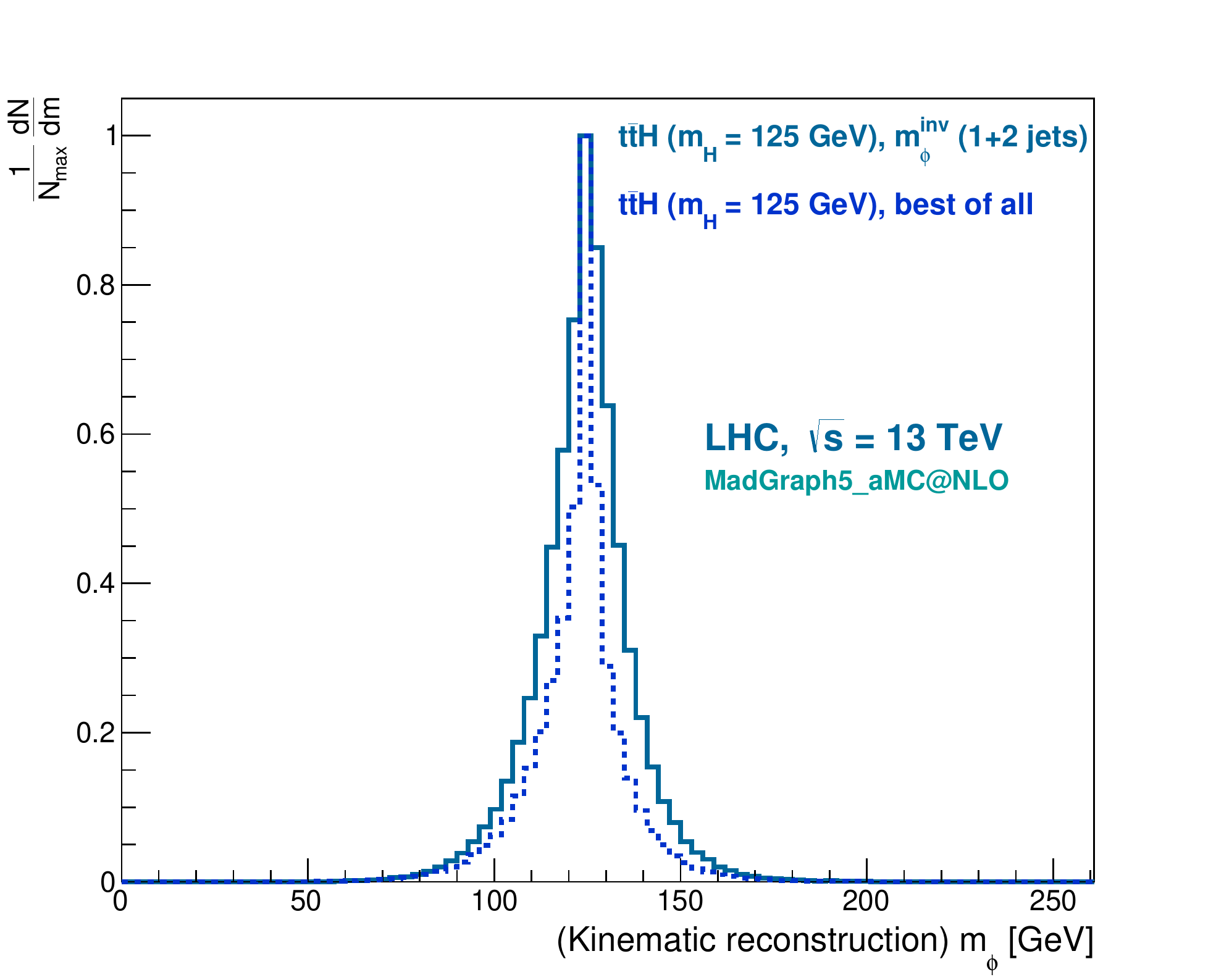}
		\caption{SM Higgs boson mass distribution after kinematic reconstruction, for $m_\phi= 125$~GeV. The solid line shows the 
		best invariant Higgs mass from one or two jets i.e., $m^{inv}_\phi$ (1+2 jets) = $m^{inv}_\phi$ (1 jet) or $m^{inv}_\phi$ (2 jets), and the dashed line represents the best of all methods (best of all), $\phi=H$.}
		\label{fig:mHiggsSM}
	\end{center}
\end{figure}

\begin{figure}[h!]
	\begin{center}
		\hspace*{-1mm}\includegraphics[height = 6.6cm]{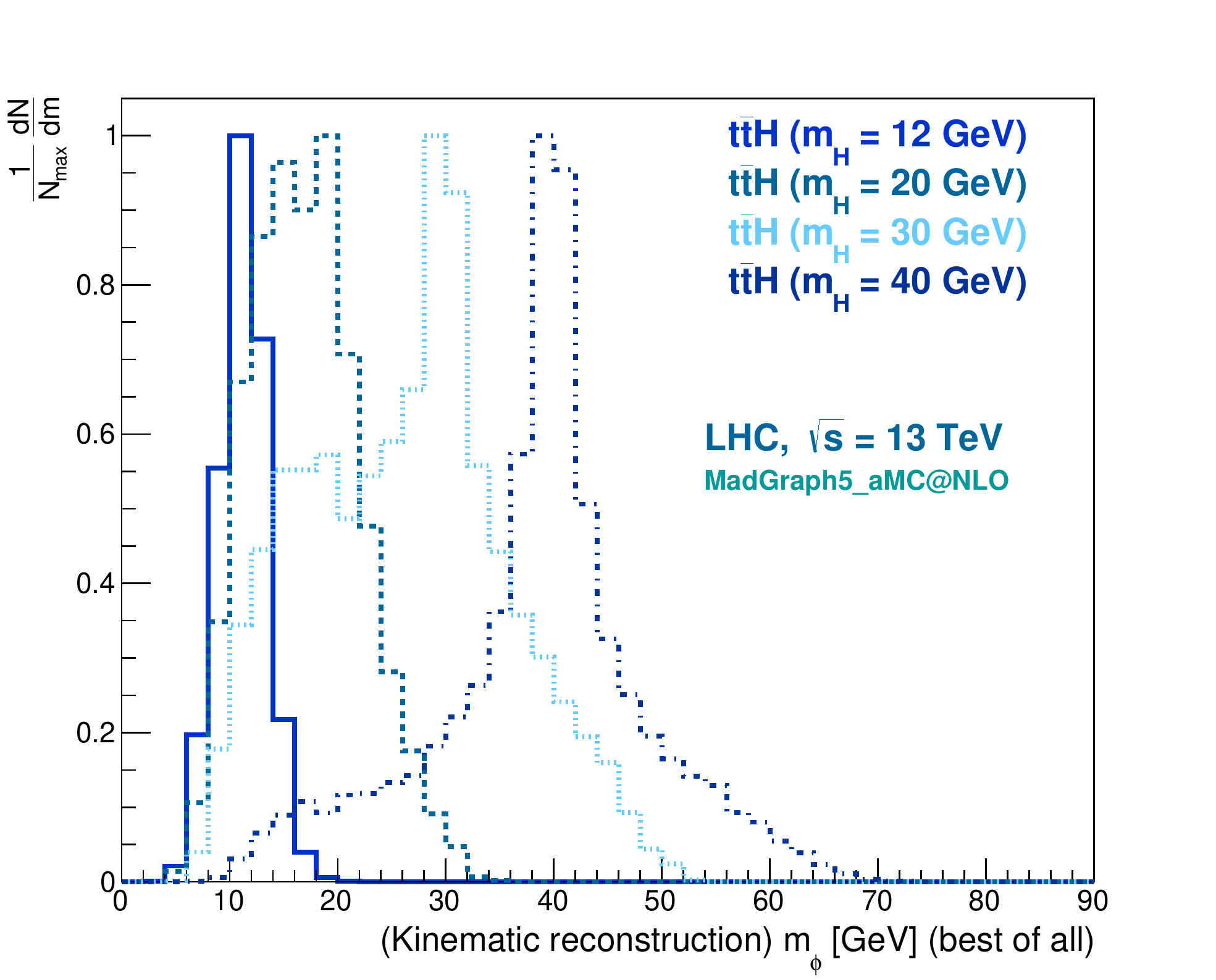}
		\hspace*{-4mm}\includegraphics[height = 6.6cm]{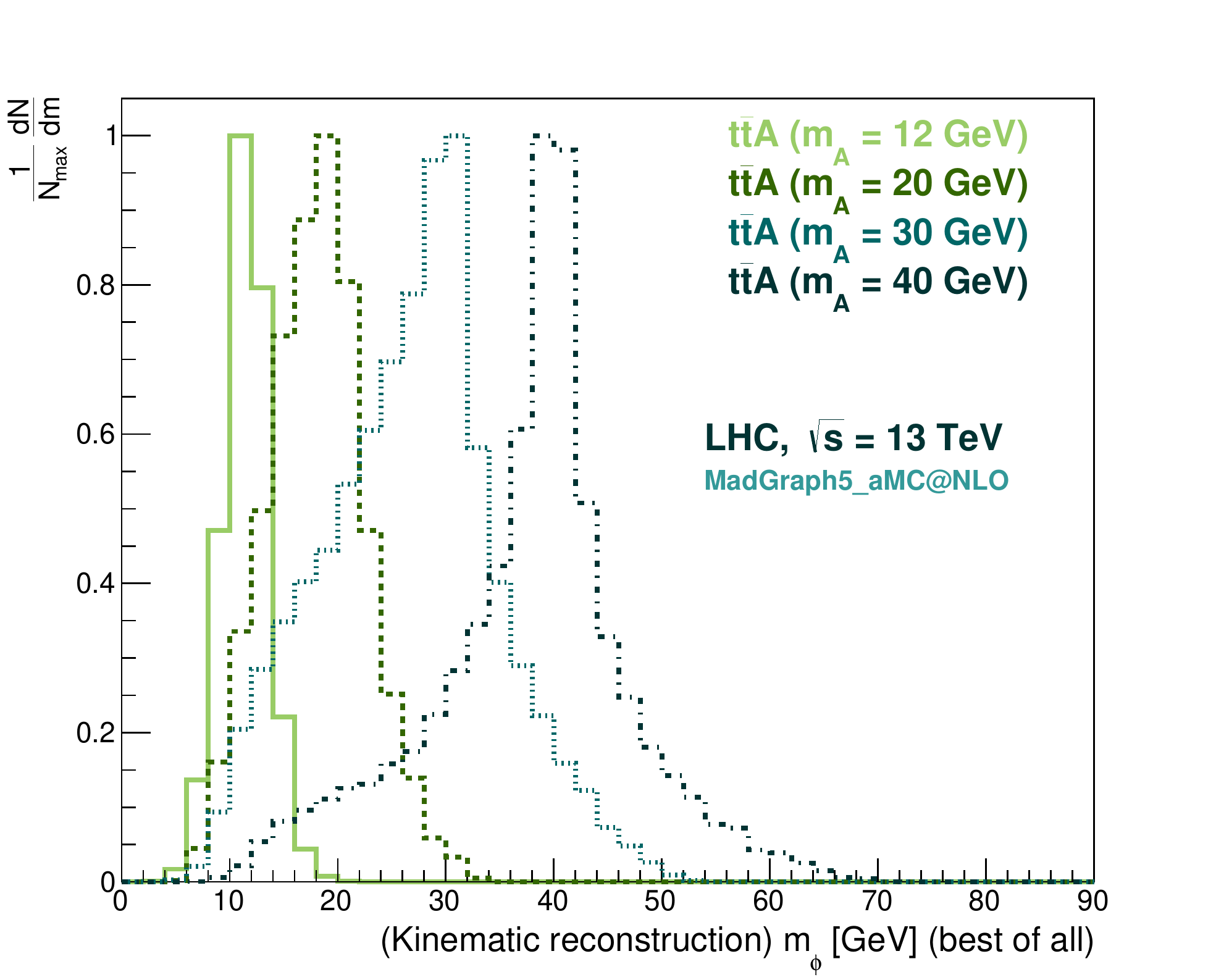}
		\caption{Higgs mass distributions after kinematic reconstruction, for $m_\phi= 12, 20, 30$ and 40~GeV, for the scalar (left) and pseudoscalar (right) cases. These are the distributions that show the best of all methods in each event.}
		\label{fig:bestMasses}
	\end{center}
\end{figure}

The efficiency is defined as the percentage of total events that survive a reconstruction method, with all its cuts. The efficiency for each method is shown in Table~\ref{table:eff_massRec}, for both the scalar and pseudoscalar cases, and for each of the Higgs masses generated. 
The percentages shown are relative to the total number of events that survived the pre-selection and the kinematic reconstruction. The vast majority of the events (more than 90\%) is best reconstructed by matching only one jet to the $\phi$ boson, for masses below 20~GeV. For $m_\phi = 30$~GeV, we still have more than 50\% of the events being better reconstructed by that method. Its efficiency drops below 50\% somewhere between $m_\phi = 30$-40~GeV. Moreover, the efficiency when only one jet is considered is consistently higher for the pseudoscalar case. All these observations are consistent with what had already been shown in Figure~\ref{fig:deltaRs}, and confirm the need for the new Higgs reconstruction that we present here, as a result of the overlap of the $\phi$ boson decay products in the low mass regime. 
For further comparisons between the different methods used to reconstruct the Higgs boson mass, we show in Figure~\ref{fig:1vs2jets} the invariant Higgs boson mass reconstructed with one jet or two jets, for the CP-odd signals. In Figure~\ref{fig:2jets_2methods}, we compare the Higgs mass reconstructed from the best one or two jets invariant mass, with the mass distribution obtained from the best of all methods in each event, for $m_\phi = 40$~GeV and both CP-even and CP-odd signals. 
An improvement in the mass resolution is clearly noticeable. For instance, the full width at half maximum (FWHM) is reduced by more than half, from 
12~GeV (13~GeV) to 5~GeV (6~GeV), for the scalar (pseudoscalar) case, when the best of all methods is used. For completeness, we show in Figure~\ref{fig:mHiggsSM} the mass distribution of the SM Higgs boson ($m_H =$ 125~GeV) when reconstructed using the best of the two jets or one jet invariant mass (solid line), and with the best method introduced in this paper (dashed line). The same improvement in the mass resolution of roughly a factor of two is observed. The reason for the improvement is directly related to the new mass reconstruction method, that takes into account the contribution from the energy (momentum) of one single jet and the angles of both which, experimentally, are better reconstructed. On the contrary, as an invariant mass calculation involves the information of the energy (momentum) of both jets, the energy resolution effects enter the calculation twice, degrading the reconstructed mass resolution. In Figure~\ref{fig:bestMasses}, the $\phi$ boson mass distributions that are obtained by picking the best of all methods in each event are shown, again for the CP-even and CP-odd cases. In all figures discussed in this paragraph, the distributions are shown after kinematic reconstruction.

For the reconstruction of the $t\bar{t}$ system and correct identification of the jets coming from the top quarks decays, we rely on a multivariate analysis method using~\texttt{TMVA}~\cite{2007physics...3039H} 
to assign those jets to their parton level counterparts. Two samples labelled as signal and background were created from simulated $t \bar{t} \phi$ signal events and used for training and testing. 
While signal samples contain kinematic distributions only from the correct (parton level) association, background samples contain equivalent kinematic distributions from wrong associations. 
The following variables were used for training the methods: $\Delta R$, $\Delta \Phi$, $\Delta \theta$ and the invariant mass for the pairs $(b_t,l^+)$, $(b_t,l^-)$, $(b_t,\bar{b}_{\bar{t}})$, $(\bar{b}_{\bar{t}} , l^+)$, and $(\bar{b}_{\bar{t}}, l^-)$, where $b_t$ ($\bar{b}_{\bar{t}}$) represents the bottom (anti-bottom) quark from the top (anti-top) decay and $l^+$ ($l^-$) is the positive (negative) lepton from the $W^+$ ($W^-$) boson decay. %parton level, unico corte e |MJ1 - MJ2| < 20 
The \texttt{TMVA} method used was the Boosted Decision Tree BDTD. %with decorrelation and Adaptive boost
The jet combination chosen is the one returning the highest value of the BDTD discriminant.  In events with jet multiplicity above ten, only the ten highest $p_T$ jets are considered. 
Jet combinations also need to verify loose selections i.e., $m_{l^+ b_t} (m_{l^- \bar{b}_{\bar{t}}})< 150$~GeV and $m_{b_\phi \bar{b}_\phi} < 300$~GeV, in order to prevent reconstruction of non physical regions of the phase space.

Following the pairing of jets and leptons, the reconstruction of the $t\bar{t}$ system (which includes the neutrinos, the $W^{\pm}$ bosons and the $t$ and $\bar{t}$ quarks), is performed in the same way as in~\cite{Azevedo:2020vfw}. It uses the masses of the $W^{\pm}$ bosons ($m_W=80.4$~GeV) and the top quarks ($m_t=173$~GeV) as input constraints to the particular  combination of jets and leptons i.e., $\ell^{\pm}\oplus\nu_\ell$ and $jet\oplus\ell^{\pm}\oplus\nu_\ell$, that gives masses closer to the input values, respectively. The only difference from the previous analysis is that the likelihood function used to pick the best solution for the $t\bar{t}$ system does not take into account the reconstructed mass of the Higgs boson (see Equation~\ref{eq:likelihood}), 

\begin{equation}
L_{t\bar{t}\phi}\propto \frac{1}{p_{T_\nu} p_{T_{\bar{\nu}}} } P(p_{T_\nu})P(p_{T_{\bar{\nu}} })P(p_{T_t })P(p_{T_{\bar{t}} })P(p_{T_{t\bar{t}} })P(m_t, m_{\bar{t}}) ,
\label{eq:likelihood}
\end{equation}

\noindent
where $P(p_{T_\nu})$, $P(p_{T_{\bar{\nu}} })$, $P(p_{T_t })$, $P(p_{T_{\bar{t}} })$, $P(p_{T_{t\bar{t}} })$ are the probability distribution functions ({\it p.d.f.}s) from the transverse momenta of the neutrinos, the top quarks and the $t\bar{t}$ system, respectively. Furthermore, $P(m_{t},m_{\bar{t}})$ is the two-dimensional (2D) mass {\it p.d.f.}  of the $t\bar{t}$ pair. 
All distributions are obtained at parton level.
We have checked, after event selection, that the reconstruction efficiency varies from 52\% (45\%) to 54\% (50\%), for scalars (pseudoscalars) corresponding to $\phi$ masses in the range 12~GeV to 40~GeV. 

\begin{figure*}
\vspace*{-1cm}	
\begin{center}
		\begin{tabular}{ccc}
			\hspace*{-3mm}\epsfig{file=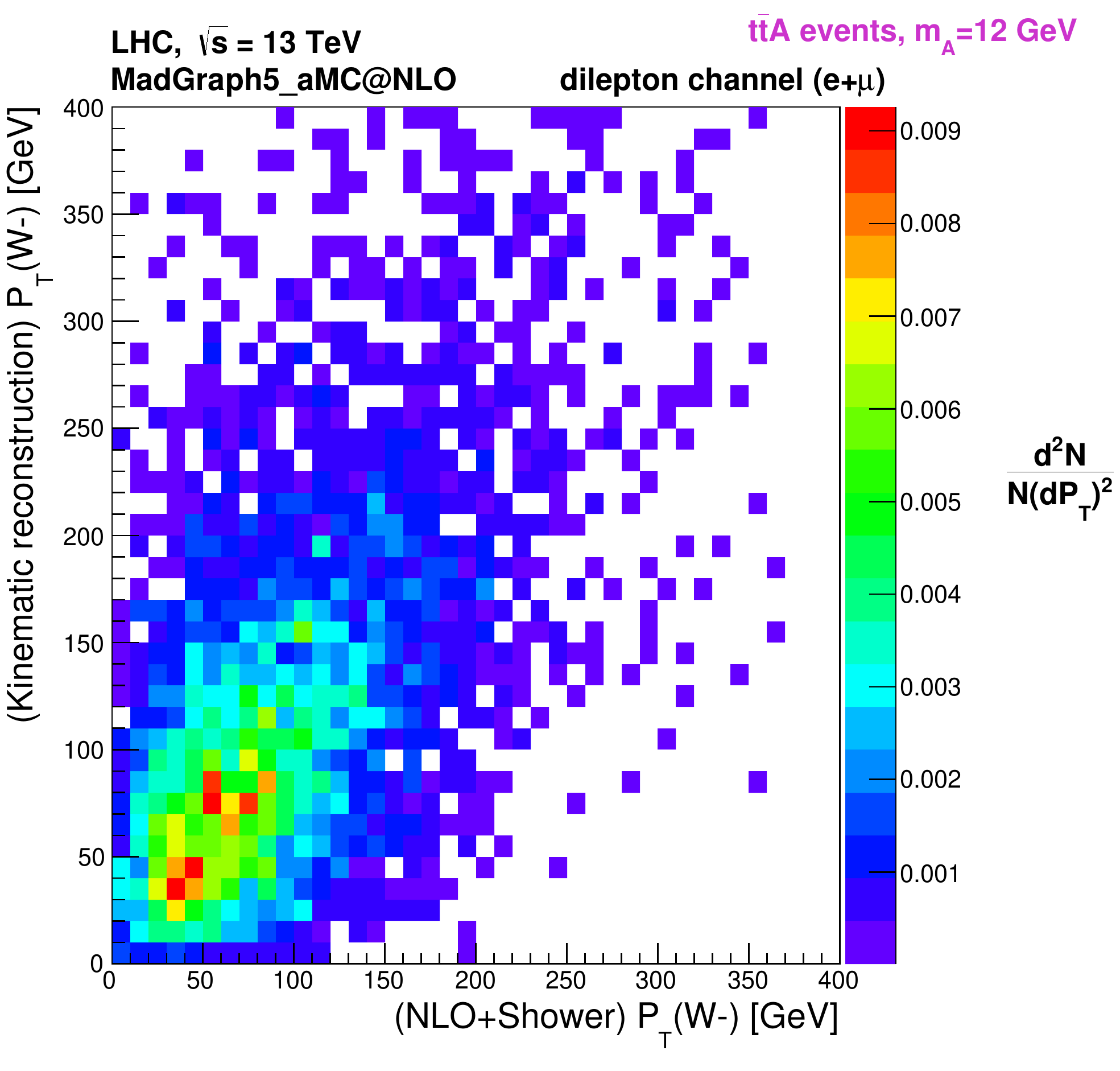,height=4.5cm,width=7.8cm,clip=} \epsfig{file=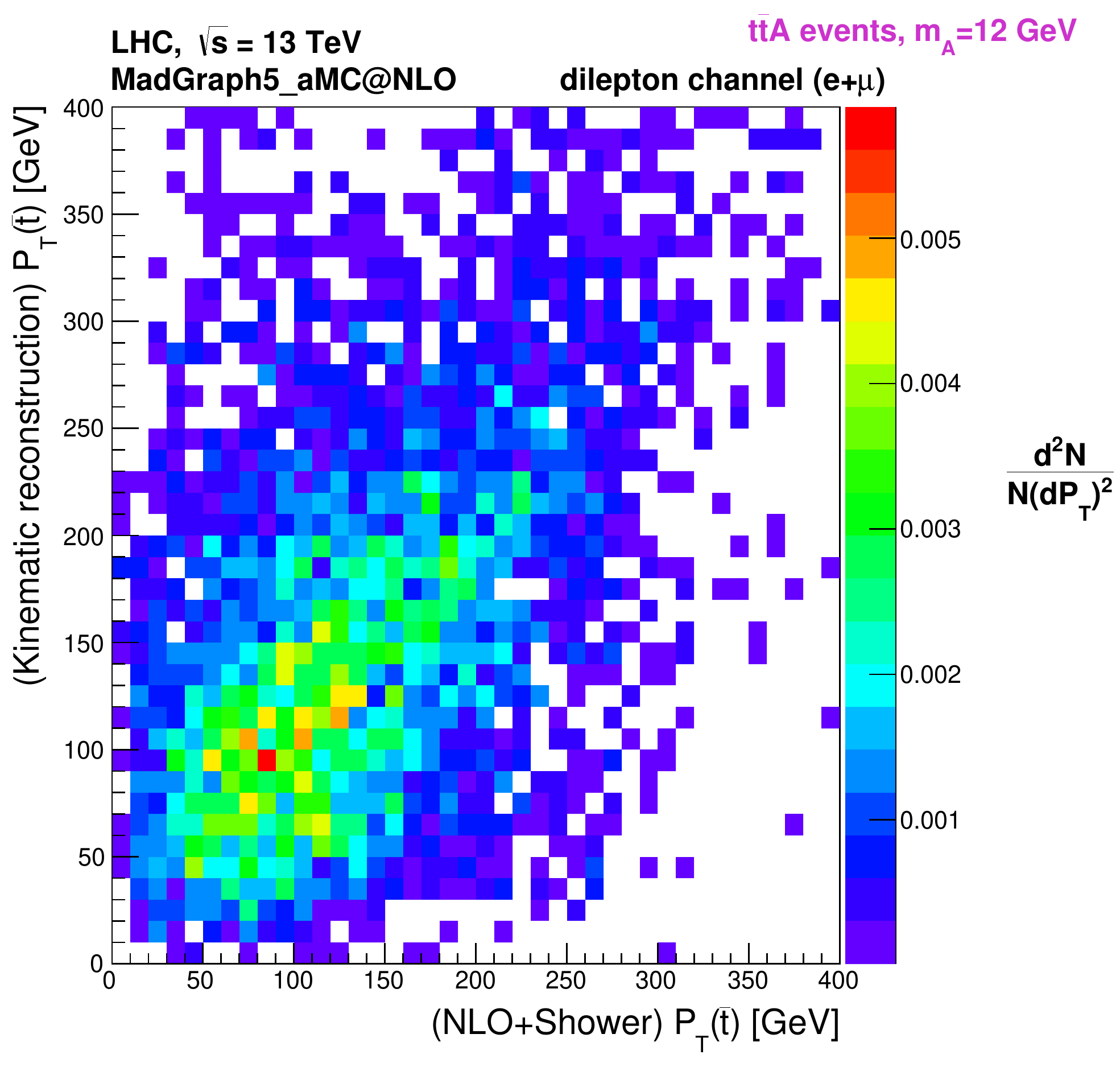,height=4.5cm,width=7.8cm,clip=} \\%[-2mm]
			\hspace*{-3mm}\epsfig{file=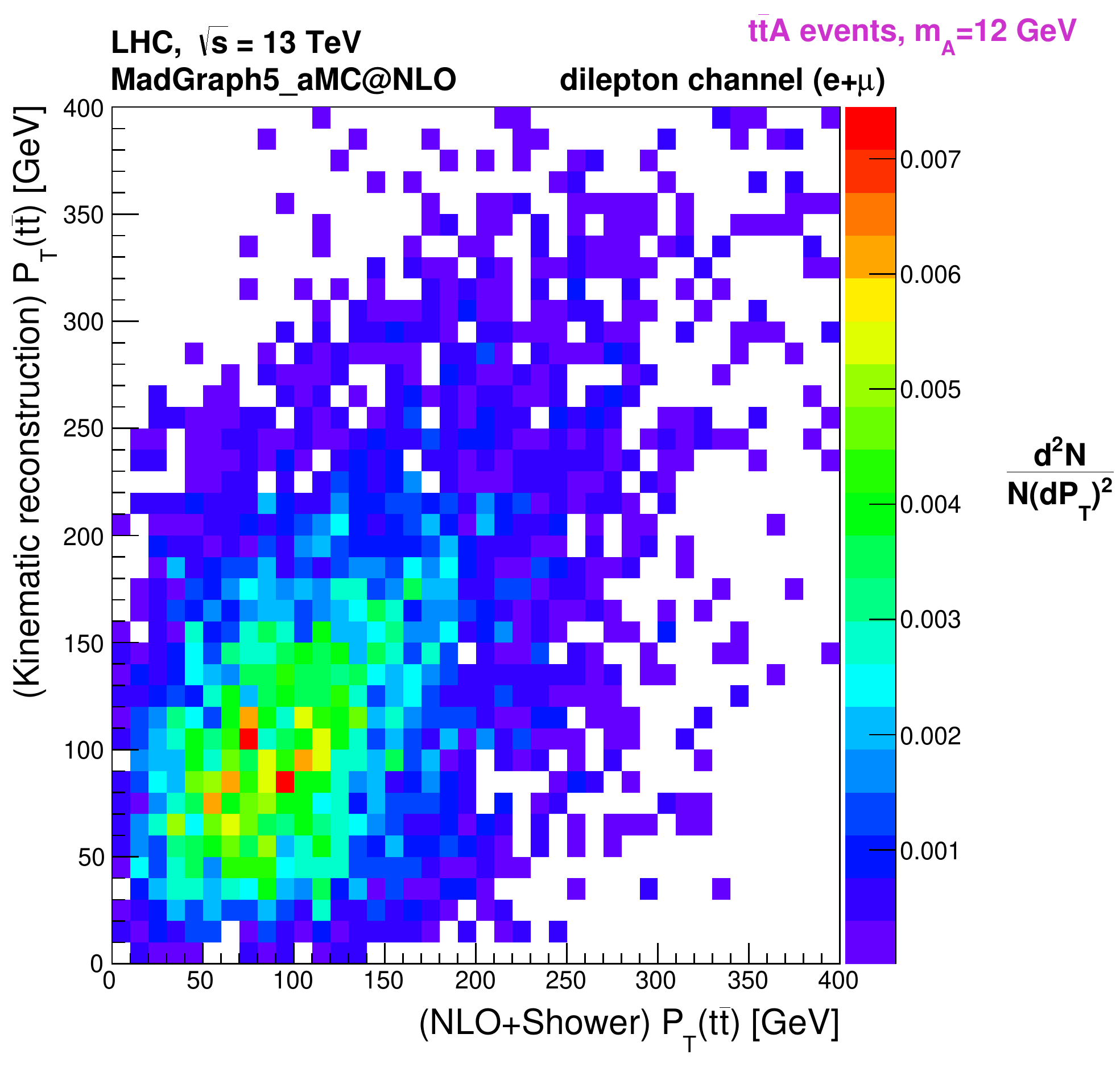,height=4.5cm,width=7.8cm,clip=} \epsfig{file=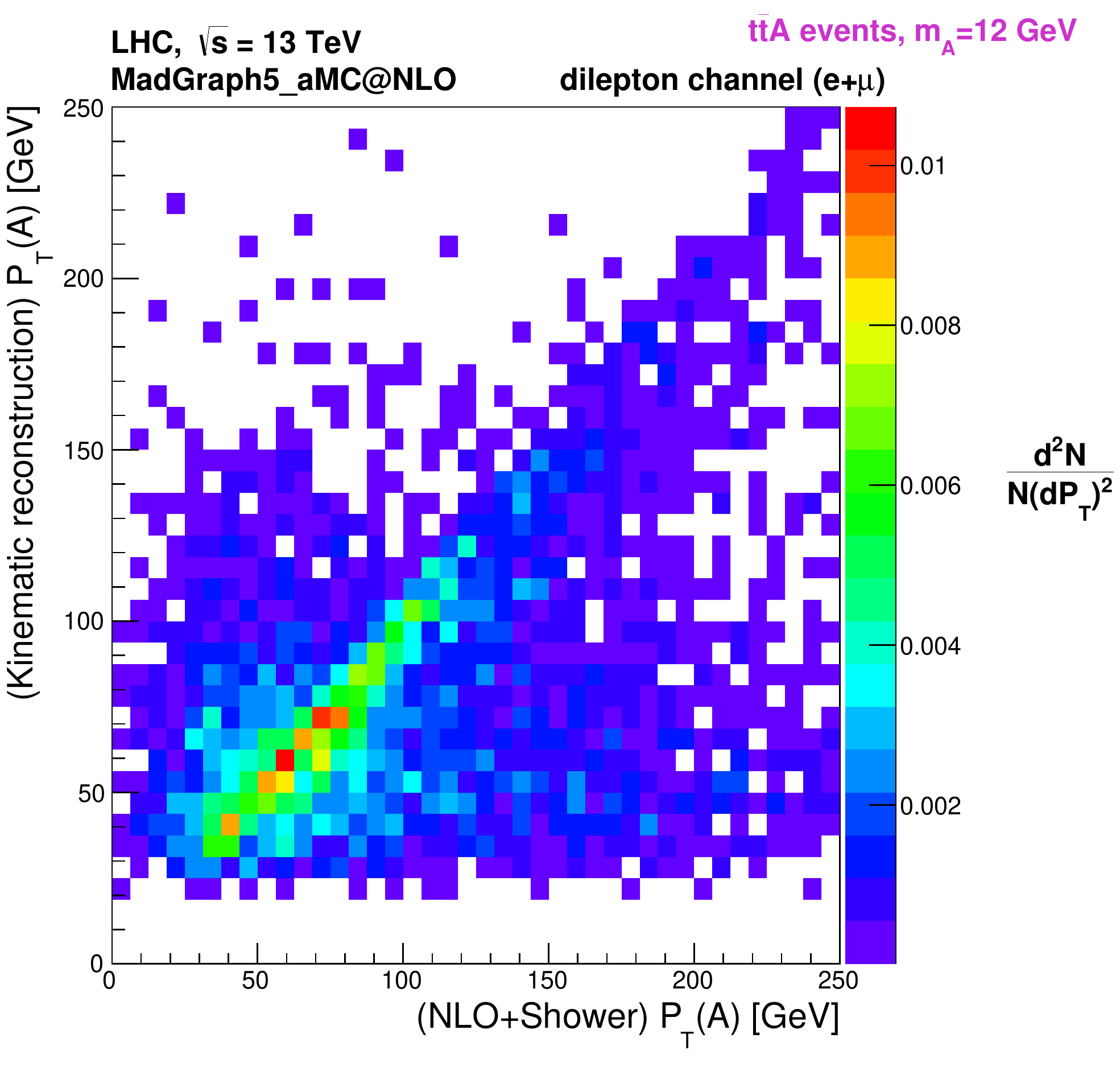,height=4.5cm,width=7.8cm,clip=} \\[-2mm]
		\end{tabular}
		\caption{Two-dimensional distributions of the transverse momentum ($p_T$) in $t\bar{t}A$ events. Variables at NLO+Shower ($x$-axis) are represented against corresponding ones after kinematic reconstruction ($y$-axis). The $p_T$ of the  $W^-$ (top-left), of the $\bar{t}$ quark (top-right), of the $t\bar{t}$ system (bottom-left) and of $A$ (bottom-right), are shown. All distributions are shown for a Higgs mass of 12~GeV.}
		\label{fig:genexp2D}
	\end{center}
\end{figure*}

Figure~\ref{fig:genexp2D} shows two-dimensional $p_T$ distributions of the $W^-$ (top-left), the anti-top quark (top-right), the $t\bar{t}$ system (bottom-left) and the Higgs boson (bottom-right) after kinematic reconstruction of $t\bar{t}A$ events, for $m_A = 12$~GeV. Similar distributions were obtained for the $W^+$ and top quark. The correlation between the parton level ($x$-axis) and reconstructed ($y$-axis) $p_T$ distributions is clearly visible, showing that the kinematic reconstruction, even without any optimisation, effectively recovers the properties of the events and, in particular, the Higgs boson. The same behaviour is observed for the $t\bar{t}H$ signals, as well as for the other scalar boson masses considered. The choice of the 12~GeV case was made to show that even for the lowest Higgs mass, the reconstruction is possible. 

\begin{figure*}
\vspace*{-0.5cm}	
	\begin{center}
		\begin{tabular}{ccc}
			\hspace*{-3mm}\epsfig{file=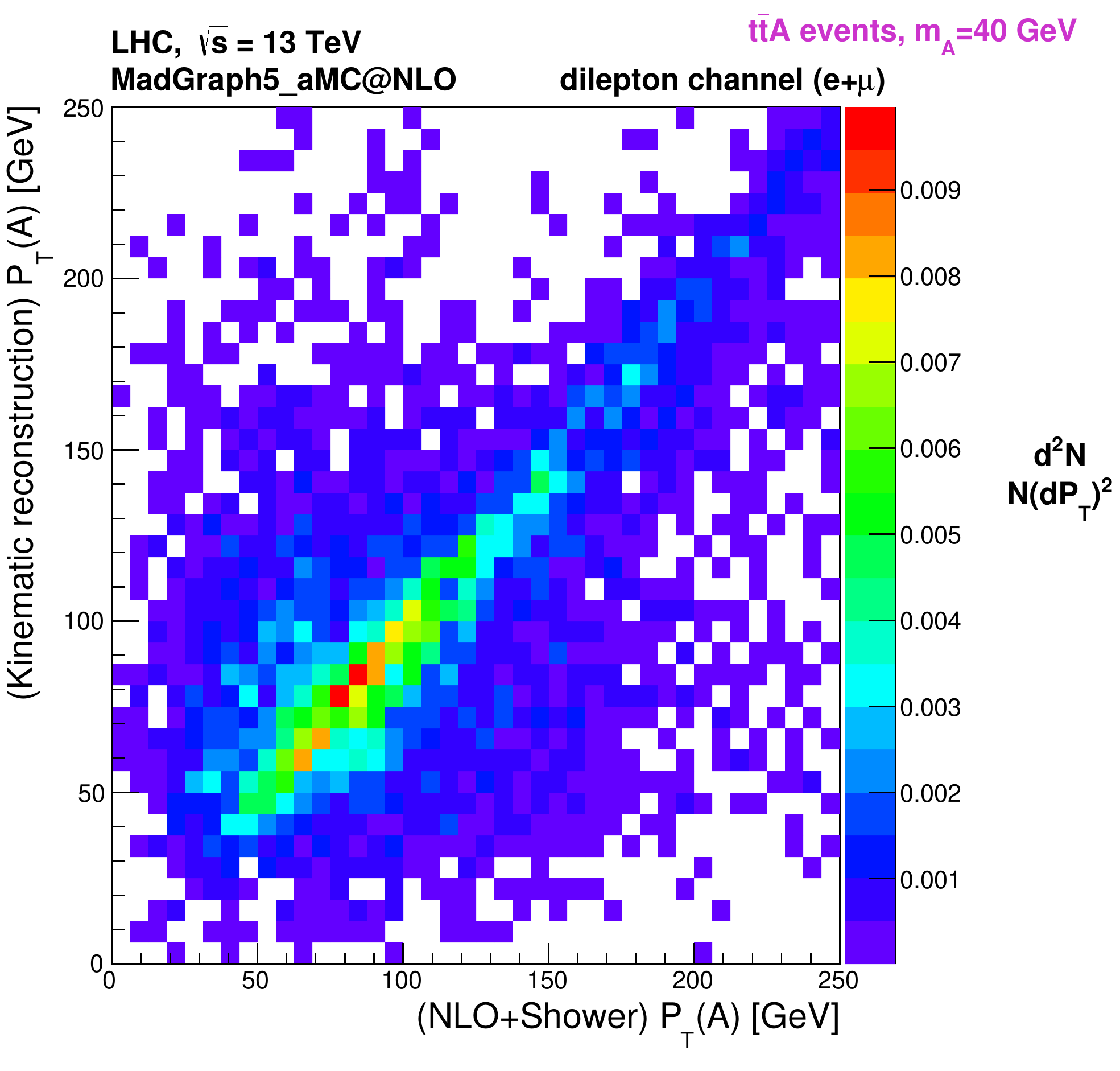,height=4.5cm,width=7.8cm,clip=} \epsfig{file=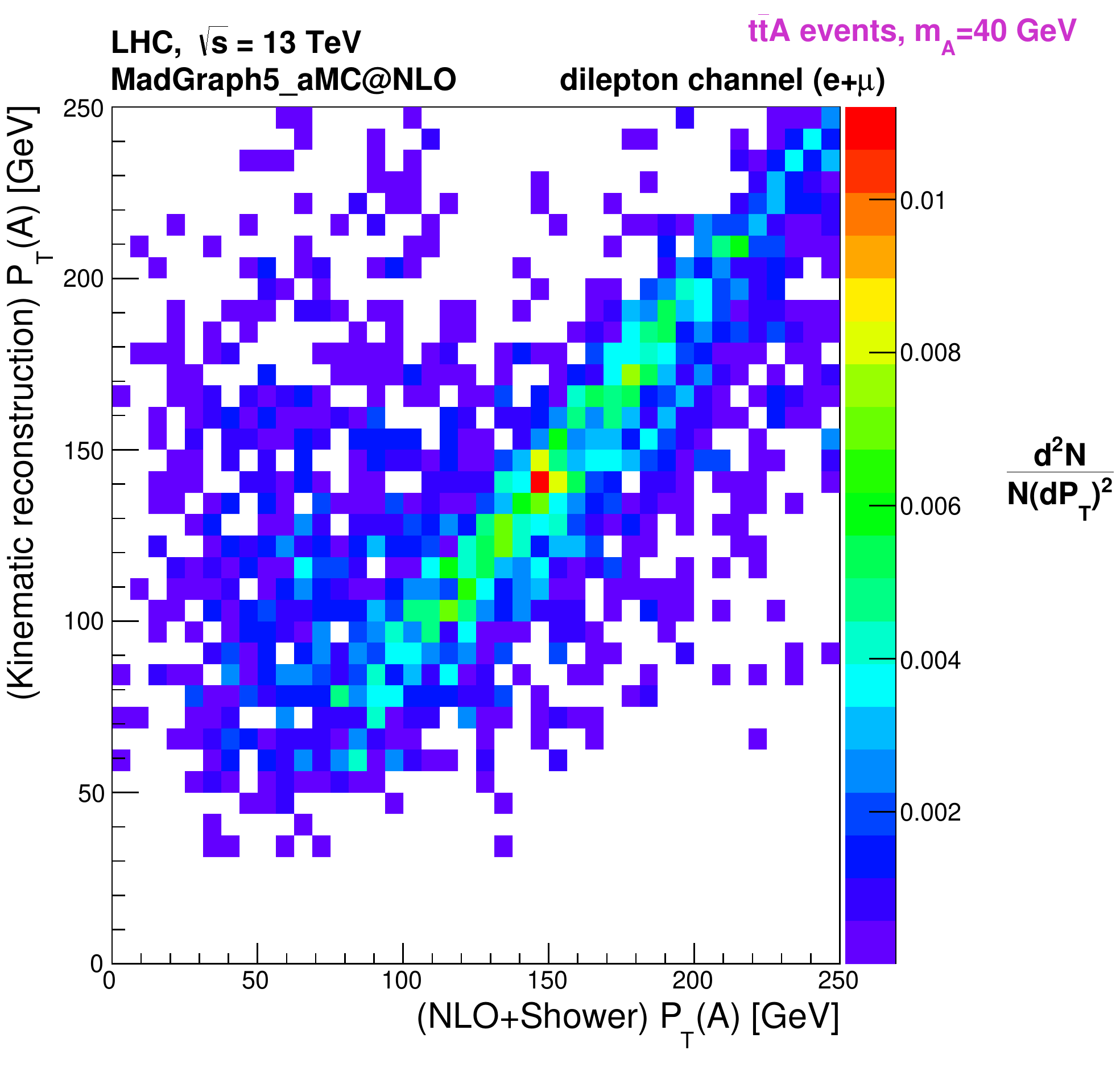,height=4.5cm,width=7.8cm,clip=} \\%[-2mm]
			\hspace*{-3mm}\epsfig{file=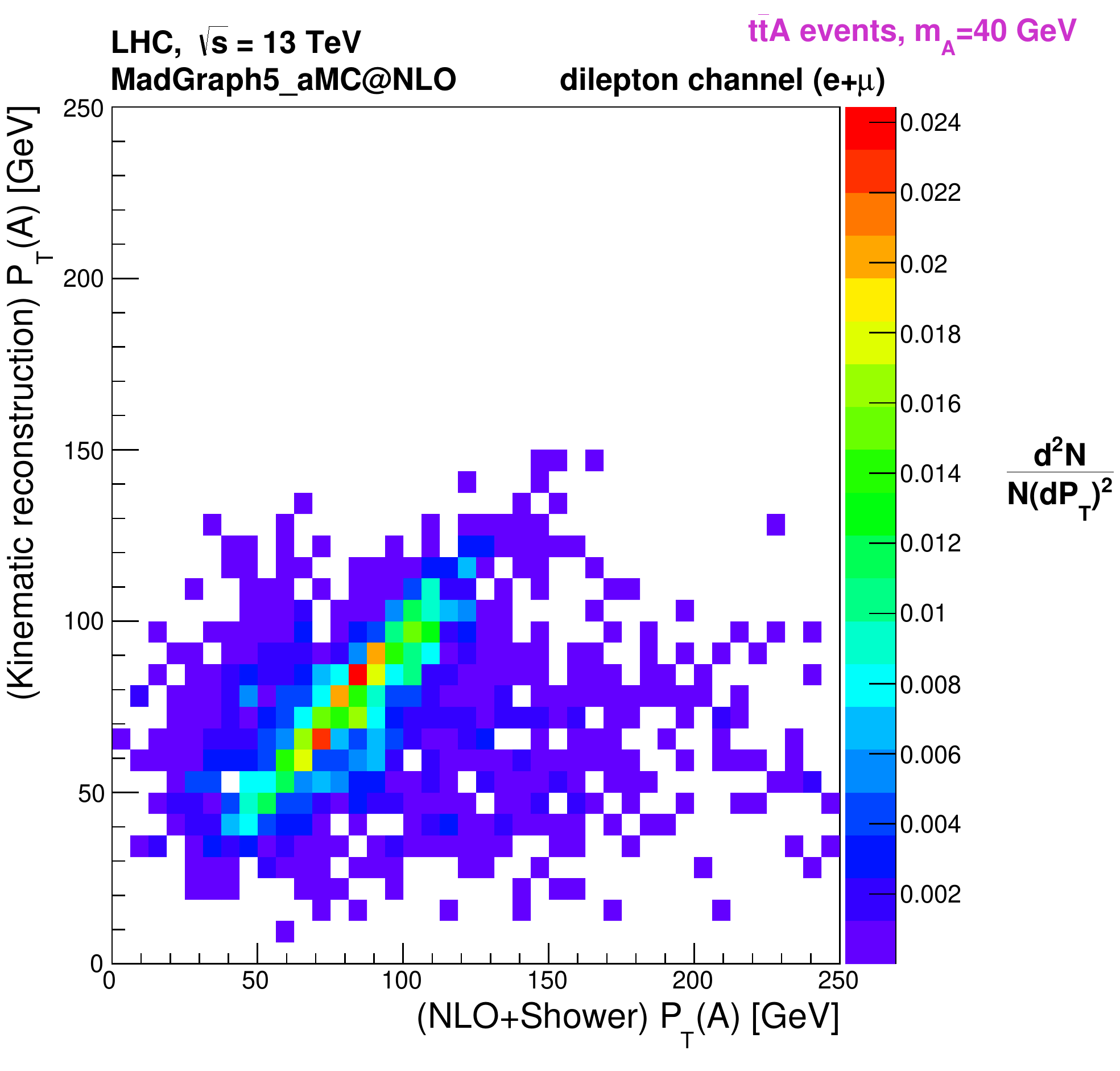,height=4.5cm,width=7.8cm,clip=} \epsfig{file=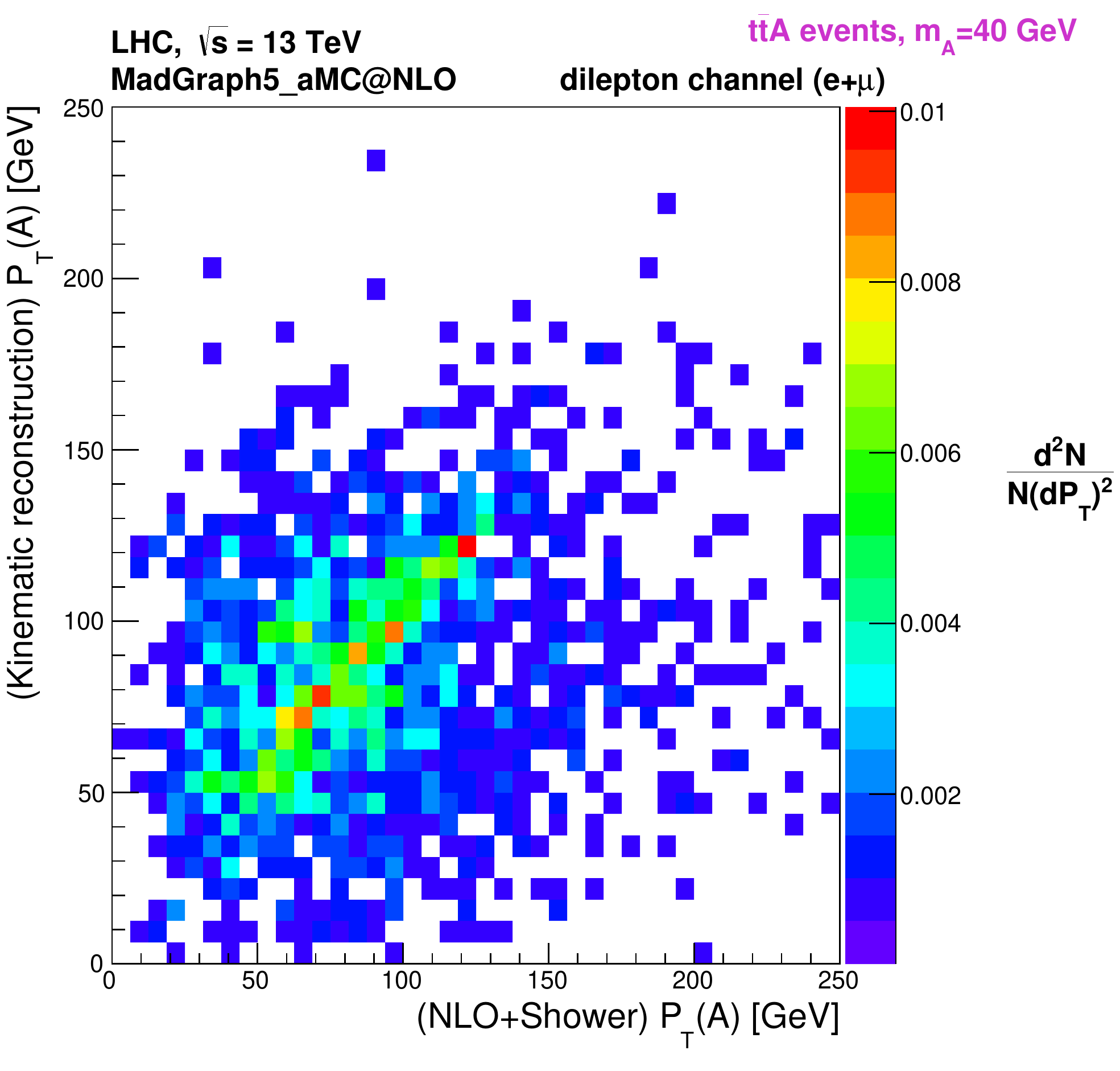,height=4.5cm,width=7.8cm,clip=} \\[-2mm]
		\end{tabular}\\[0mm]
		\vspace*{-3.9cm}\hspace*{-4.5cm}\textbf{\textcolor{black}{$m^{inv}_\phi$(2 jets)}} \hspace*{6.3cm}\textbf{\textcolor{black}{$m^{(1)}_\phi$}} \\[0mm]
		\vspace*{-2.3cm}\hspace*{9.5cm}\textbf{\textcolor{black}{$m^{inv}_\phi$(1 jet)}} \\[0mm]
		\vspace*{4.9cm}
		\caption{Two-dimensional distributions of the $\phi=A$ transverse momentum, $p_T$(A), in $t\bar{t}A$ events. The $p_T$(A) at NLO+Shower ($x$-axis) is represented 
		against $p_T$(A) after kinematic reconstruction ($y$-axis). The results for  the $\phi$ boson mass reconstruction methods are shown for the best of all methods (upper-left), 
		the invariant mass from 1 jet only (upper-right), the invariant mass from 2 jets (lower-left) and $m^{(1)}_\phi$ (lower-right). All distributions are shown for a Higgs mass of 40~GeV.}
		\label{fig:genexp2D_4methods}
	\end{center}
\end{figure*}

In Figure~\ref{fig:genexp2D_4methods}, we show the Higgs reconstructed $p_T$ versus the parton level value for the best of all methods (top-left) and for each one of the methods used to reconstruct the Higgs boson (remaining plots), for events from a pure pseudoscalar signal with $m_A = 40$~GeV. For the plots representing only one method, only the events where that method has been picked up as the best are considered. %e mais a "pureza" que a eficiencia que aumenta; O metodo de reconstruir a massa do higgs vais buscar os jatos que dao a melhor massa. Mas esses jatos 
%podem não ser os melhores, e ser uma coincidencia darem boa massa, mas o que vemos nos plots 2d e que os jatos que dao as melhores massas também são os melhores em termos de cinematica. 

Regardless of the method used, a visible correlation between the parton and reconstructed levels is still observed. Furthermore, each method tends to cover a different $p_T$ region, thus, choosing the best of all methods in each event allows to cover a larger number of solutions than each individual method, increasing the efficiency of the reconstruction.

%\clearpage

%%%%%%%%%%%%%%%. Full Event Selection
\section{Full event selection \label{sec:selection}}
\hspace{\parindent} %forca identacao

\begin{figure}[h!]
	\begin{center}
		\begin{tabular}{ccc}
			\hspace*{-3mm}
			\includegraphics[height=7.cm]{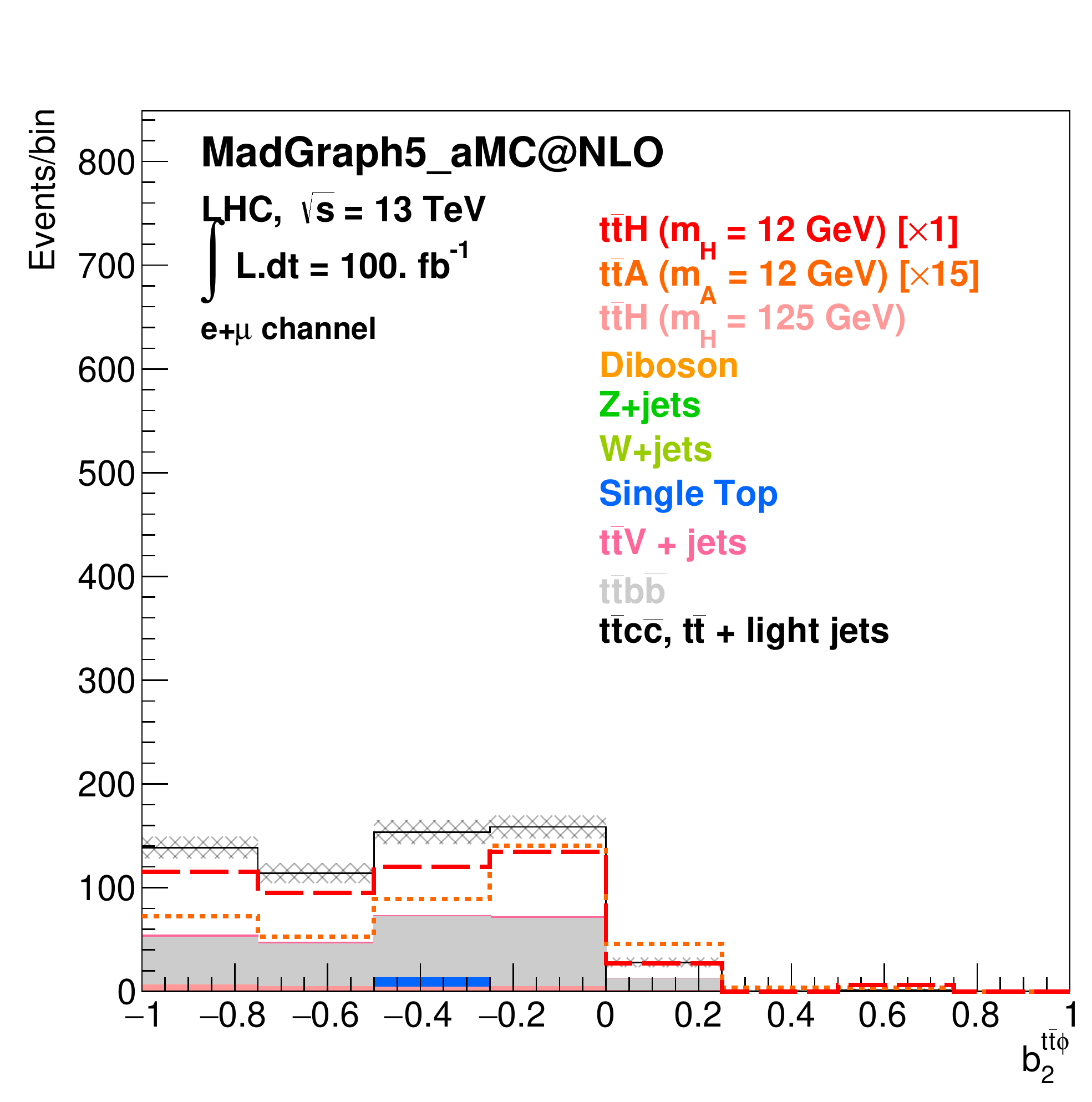}
			\includegraphics[height=7.cm]{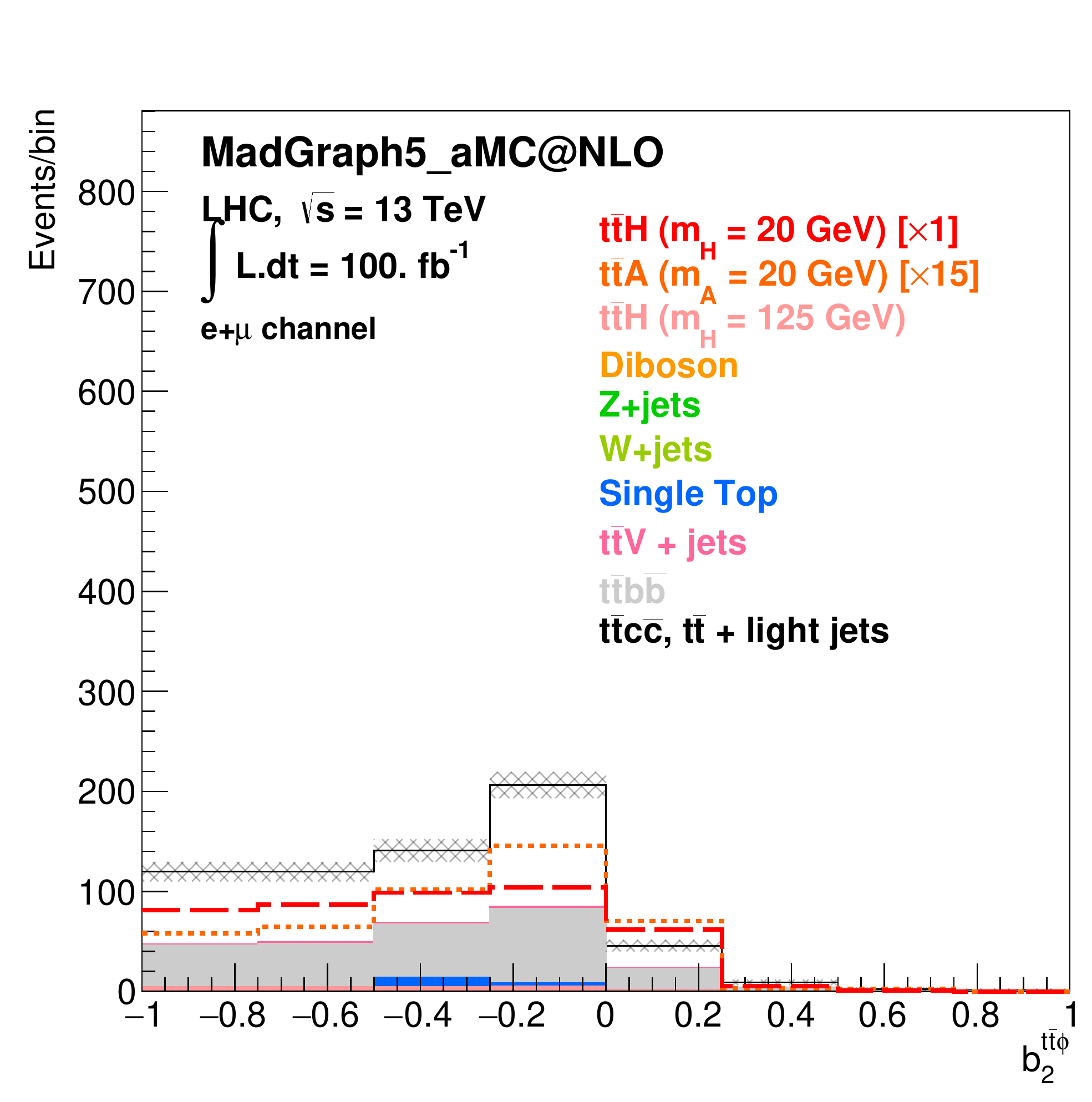}
			\\
			\hspace*{-3mm}
			\includegraphics[height=7.cm]{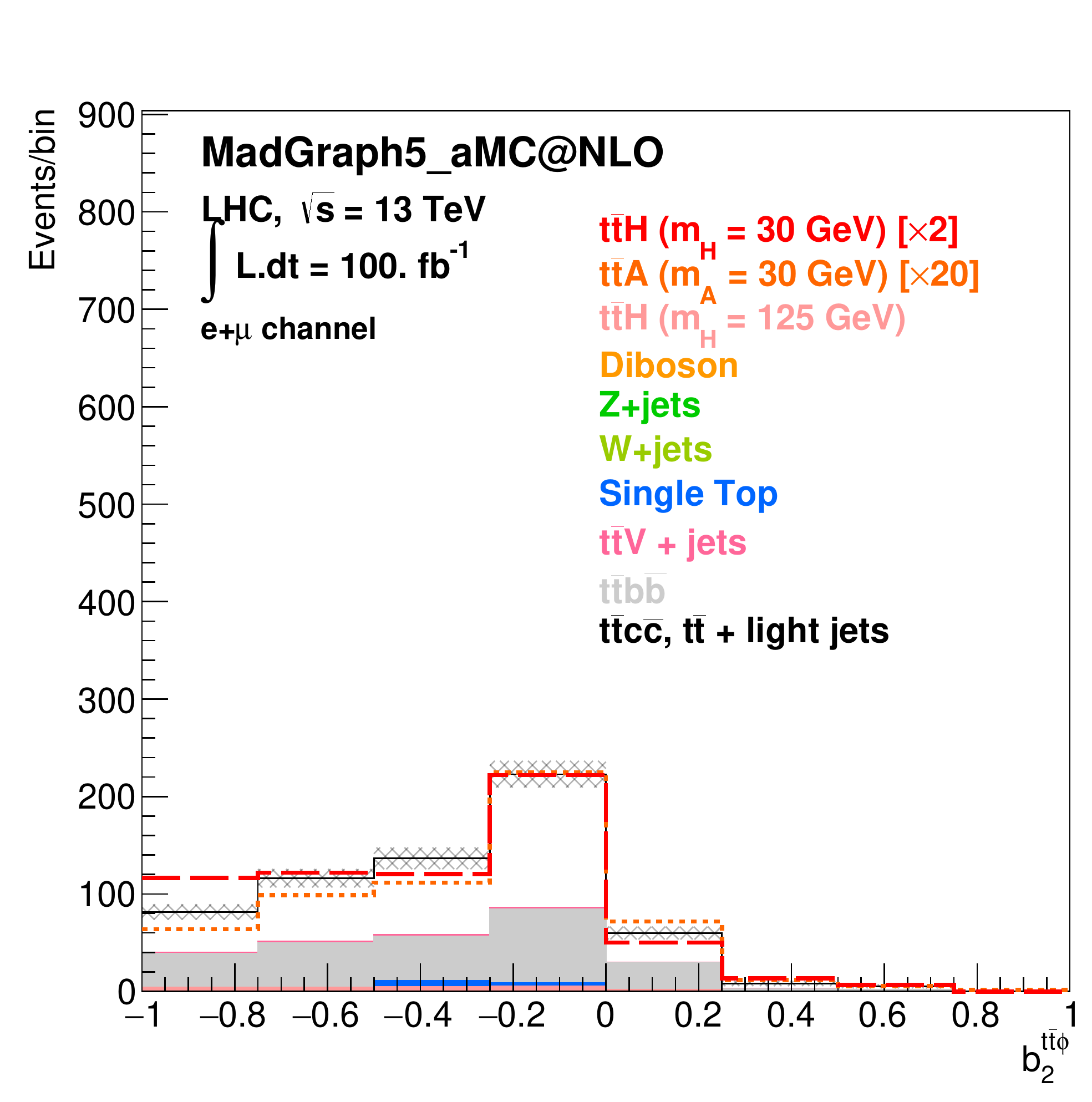}
			\includegraphics[height=7.cm]{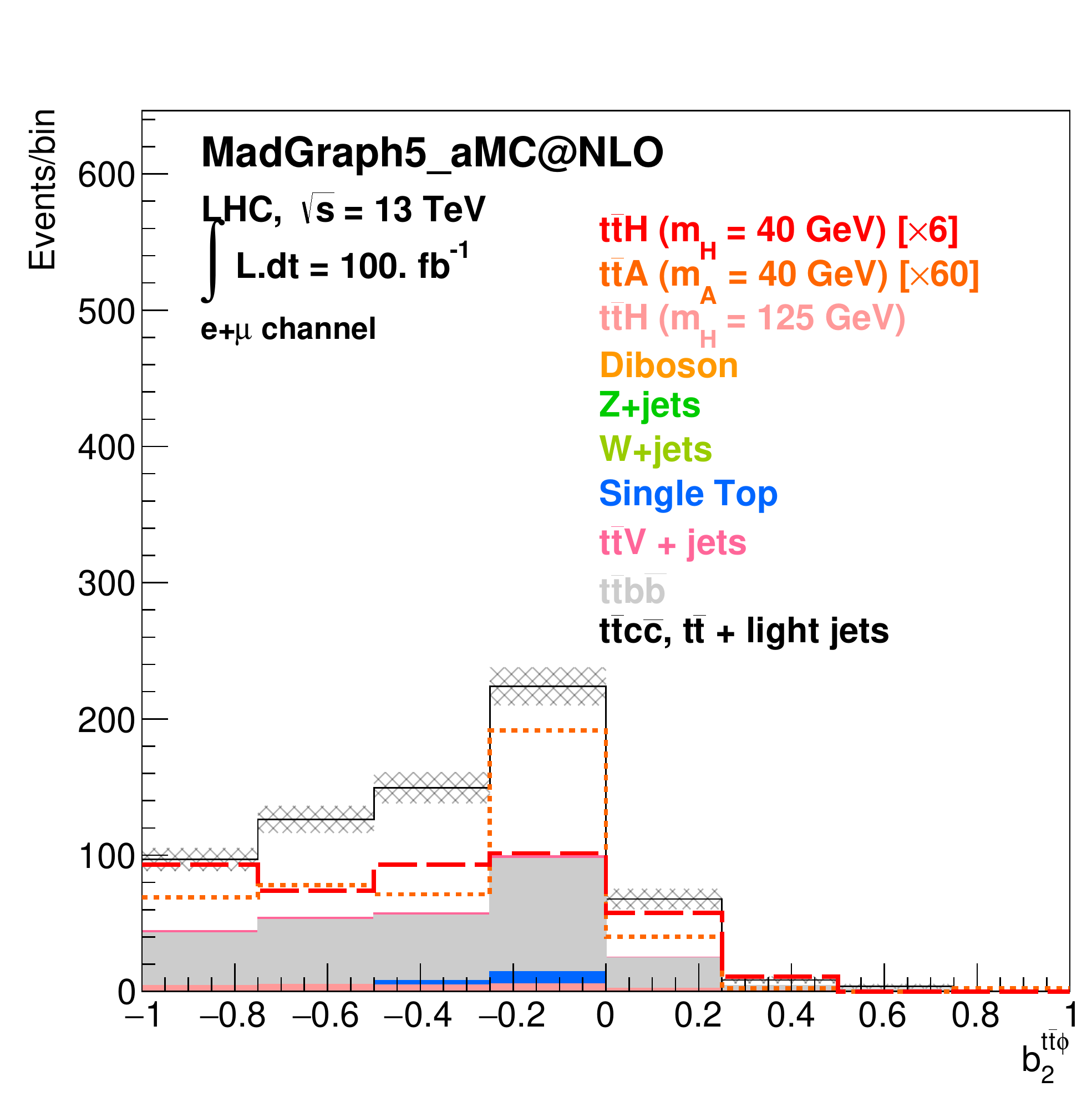}
		\end{tabular}
		\caption{Expected $b_2^{t\bar{t}\phi}$ distributions for the background and signal events, for a luminosity of 100 fb$^{-1}$. Kinematic reconstruction and final selection cuts are considered. Upper-left: distribution for $m_\phi = 12$~GeV. The $t\bar{t}A$ signal is increased by a factor of 15. Upper-right: distribution for $m_\phi = 20$~GeV. The $t\bar{t}A$ signal is increased by a factor of 15. Lower-left: distribution for $m_\phi = 30$~GeV. The $t\bar{t}H$ and $t\bar{t}A$ signals are increased by factors of 2 
		and 20, respectively. Upper-right: distribution for $m_\phi = 40$~GeV. The $t\bar{t}H$ and $t\bar{t}A$ signals are increased by factors of 6 and 60, respectively.}
		\label{fig:stackb2b4}
	\end{center}
\end{figure}

Following the kinematical reconstruction described in the previous section, we applied additional selection criteria to the events. These cuts define what we call the {\it final selection}. 
The first one was implemented to reject opposite charge dilepton events from  the $Z$ + jets background, by requiring the invariant mass of the dilepton system ($m_{\ell^+ \ell^-}$) to be outside a 10~GeV window around the $Z$ boson mass ($m_Z=91$~GeV). The second one selects events with at least 3 $b$-tagged jets. In Figure~\ref{fig:stackb2b4}, the expected number of events that survive the full selection criteria (pre-selection cuts, kinematical reconstruction and final selection cuts), for the different SM backgrounds, is shown at the LHC and for an integrated luminosity of 100~fb$^{-1}$. Those distributions are compared to the CP-even and CP-odd signals, for $m_\phi = 12$ (top-left), 20 (top-right), 
30 (bottom-left) and 40~GeV (bottom-right). More details on the background composition can be found in~\cite{Azevedo:2020vfw}. Only for representation purposes, signals from the Higgs bosons have been rescaled by $x_{scale}$ factors in Figure~\ref{fig:stackb2b4} (labelled by the $[\times x_{scale}]$ factors in the plots), that range from 1 (15) to 6 (60)  for the scalar (pseudoscalar) Higgs boson. As was seen previously, the dominant backgrounds are essentially coming from $t\bar{t}$ processes, with a particular important contribution from $t\bar{t}b\bar{b}$. 
No events from $pp\rightarrow t\phi j$ survived the final selection for the $\phi$ mass range considered in this paper. %It should be stressed at this point that, when comparing the $b_2^{t\bar{t}\phi}$ distributions shapes obtained here, with the ones from~\cite{Azevedo:2020vfw}, for the  $m_{\phi}$ = 40~GeV case, it seems the new reconstruction applied in this paper allows to get a better resolution on those particular distributions. 

%%%%%%%%%%%%%%%. Results and Discussion
\section{Results and discussion \label{sec:results}}
\hspace{\parindent} 

The results are presented as confidence levels (CLs) for the exclusion of the SM with a contribution from a new Higgs boson $\phi$ with mixed scalar and pseudoscalar couplings (CP-mixed case), assuming the SM hypothesis. The CLs are computed for fixed LHC integrated luminosities ($L$) of 300~fb$^{-1}$ and 3000~fb$^{-1}$. 
Four mass values of the $\phi$ boson are considered, in the range 12-40~GeV, extending further the results obtained in~\cite{Azevedo:2020vfw} to the more challenging lower mass region. 
Assuming on-shell decays, only the mass range $m_\phi > 2 \, m_b \approx 9.4$~GeV is kinematically accessible for $\phi\to b \bar b$ ($m_b$ is the bottom-quark mass). 
We also exclude a narrow mass window around the $\Upsilon$ states, between 8.5 and 11~GeV, which is why the lowest mass considered in this paper is 12~GeV. 
For even lower masses, a new analysis with a different final state has to be used. 

The CLs are shown as contour plots in the ($\kappa$, $\tilde{\kappa}$) 
2D plane (with $\kappa = \kappa_t \cos \alpha$ and $\tilde{\kappa} = \kappa_t \sin \alpha$), which was scanned using steps of 0.05 (0.02) on the values of  $\kappa$ and $\tilde{\kappa}$ in the range [-1.50, 1.50] ([-1.00, 1.00]) for $L$ = 300~fb$^{-1}$ ($L$ = 3000~fb$^{-1}$). The $b_2$ and $b_4$ distributions are used to set the CLs evaluated in both the LAB and $t\bar t \phi$ centre-of-mass systems, for comparison. 
The contribution of all SM backgrounds is taken into account, normalised to the LHC luminosity, as well as the different signal hypotheses.
The CL is given as one minus the $p$-value, under the signal hypothesis, for observing the test-statistic value that is expected (median) in the SM hypothesis.
The test-statistic used is the logarithm of the ratio between likelihoods of the signal and SM hypotheses, and the computation of $p$-values and medians is done using an ensemble of toy experiments.
%follows the prescription set by~\cite{Read:2002hq, Junk:1999kv}.
 
\begin{figure}[h!]
\begin{center}
	\begin{tabular}{ccc}
	\hspace*{-5mm}\includegraphics[height =4.cm]{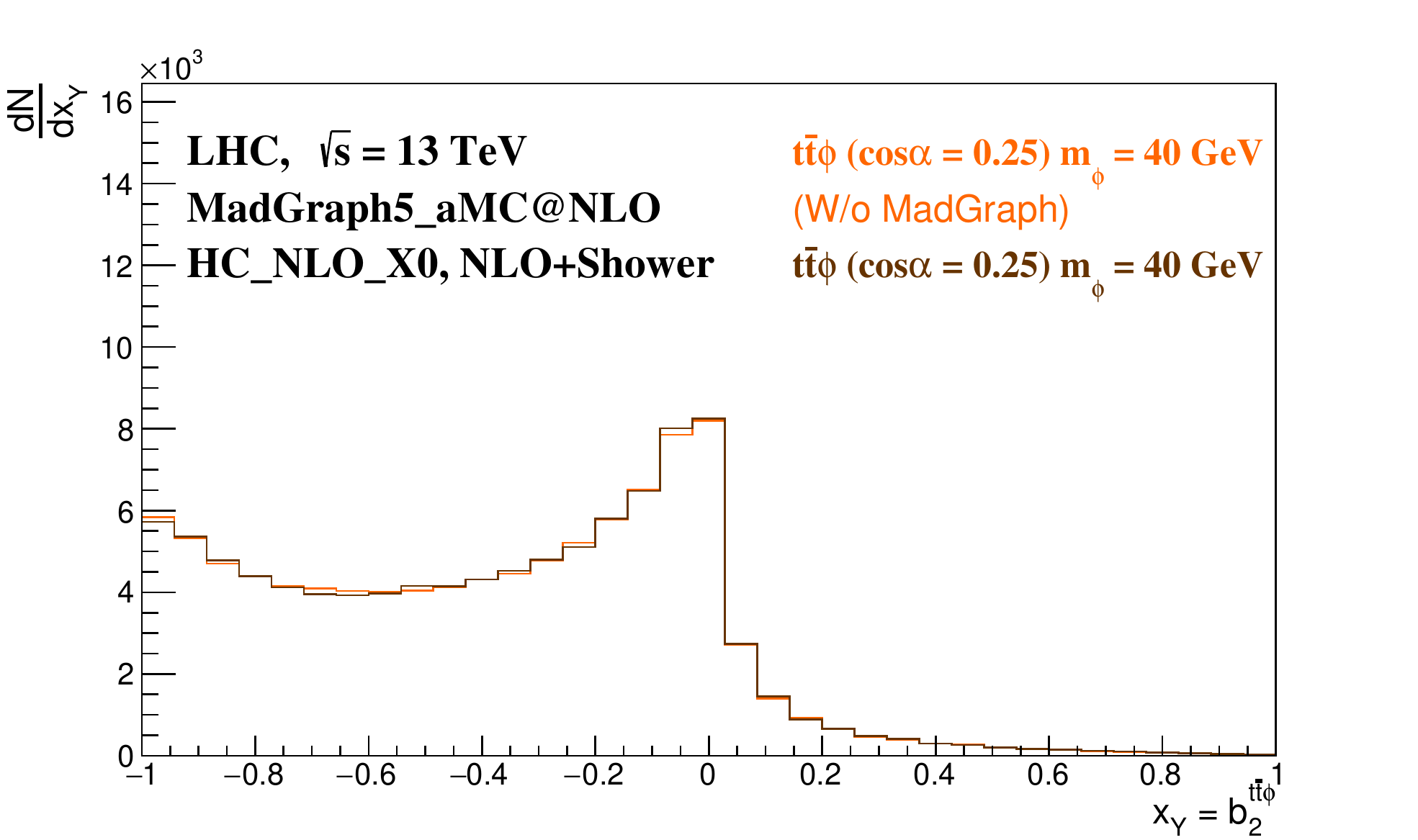}
	\hspace*{-5mm} \includegraphics[height = 4.cm]{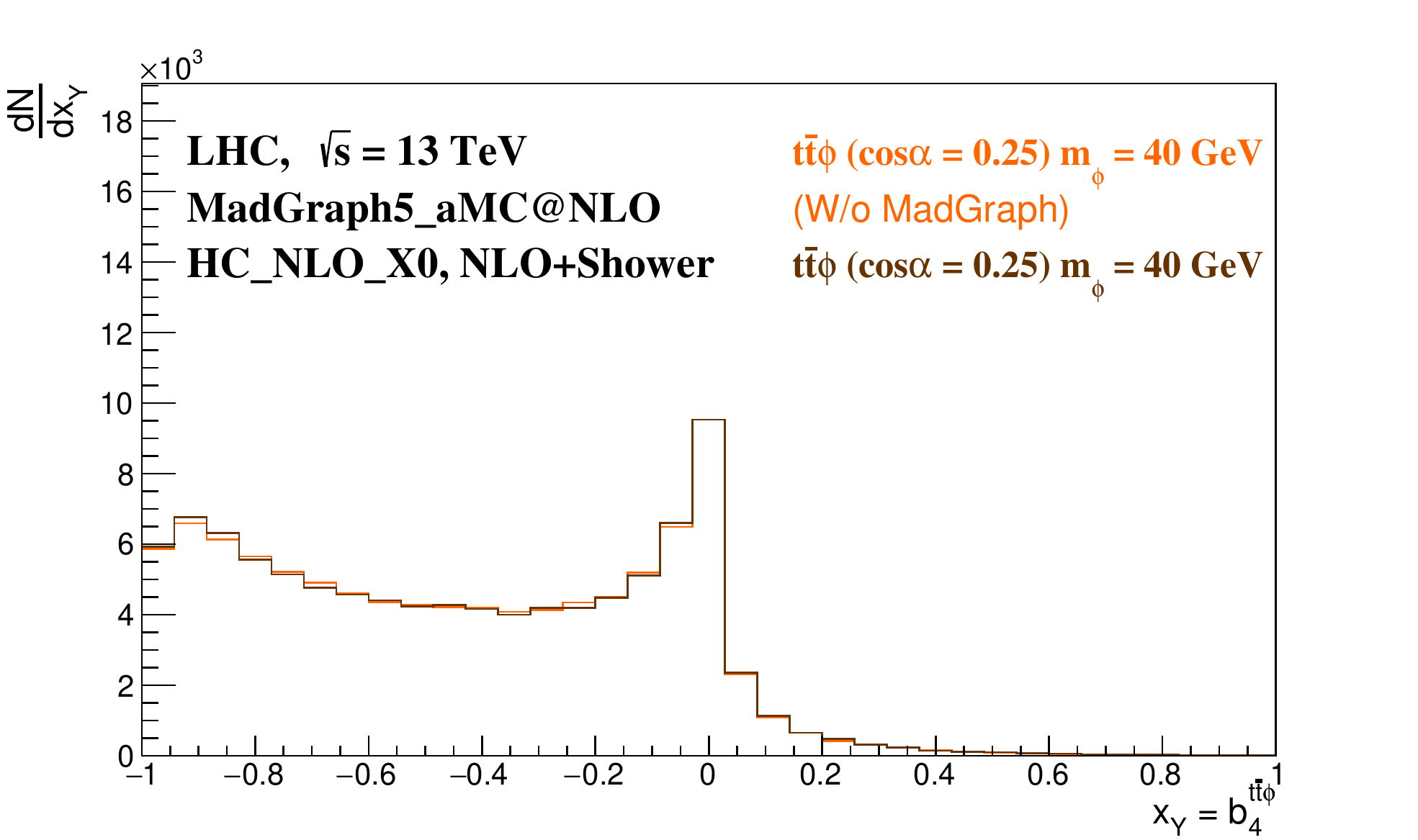} \\ \hspace*{-5mm} \includegraphics[height = 4.cm]{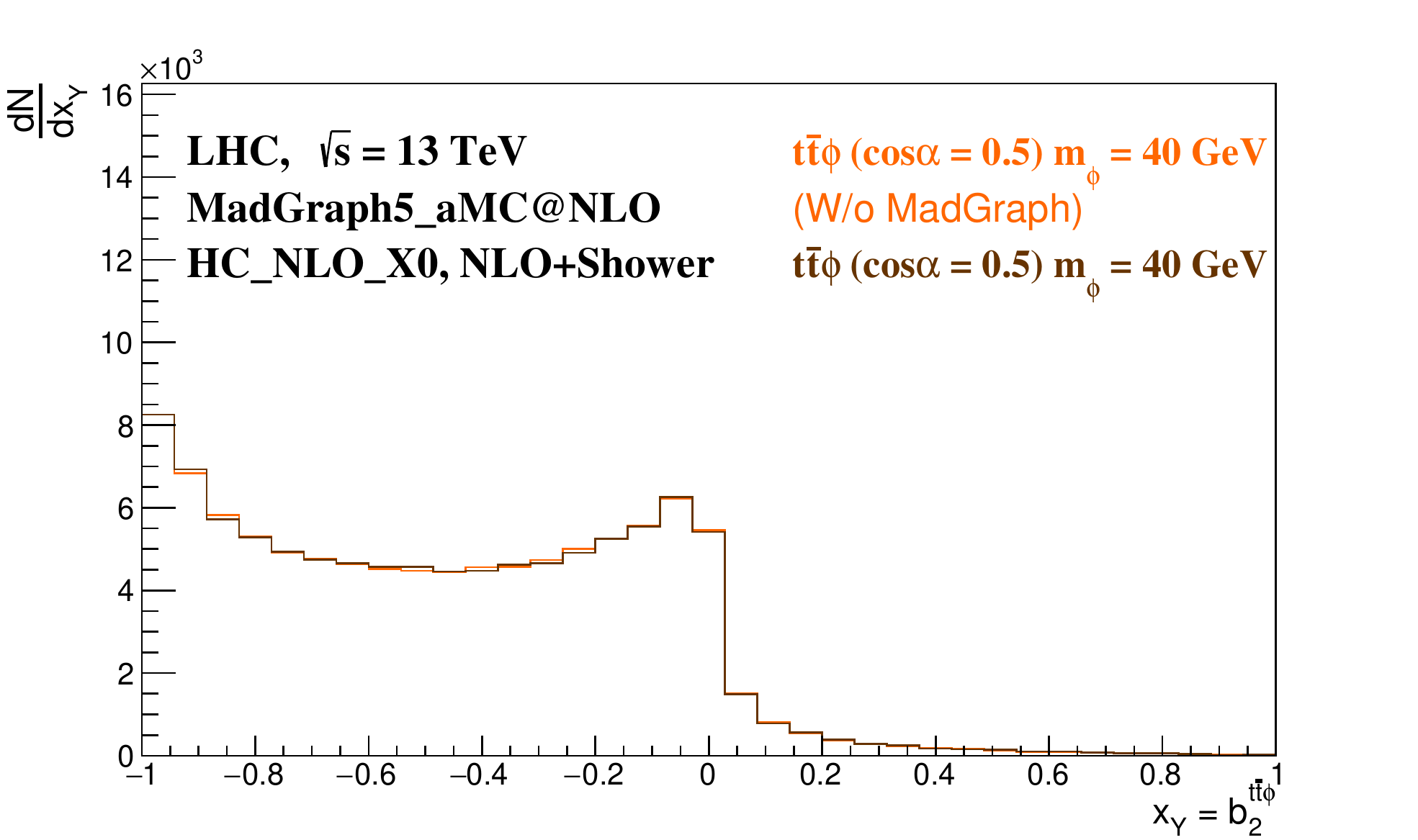} 
	\hspace*{-5mm} \includegraphics[height = 4.cm]{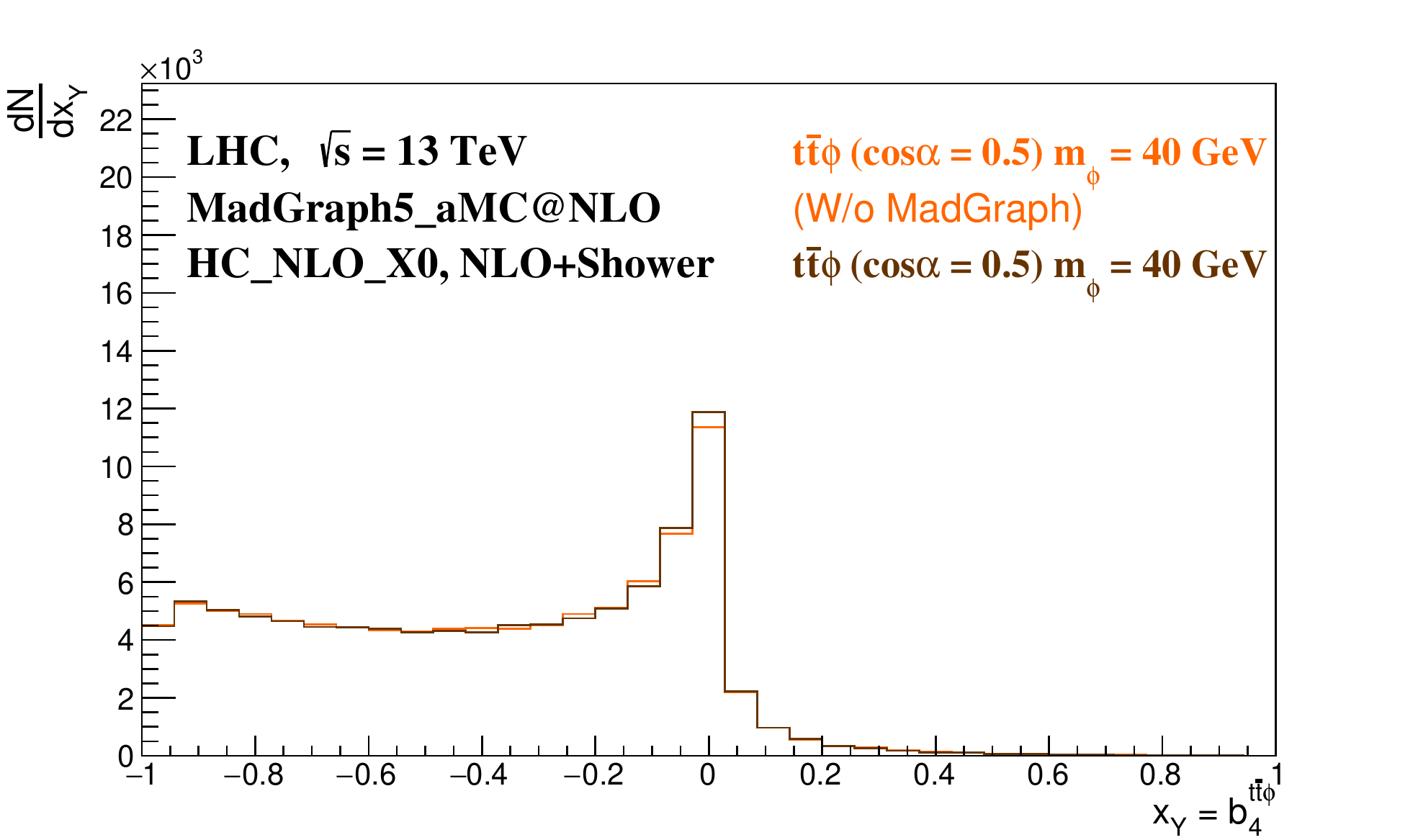} \\
	\hspace*{-5mm} \includegraphics[height = 4.cm]{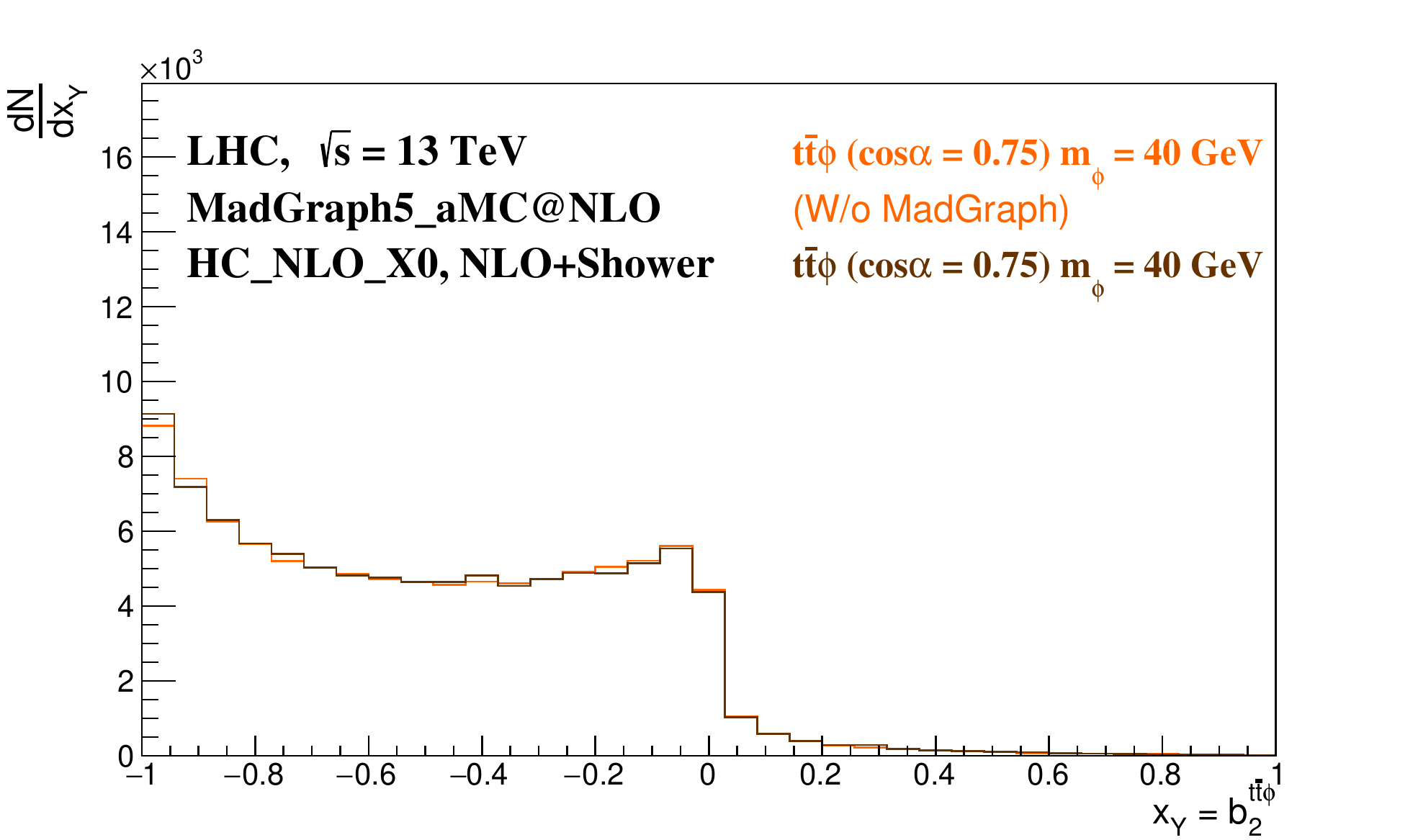} 
	\hspace*{-5mm} \includegraphics[height = 4.cm]{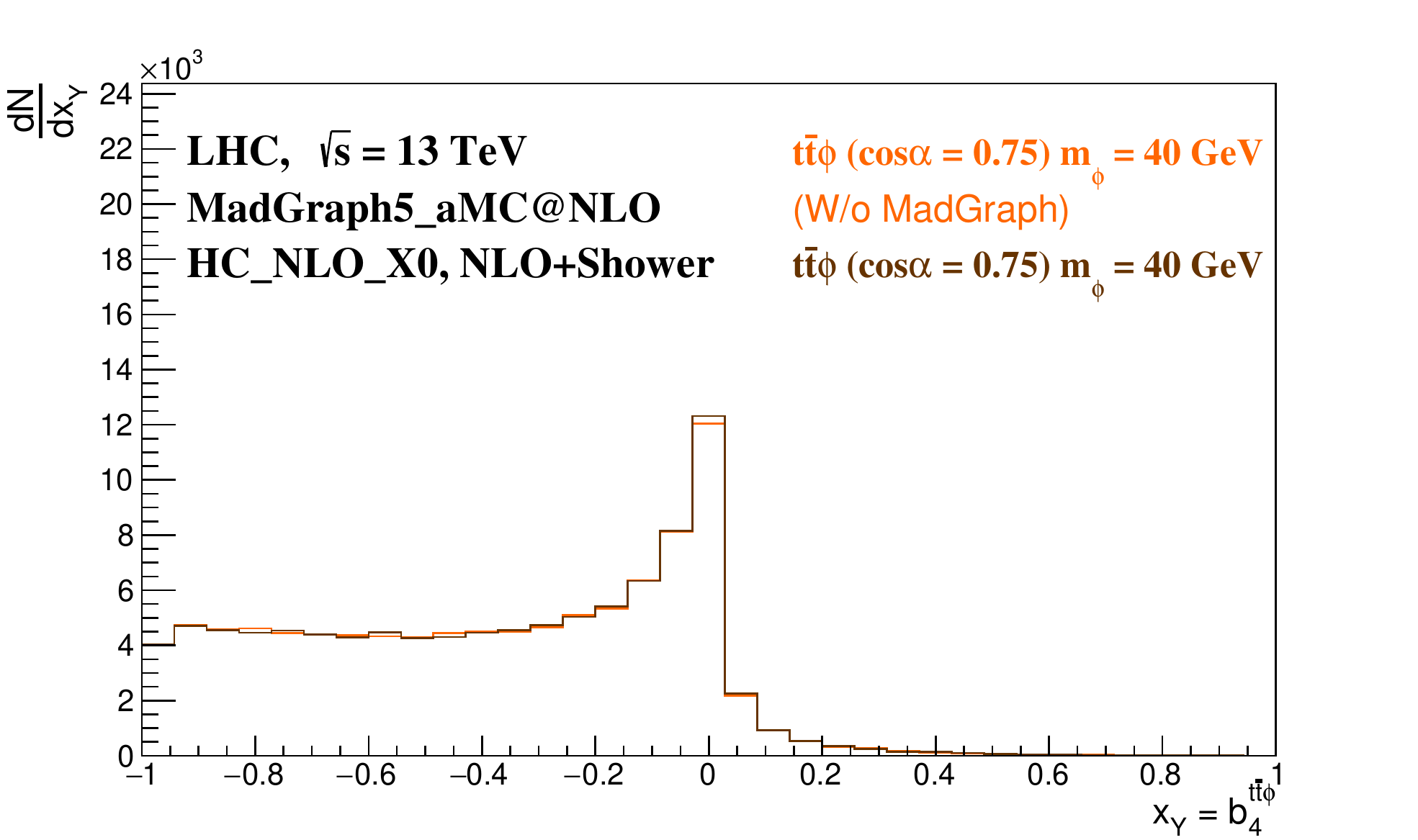}
	\end{tabular}
	\caption{Parton level $b_2^{t\bar{t}\phi}$ (left) and $b_4^{t\bar{t}\phi}$ (right) distributions at NLO+Shower, for $m_\phi= 40$~GeV, and $\cos \alpha = 0.25$ (top), 0.5 (middle) and 0.75 (bottom).}
	\label{fig:mixed-Xsec}
\end{center}
\end{figure}

Before discussing the full impact of the results obtained in the low mass region, it is convenient to notice that the total cross section for CP-mixed signals can be evaluated using 

\begin{equation}
\begin{aligned}
\sigma_\text{CP-mixed} = \sigma_\text{CP-even} \, \, \kappa^2 + \sigma_\text{CP-odd} \, \, \tilde{\kappa}^2 \, \, , \\
\end{aligned}
\label{eq:mixed}
\end{equation}
where $\sigma_\text{CP-mixed}$, $\sigma_\text{CP-even}$ and $\sigma_\text{CP-odd}$ correspond to the signal cross section for the CP-mixed, CP-even and CP-odd cases, respectively. 
The validity of Equation~\ref{eq:mixed} was verified by looking at several differential distributions for the signal events, where the mass of the $\phi$ boson and the CP-angle $\alpha$ were varied. For each angle $\alpha$, we compared those distributions when the CP-mixed signals were generated directly using \texttt{MadGraph5\_aMC@NLO}, and also by using Equation~\ref{eq:mixed} to compute the number of events in the CP-mixed case, from the CP-even and -odd samples. Both approaches gave similar results. 
This can be seen in Figure~\ref{fig:mixed-Xsec}, where we show in brown the distributions using \texttt{MadGraph5\_aMC@NLO}, and in orange the ones obtained from Equation~\ref{eq:mixed} (labelled W/o MadGraph in the plots), for $m_\phi = 40$~GeV and $\cos \alpha = 0.25, 0.5$ and $0.75$. 

\begin{figure}[h!]
        \vspace*{-2cm}
	\begin{center}
		\begin{tabular}{ccc}
			\hspace*{-9mm}
			\includegraphics[height=4.5cm]{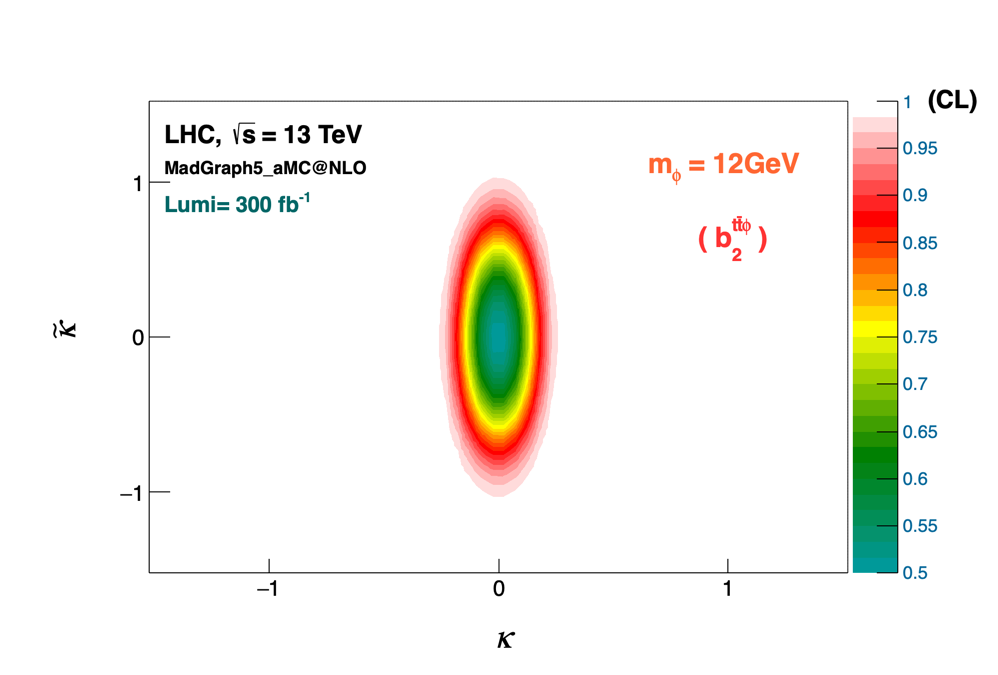}
			\hspace*{-2mm}\includegraphics[height=4.5cm]{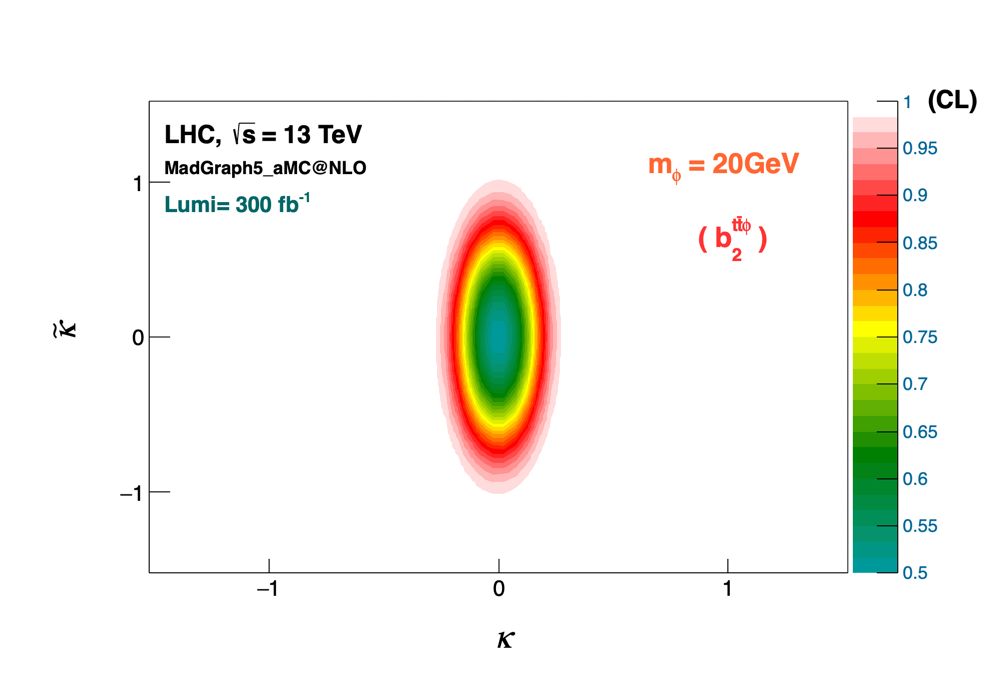} \\[-2mm]
			\hspace*{-9mm}
			\includegraphics[height=4.5cm]{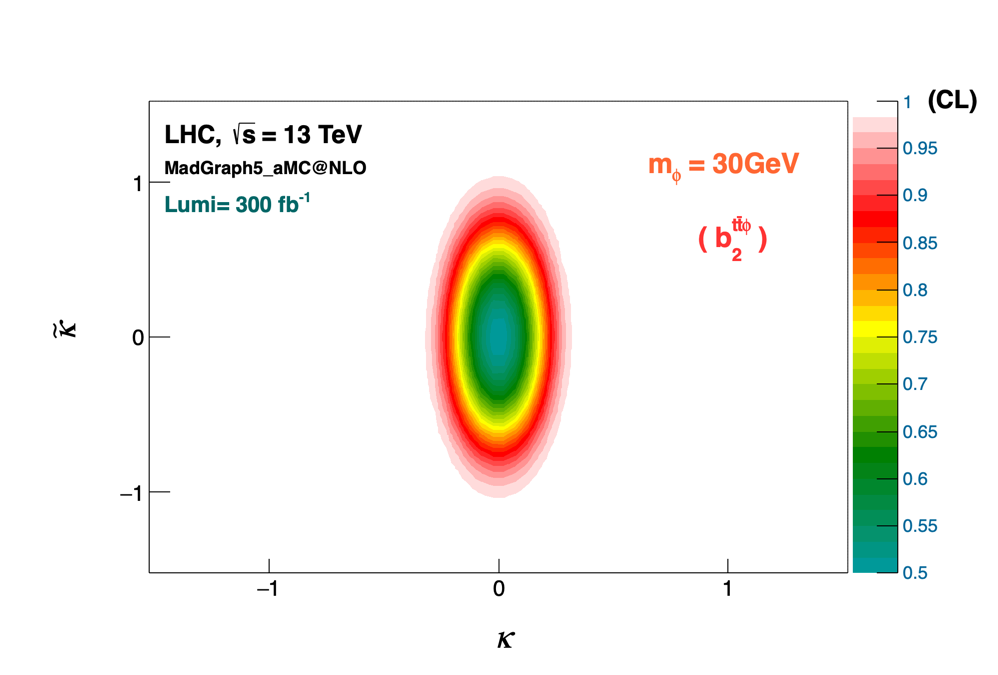}
			\hspace*{-2mm}\includegraphics[height=4.5cm]{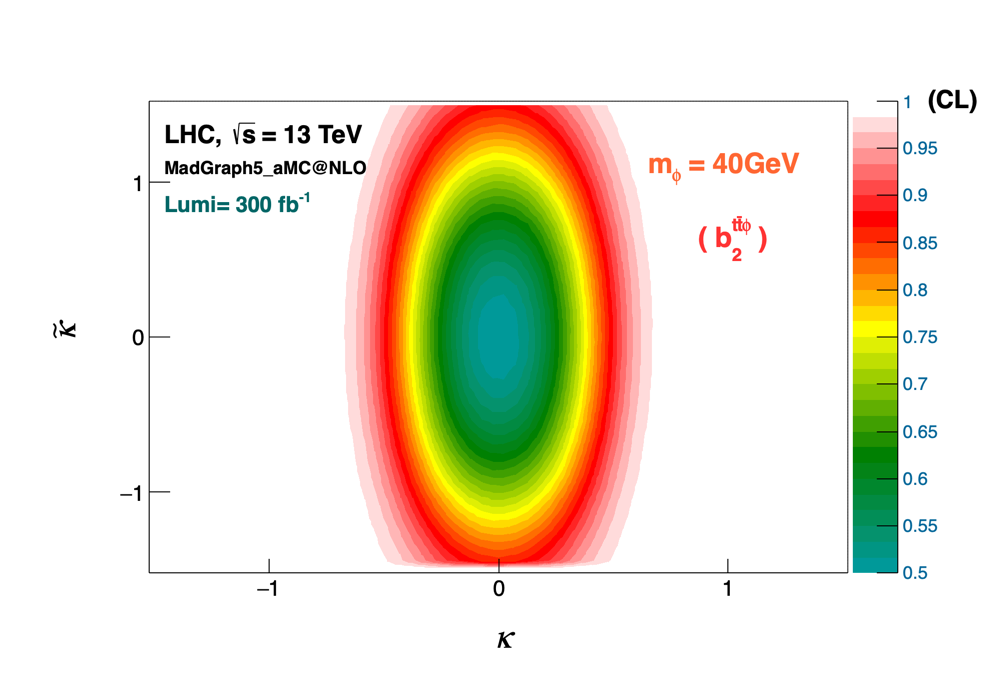}
		\end{tabular} \\[-2mm]
		\caption{Two-dimensional CLs for the $b_2^{t\bar{t}\phi}$ variable, and a fixed luminosity of 300~fb$^{-1}$. The $\phi$ boson masses represented are: 12~GeV (top-left), 20~GeV (top-right), 30~GeV (bottom-left), and 40~GeV (bottom-right).}
		\label{fig:CL_300_b2tth}
	\end{center}
\end{figure}

\begin{figure}
        \vspace*{-2cm}
	\begin{center}
		\begin{tabular}{ccc}
			\hspace*{-9mm}
			\includegraphics[height=4.5cm]{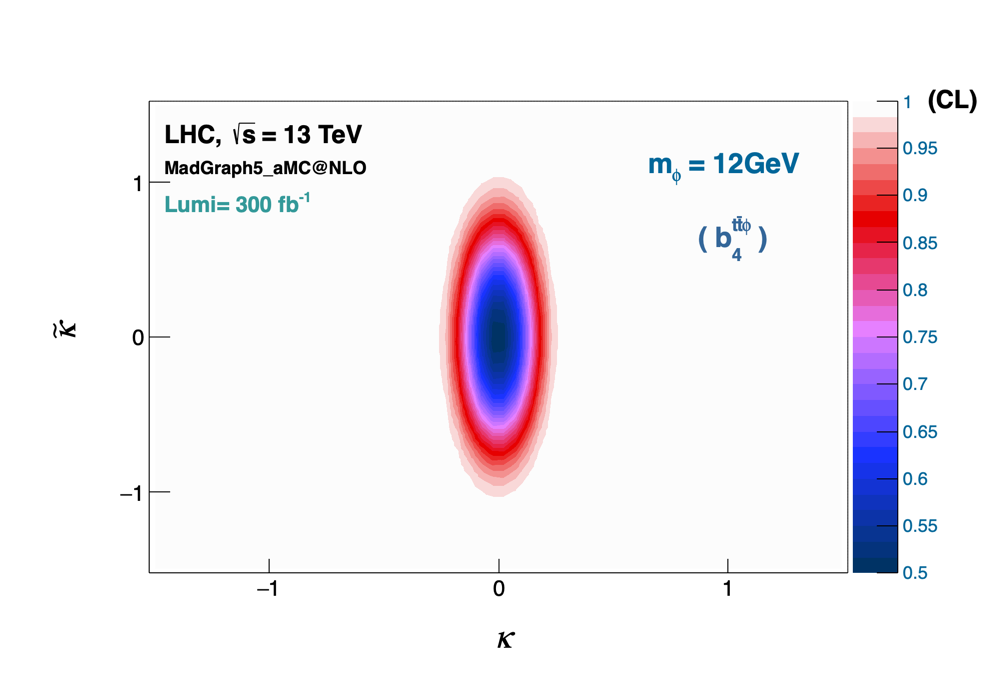}
			\hspace*{-2mm}\includegraphics[height=4.5cm]{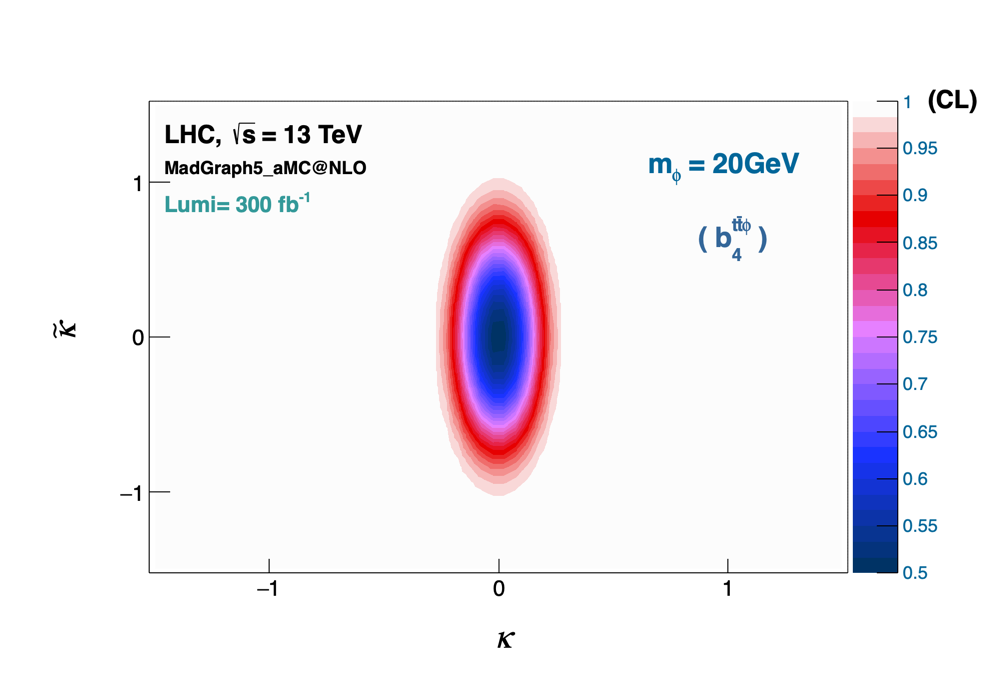} \\[-2mm]
			\hspace*{-9mm}
			\includegraphics[height=4.5cm]{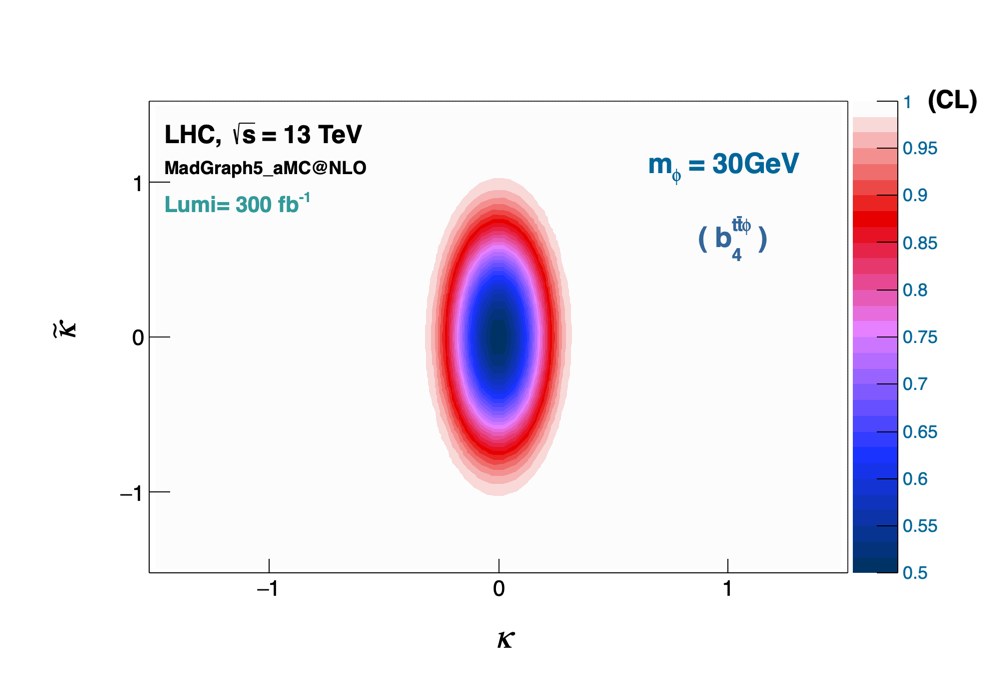}
			\hspace*{-2mm}\includegraphics[height=4.5cm]{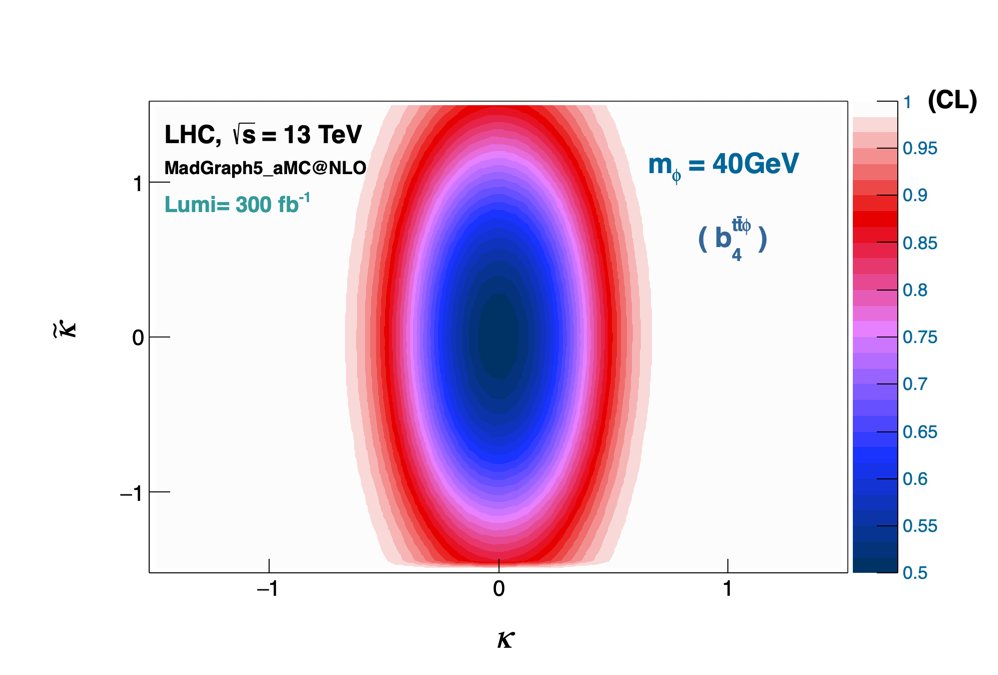}
		\end{tabular} \\[-2mm]
		\caption{Two-dimensional CLs for the $b_4^{t\bar{t}\phi}$ variable, and a fixed luminosity of 300~fb$^{-1}$. The $\phi$ boson masses represented are: 12~GeV (top-left), 20~GeV (top-right), 30~GeV (bottom-left), and 40~GeV (bottom-right).}
		\label{fig:CL_300_b4tth}
	\end{center}
\end{figure}

\begin{figure}
        \vspace*{-2cm}
	\begin{center}
		\begin{tabular}{ccc}
			\hspace*{-9mm}
			\includegraphics[height=4.5cm]{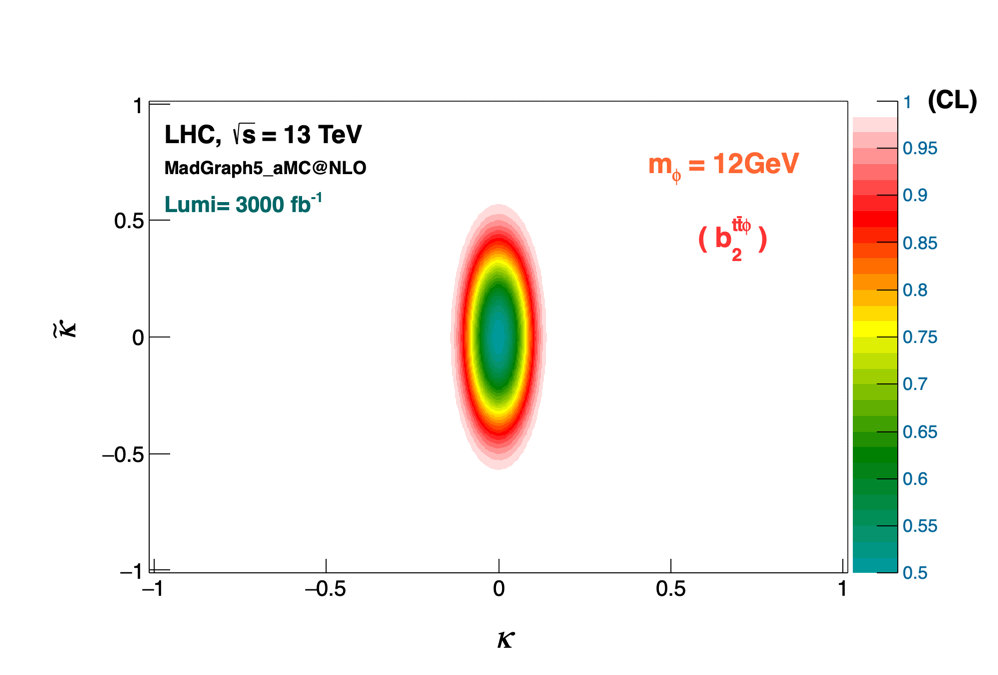}
			\hspace*{-2mm}\includegraphics[height=4.5cm]{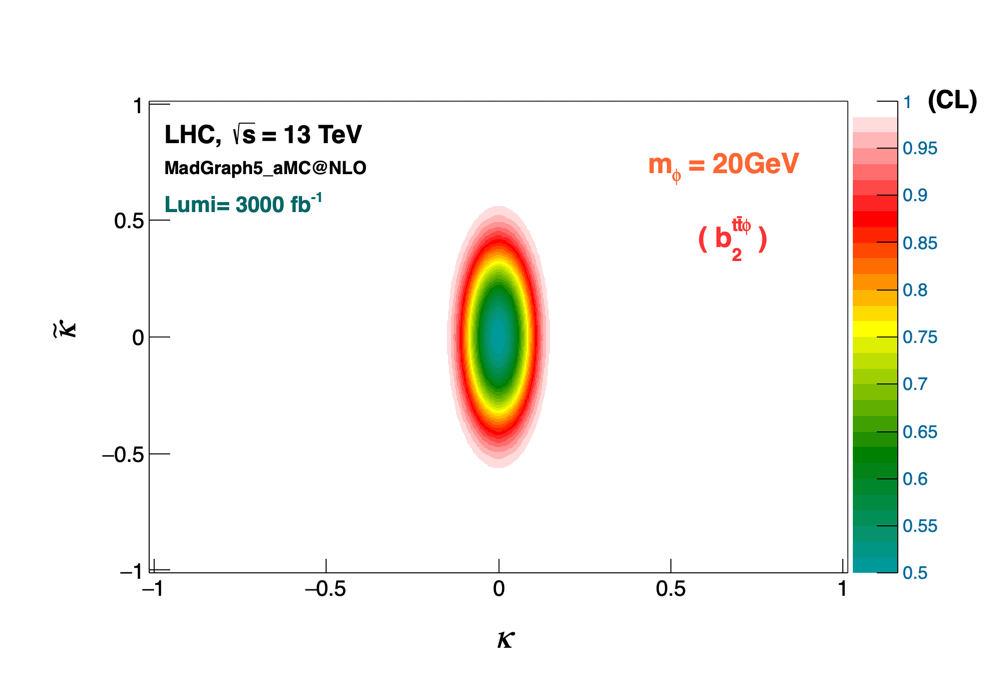} \\[-2mm]
			\hspace*{-9mm}
			\includegraphics[height=4.5cm]{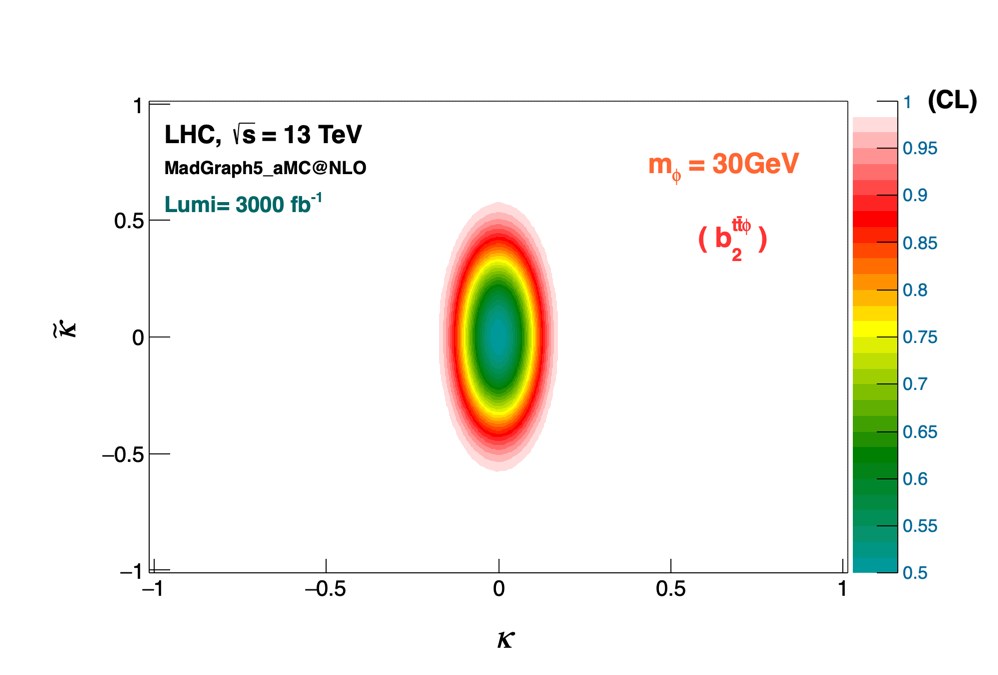}
			\hspace*{-2mm}\includegraphics[height=4.5cm]{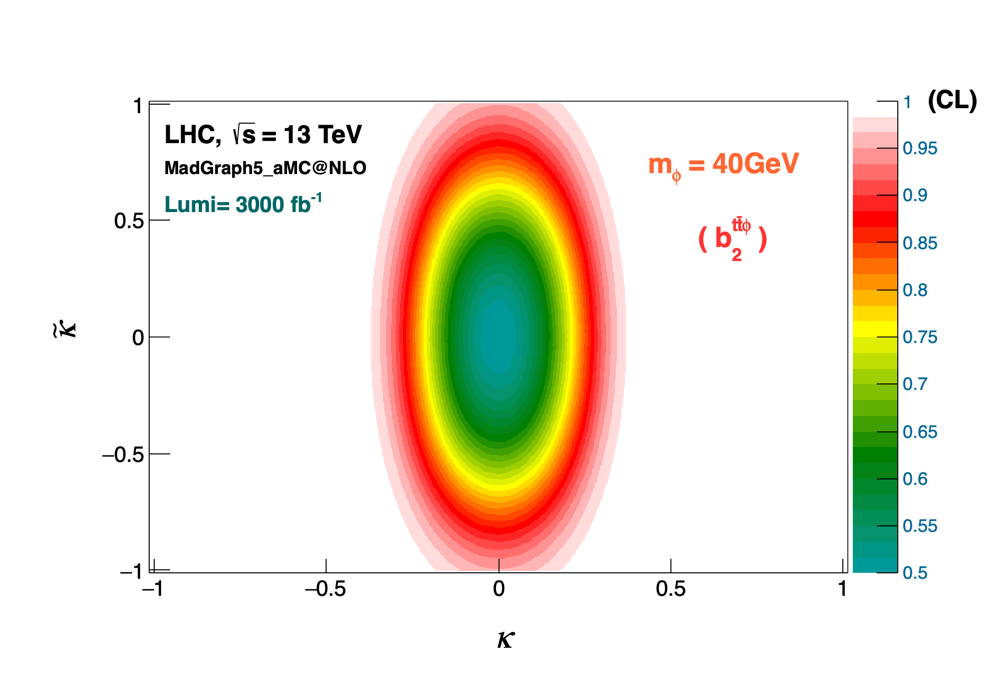}
		\end{tabular} \\[-4mm]
		\caption{Two-dimensional CLs for the $b_2^{t\bar{t}\phi}$ variable, and a fixed luminosity of 3000~fb$^{-1}$. The $\phi$ boson masses represented are: 12~GeV (top-left), 20~GeV (top-right), 30~GeV (bottom-left), and 40~GeV (bottom-right).}
		\label{fig:CL_3000_b2tth}
	\end{center}
\end{figure}

\begin{figure}
        \vspace*{-2cm}
	\begin{center}
		\begin{tabular}{ccc}
			\hspace*{-9mm}
			\includegraphics[height=4.5cm]{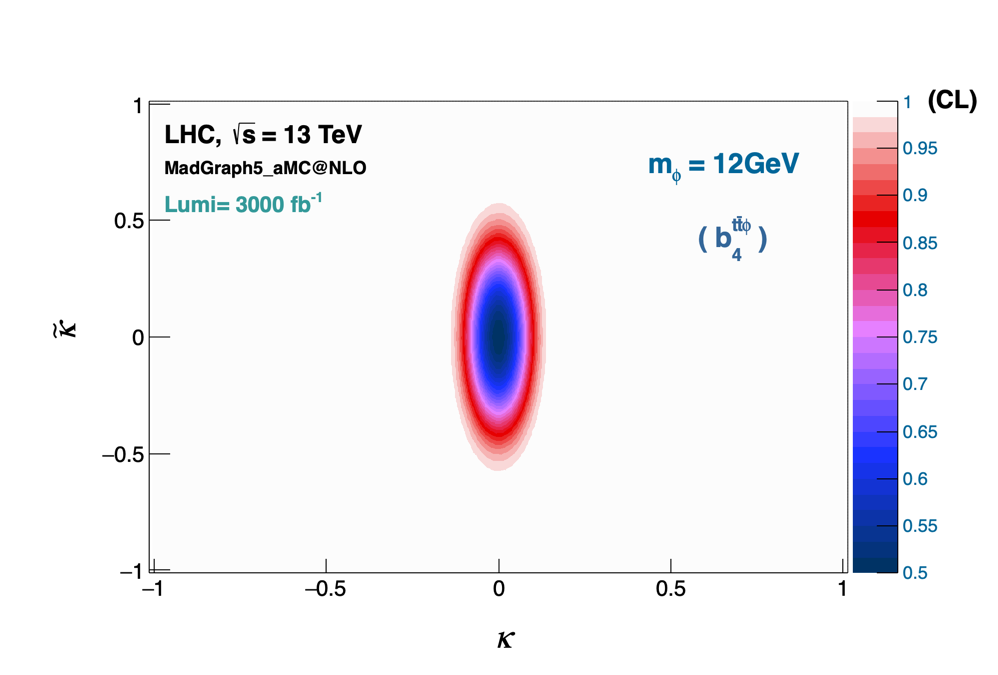}
			\hspace*{-2mm}\includegraphics[height=4.5cm]{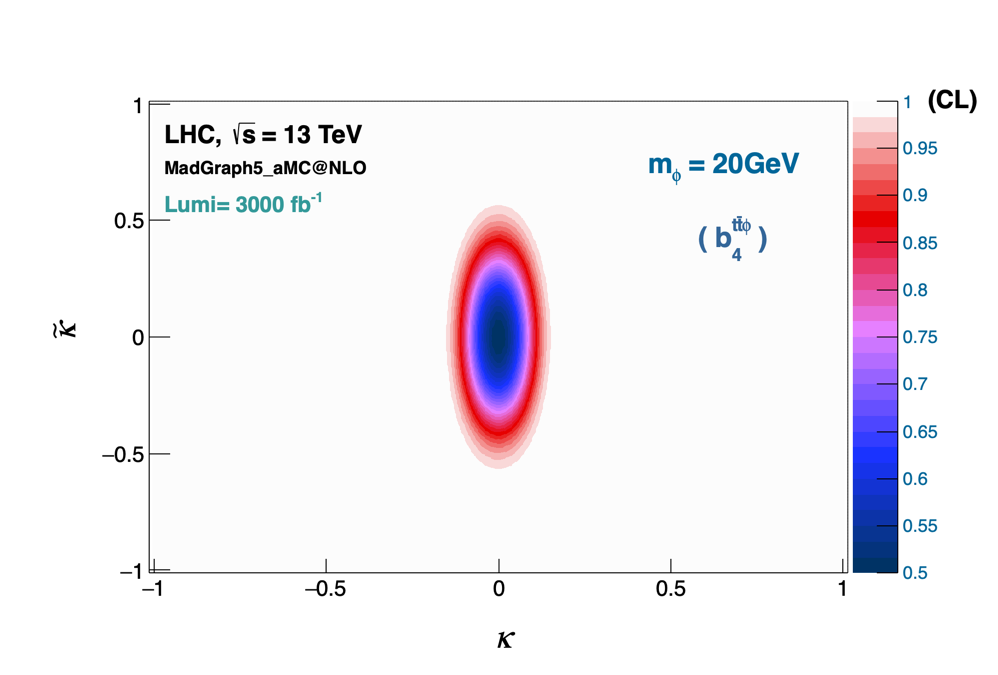} \\[-2mm]
			\hspace*{-9mm}
			\includegraphics[height=4.5cm]{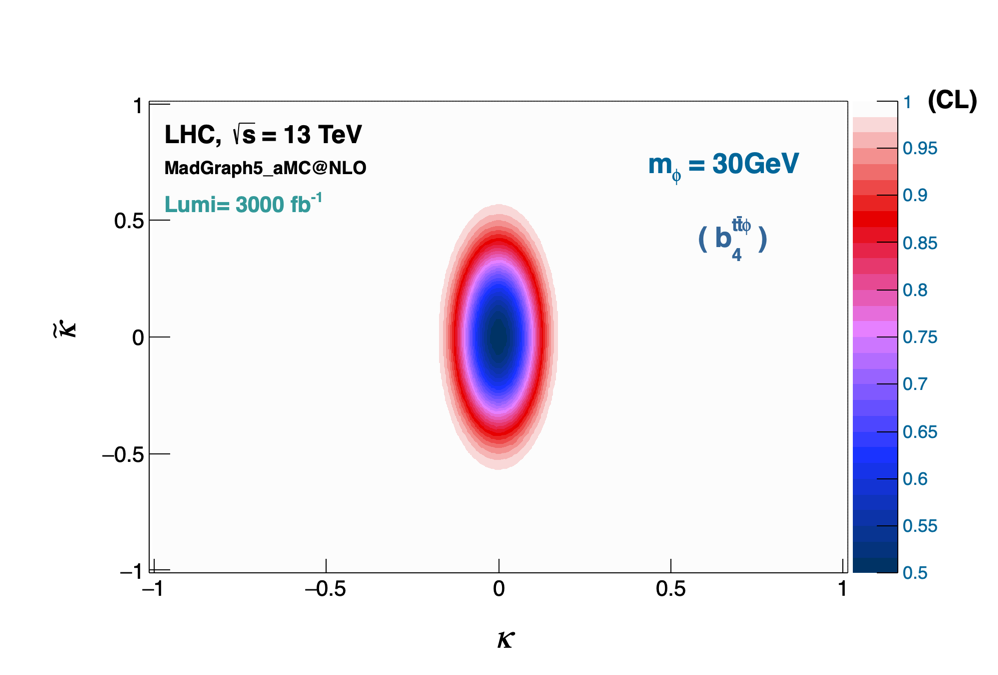}
			\hspace*{-2mm}\includegraphics[height=4.5cm]{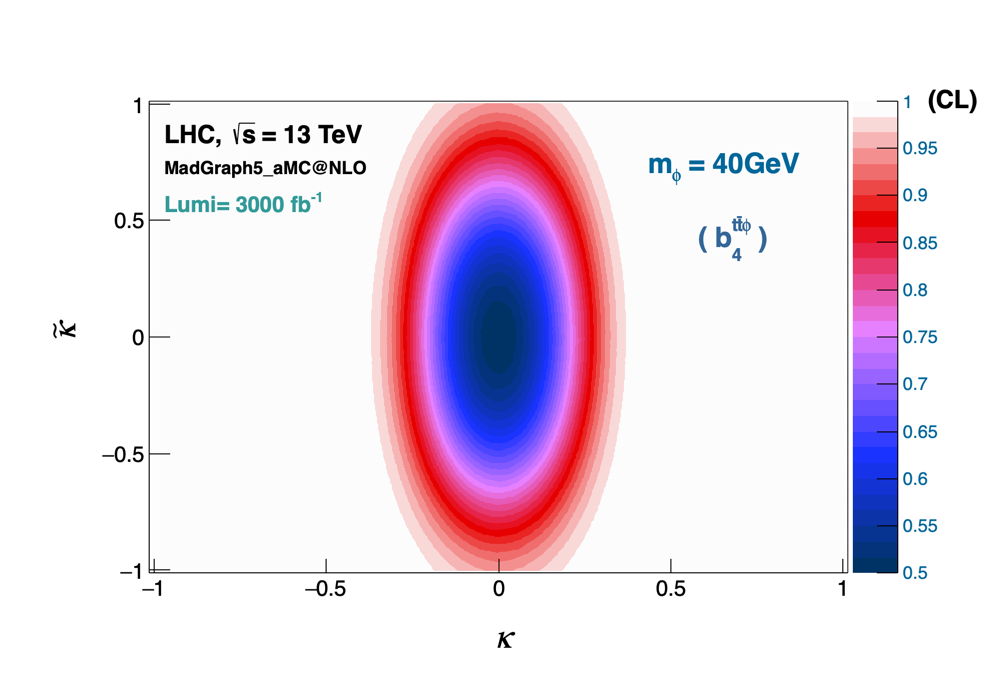}
		\end{tabular} \\[-2mm]
		\caption{Two-dimensional CLs for the $b_4^{t\bar{t}\phi}$ variable, and a fixed luminosity of 3000~fb$^{-1}$. The $\phi$ boson masses represented are: 12~GeV (top-left), 20~GeV (top-right), 30~GeV (bottom-left), and 40~GeV (bottom-right).}
		\label{fig:CL_3000_b4tth}
	\end{center}
\end{figure}

In Figures~\ref{fig:CL_300_b2tth} and \ref{fig:CL_300_b4tth}, we show the exclusion CLs contour lines in the 2D plane ($\kappa$, $\tilde{\kappa}$), for $m_\phi = 12, 20, 30$ and 40~GeV, and a luminosity of 300~fb$^{-1}$, for the variables $b_2^{t\bar{t}\phi}$ and $b_4^{t\bar{t}\phi}$, respectively. The $b_2$ and $b_4$ distributions in the centre-of-mass of the $t\bar{t}\phi$ system gave better exclusion levels than the ones computed in the laboratory frame. This is the reason why we only show the former. In Figures~\ref{fig:CL_3000_b2tth} and \ref{fig:CL_3000_b4tth},
the same information is represented, but for a luminosity of 3000~fb$^{-1}$, the full luminosity expected at the end of the High Luminosity phase of the LHC (HL-LHC).     



\clearpage

The CLs get progressively better as the Higgs mass decreases, which was to be expected since the $t\bar{t}\phi$ cross-section increases. Moreover, the CP-even component of the coupling is always more constrained than the CP-odd component, for a given CL. That difference is related to the different behaviour of the cross sections - the CP-even cross sections rises more steeply as the scalar mass decreases than the CP-odd one.
In Table \ref{table:exclusion_limits}, we show the exclusion limits for the top quark Yukawa couplings to the $\phi$ boson, for the range of Higgs masses considered in this paper. 
The limits are shown for the variables $b_2^{t\bar{t}\phi}$ and $b_4^{t\bar{t}\phi}$, at confidence levels of 68\% and 95\%, for the expected luminosity at the end of the HL-LHC. 
The ranges of values represented for $\kappa/\tilde{\kappa}$ are the ones that cannot be excluded for the CL indicated. The lowest exclusion limits that we expect at 95\% CL for the pair ($|\kappa|$, $|\tilde{\kappa}|$) are, approximately, (0.10, 0.50). For $m_\phi =$ 40~GeV, the same limit increases by a factor of roughly 3 (2) for the CP-even (CP-odd) coupling constant. Furthermore, most of the results are similar for both variables considered.

\begin{table}[hbt!]
	\renewcommand{\arraystretch}{1.3}
	\begin{center}
		\begin{tabular}{|c|c|cc|cc|}
			
			%\toprule
			\hline
			%\multirow{3}{3.5cm}{\centering $L$ = 3000~fb$^{-1}$}
			\multicolumn{2}{|c|}{}  & 
			\multicolumn{2}{c}{Exclusion Limits} & \multicolumn{2}{|c|}{Exclusion Limits} \\ [-1mm]
			\multicolumn{2}{|c|}{$L$ = 3000~fb$^{-1}$}  & \multicolumn{2}{c}{from $b_2^{t\bar{t}\phi}$} & \multicolumn{2}{|c|}{from $b_4^{t\bar{t}\phi}$} \\
			
			%\midrule 
			\multicolumn{2}{|c|}{}  & (68\% CL)	& (95\% CL) & (68\% CL) & (95\% CL)  \\ \hline 
			              
			\hspace*{-3mm} \multirow{2}{2.5cm}{\centering $m_\phi$ = 12~GeV} & 	    	          	
			$\kappa \in$ & [-0.05, +0.05]  &   [-0.11, +0.11] & [-0.05, +0.05]  &   [-0.11, +0.11]   \\               
			 & $\tilde{\kappa} \in$ & [-0.26, +0.26] & [-0.50, +0.50]  & [-0.26, +0.26] & [-0.50, +0.50]  
			\\ \hline
			               
			\hspace*{-3mm} \multirow{2}{2.5cm}{\centering $m_\phi$ = 20~GeV} &    	          
			$\kappa \in$   & [-0.07, +0.07]  &   [-0.13, +0.13]  	& [-0.07, +0.07]  & [-0.13, +0.13]   \\               
			& $\tilde{\kappa} \in$ & [-0.26, +0.26]  &   [-0.49, +0.49]  	&   [-0.26, +0.26]  &   [-0.50, +0.50]  
			\\ \hline
			             
			\hspace*{-3mm} \multirow{2}{2.5cm}{\centering $m_\phi$ = 30~GeV}    		&    	          
			$\kappa \in$ & [-0.07, +0.07]  &   [-0.14, +0.14]  	&              [-0.07, +0.07]  &   [-0.14, +0.14]   \\               
			&    	$\tilde{\kappa} \in$ & [-0.26, +0.20]  &   [-0.50, +0.50]  	&   [-0.26, +0.26]  &   [-0.50, +0.50]  \\ \hline 
			             
			\hspace*{-3mm} \multirow{2}{2.5cm}{\centering $m_\phi$ = 40~GeV}    		&    	          
			$\kappa \in$ & [-0.17, +0.17]  &   [-0.32, +0.32]  	&        [-0.17, +0.17]  &   [-0.32, +0.32]   \\               
			&    	$\tilde{\kappa} \in$ & [-0.53, +0.53]  &   [-1.00, +1.00]  	 &     [-0.53, +0.53]  &   [-1.01, +1.01]  
			\\   
			\hline          
			%\bottomrule
		\end{tabular}
		\caption{Exclusion limits for the $t\bar{t}\phi$ CP-couplings as a function of the $\phi$ boson mass, and a fixed luminosity of 3000~fb$^{-1}$. The limits are shown at confidence levels of 68\% and 95\%, for the variables $b_2^{t\bar{t}\phi}$ and $b_4^{t\bar{t}\phi}$.}
		\label{table:exclusion_limits}
	\end{center}
\end{table}

%%%%%%%%%%%%%%%%%%%%%%%%%%%%%%%%%%%%%%%%%%%%%%%%%%%%%%
\section{Interpretation in the framework of the C2HDM}
\label{sec:C2HDM}
\hspace{\parindent} %forca identacao

Let us now proceed to understand how these results affect our benchmark model, the C2HDM. We start with a very brief review of the model just to fix the notation and refer the reader to ref.~\cite{ Fontes:2017zfn} for details, including how theoretical and experimental constraints affect the model.
The scalar potential breaks CP explicitly and is invariant under the 
$\mathbb{Z}_2$ symmetry $\Phi_1 \to \Phi_1, \Phi_2 \to -\Phi_2$, softly broken by the $m_{12}^2$ term,
\beq
V &=& m_{11}^2 |\Phi_1|^2 + m_{22}^2 |\Phi_2|^2
- \left(m_{12}^2 \, \Phi_1^\dagger \Phi_2 + h.c.\right)
+ \frac{\lambda_1}{2} (\Phi_1^\dagger \Phi_1)^2 +
\frac{\lambda_2}{2} (\Phi_2^\dagger \Phi_2)^2 \nonumber \\
&& + \lambda_3
(\Phi_1^\dagger \Phi_1) (\Phi_2^\dagger \Phi_2) + \lambda_4
(\Phi_1^\dagger \Phi_2) (\Phi_2^\dagger \Phi_1) +
\left[\frac{\lambda_5}{2} (\Phi_1^\dagger \Phi_2)^2 + h.c.\right] \; ,
\eeq
where $\Phi_i$ $(i=1,2)$ develop the real vacuum expectation values (VEVs) $v_{1}$ and $v_{2}$.
The only complex parameters in the potential are $m_{12}^2$ and $\lambda_5$. The ratio of the VEVs is $\tan \beta \equiv \frac{v_2}{v_1} $
and the rotation matrix from gauge to mass eigenstates is given by
\beq
\left( \begin{array}{c} H_1 \\ H_2 \\ H_3 \end{array} \right) = R
\left( \begin{array}{c} \rho_1 \\ \rho_2 \\ \rho_3 \end{array} \right)
\; ,
\label{eq:c2hdmrot}
\eeq
with
\be
R =
\left(
\begin{array}{ccc}
	c_1 c_2 & s_1 c_2 & s_2\\
	-(c_1 s_2 s_3 + s_1 c_3) & c_1 c_3 - s_1 s_2 s_3  & c_2 s_3\\
	- c_1 s_2 c_3 + s_1 s_3 & -(c_1 s_3 + s_1 s_2 c_3) & c_2 c_3
\end{array}
\right)\, ,
\label{matrixR}
\ee
where $s_i = \sin{\alpha_i}$,
$c_i = \cos{\alpha_i}$ ($i = 1, 2, 3$),
and
\be
- \pi/2 < \alpha_1 \leq \pi/2,
\hspace{5ex}
- \pi/2 < \alpha_2 \leq \pi/2,
\hspace{5ex}
- \pi/2 < \alpha_3 \leq \pi/2.
\label{range_alpha}
\ee
We choose to order the Higgs bosons such that $m_{H_1} \le m_{H_2} \le m_{H_3}$. Avoiding flavour changing neutral currents at tree-level is accomplished by extending the symmetry to the fermions resulting in four different types of Yukawa. However, since the top Yukawa couplings are the same in all four types we refrain to discuss the details of the different models. The Yukawa Lagrangian for up quarks (for all $\mathbb{Z}_2$ types) is
\beq
{{\cal L}_Y}_i = - \frac{{m_f}}{v} \bar{\psi}_f \left[ \frac{R_{i2}}{s_\beta} - i \frac{R_{i3}}{t_\beta}  \gamma_5 \right] \psi_f H_i \;, \label{eq:yuklag}
\eeq
where $\psi_f$ denotes the fermion fields with mass $m_f$, $i$ is the scalar index, $v^2=v_1^2 + v_2^2$ (fixed by the $W$ boson mass). 

In order to understand how the exclusion results discussed in the previous section translate to the parameter space of the C2HDM we first map equation~\ref{eq:yuklag} into equation~\ref{eq:higgscharacter}.,
\begin{equation}
\left\{
\begin{aligned} \kappa_t \cos \alpha & = \frac{s_1 \, c_2}{s_{\beta}}
\\ \kappa_t \sin \alpha & =  - \frac{s_2}{t_{\beta}}
\end{aligned} \right.
\qquad \qquad s_\beta ^2 \kappa_t^2  = s_1^2 c_2^2 + s_2^2 c_\beta^2 .
\end{equation}
We will just focus on the scenario where $H_1$ is the lightest scalar and the 125 GeV Higgs can be either $H_2$ or $H_3$.
Both $\kappa_t$ and $\alpha$ are free to vary in the experimental allowed region while the parameters of the C2HDM vary in their allowed ranges subject to theoretical and experimental constraints~\cite{ Fontes:2017zfn}. It is important to note that $ \sin \alpha = 0$ and $ \sin \alpha_2 = 0$ are equivalent, meaning that the CP-conserving scenario is obtained with no ambiguity. The $H_1 VV$ coupling, where V is a vector boson, is proportional to $\cos \alpha_2$ which means that the CP-odd scenario is attained for $\alpha_2 = \pi/2$. The equations for the pure CP-odd and pure CP-even scenarios are
\begin{equation}
\left\{
\begin{aligned} 
\sin \alpha = 0 & \implies \kappa_t = \pm \frac{s_1}{s_{\beta}} \, ,
\\ \cos \alpha= 0 & \implies \kappa_t = \pm \frac{s_2}{t_{\beta}} \,\,  (\text{if} \,\,  s_1=0)  \quad \text{or} \quad \kappa_t = \pm \frac{1}{t_{\beta}} \,\,  (\text{if} \,\,  c_2=0) \, , 
\end{aligned} \right.
\end{equation}
which means that the experimental exclusion of $\kappa_t$ will constrain the parameters of the C2HDM. If $\cos \alpha_2 = 0$ that limit is turned
in a constraints on $\tan \beta$ which is already forced to be above 1 by low energy physics measurements (see~\cite{ Fontes:2017zfn}).

\begin{figure}[!h]
	\begin{center}
		\begin{tabular}{ccc}
			\hspace*{-3mm}
			\includegraphics[width=5.3cm]{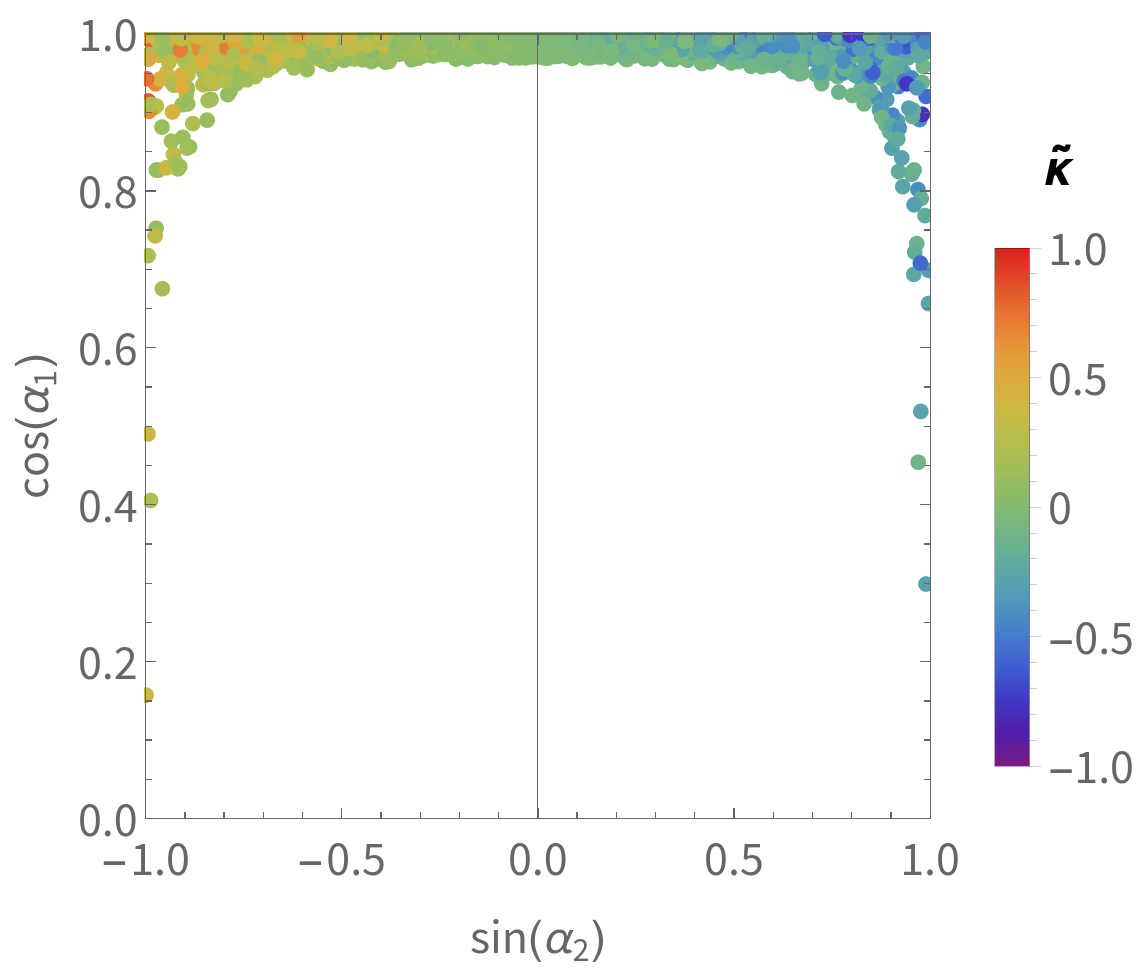}
			\includegraphics[width=5.3cm]{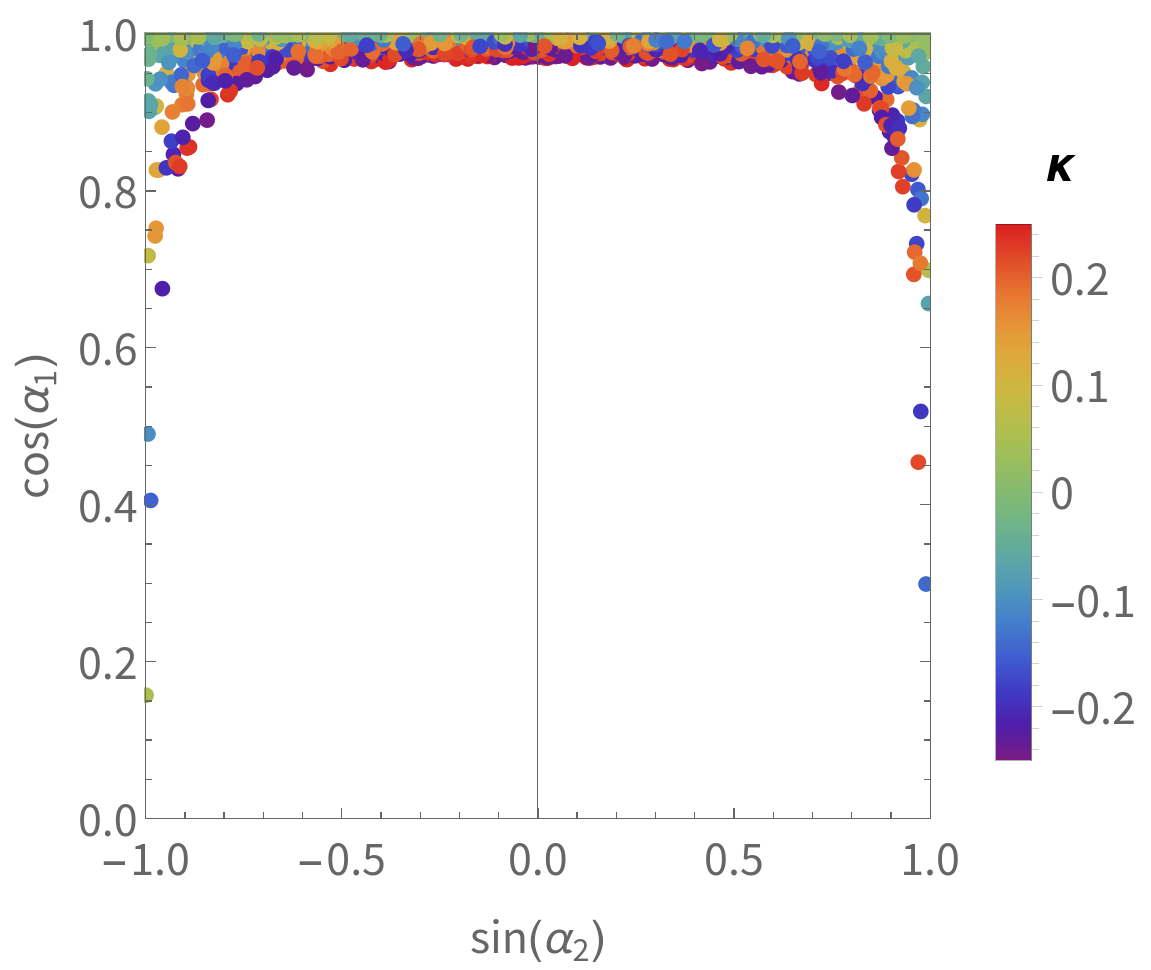}	
			\includegraphics[width=5.3cm]{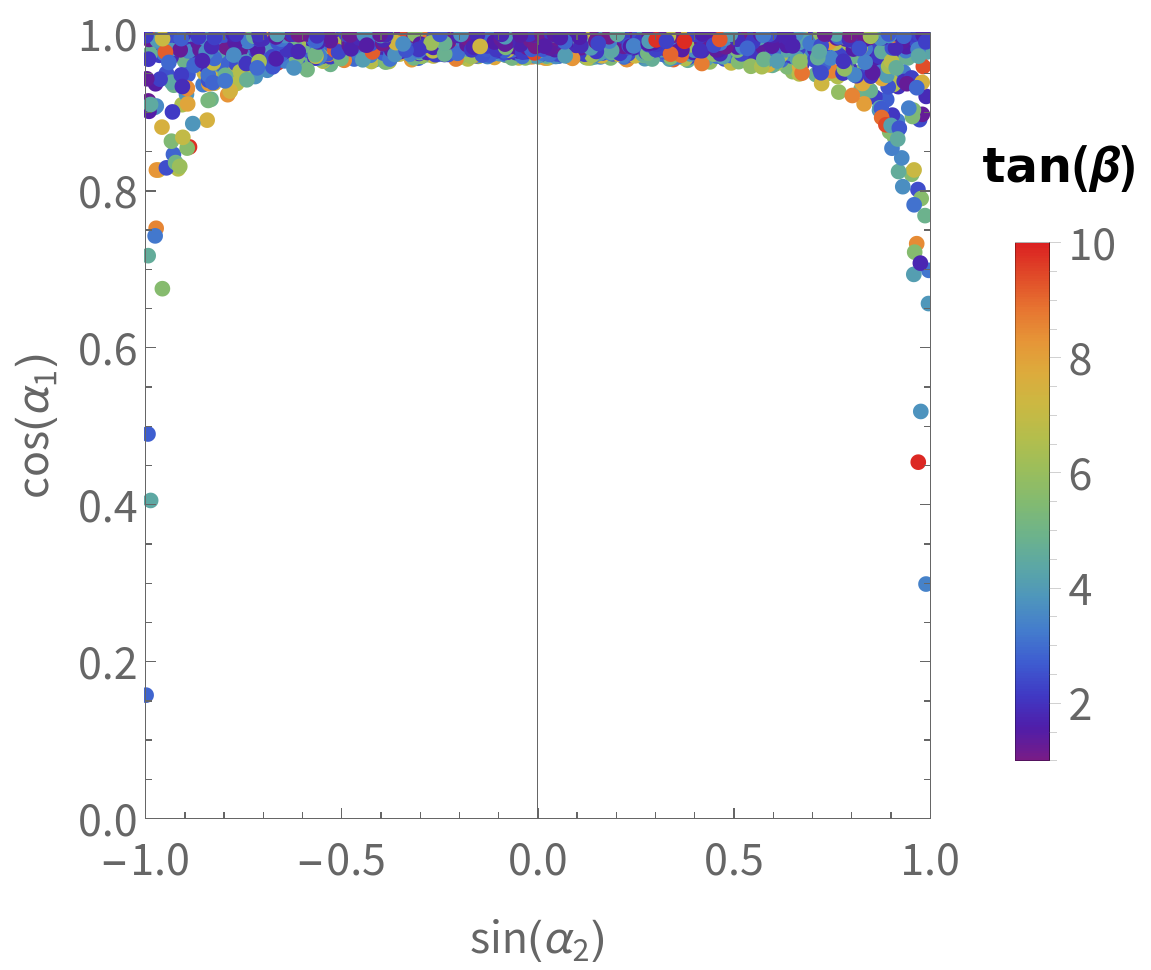}		
		\end{tabular}
		\caption{Points allowed in the plane $c_1$ vs. $s_2$ for $ |\tilde \kappa| \leq 1.0$  and $ |\kappa|  \leq 0.25$ and $1 \leq \tan \beta \leq 10$. The scalar mass is 12 GeV and the luminosity is 300 fb$^{-1}$. In the left plot we see the variation with $\tilde \kappa$, in the middle with $\kappa$ and on the right with $\tan \beta$.}
		\label{low}
	\end{center}
\end{figure}

In Figure~\ref{low}  we present the allowed points in the C2HDM parameter space ($c_1$ vs. $s_2$) for a scalar of 12 GeV using the exclusion limit
for a luminosity of 300 fb$^{-1}$. The constraints for this luminosity are $\kappa^2/a^2+\tilde{\kappa}^2/b^2\leq1$, with $a=0.25$ and $b=1$. We also choose 
the range $1 \leq \tan \beta \leq 10$ in accordance with theoretical and experimental constraints. Note that although values of $\tan \beta$ above 10 are allowed, they do not change the overall picture in the plots. 
In the left plot of Figure~\ref{low} we can see the variation with $\tilde \kappa$, in the middle plot the variation with $\kappa$ is shown and on the right panel the colour code represents the variation of $\tan \beta$.
There are just two striking features in the plots. The first one is that the constraints affect mostly the values of $\cos \alpha_1$ which are constrained to be above 0.1 but are concentrated in the region close to  $\cos \alpha_1=1$. 

\begin{figure}[!h]
	\begin{center}
		\begin{tabular}{ccc}
			\hspace*{-3mm}
			\includegraphics[width=5.3cm]{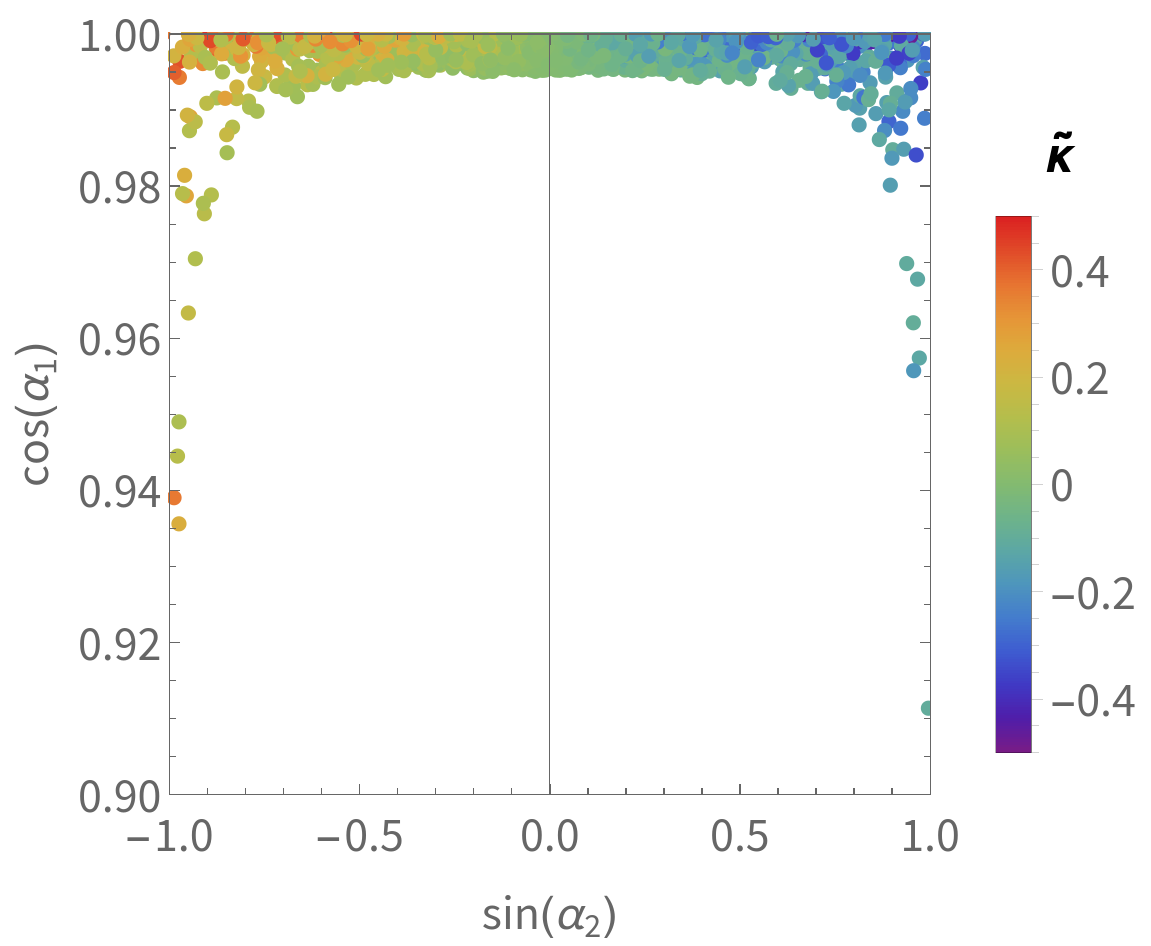}
			\includegraphics[width=5.3cm]{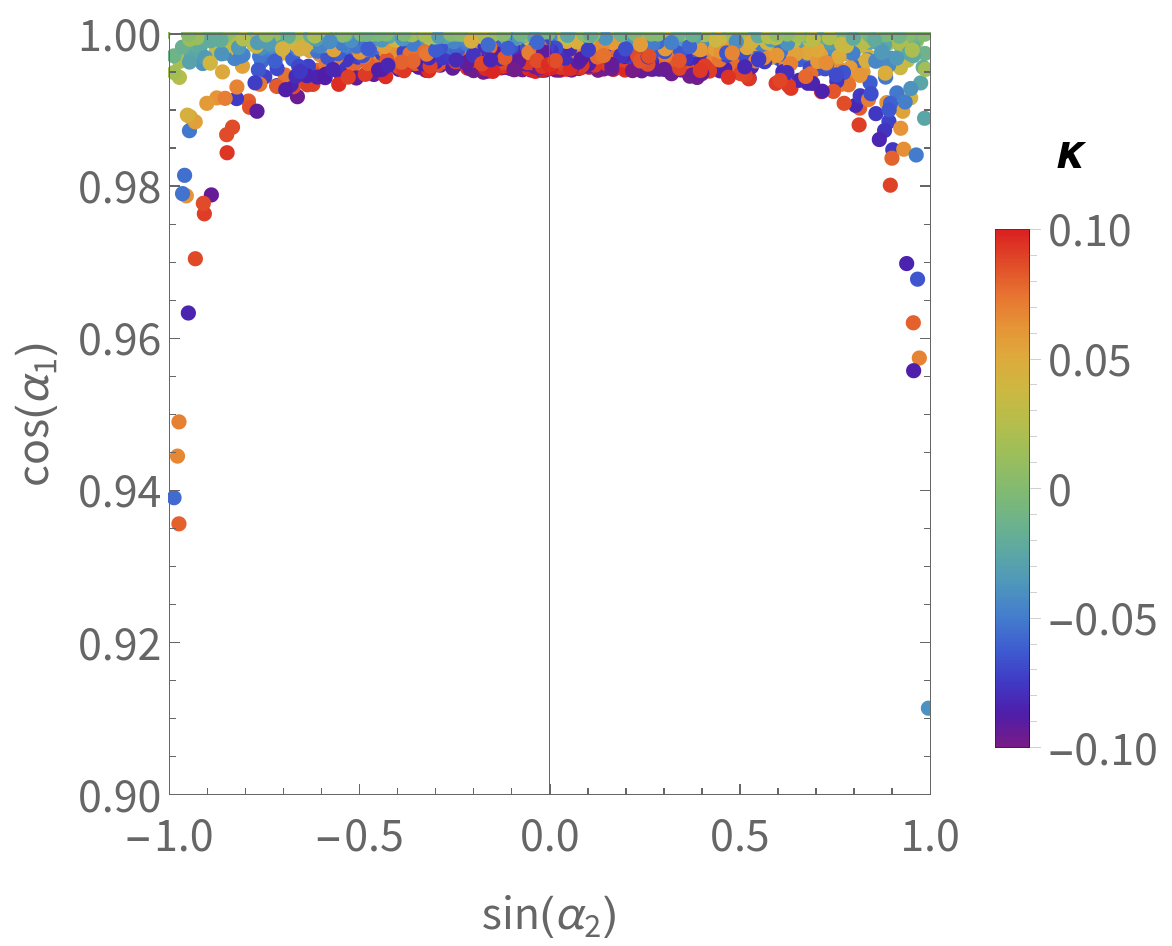}	
			\includegraphics[width=5.3cm]{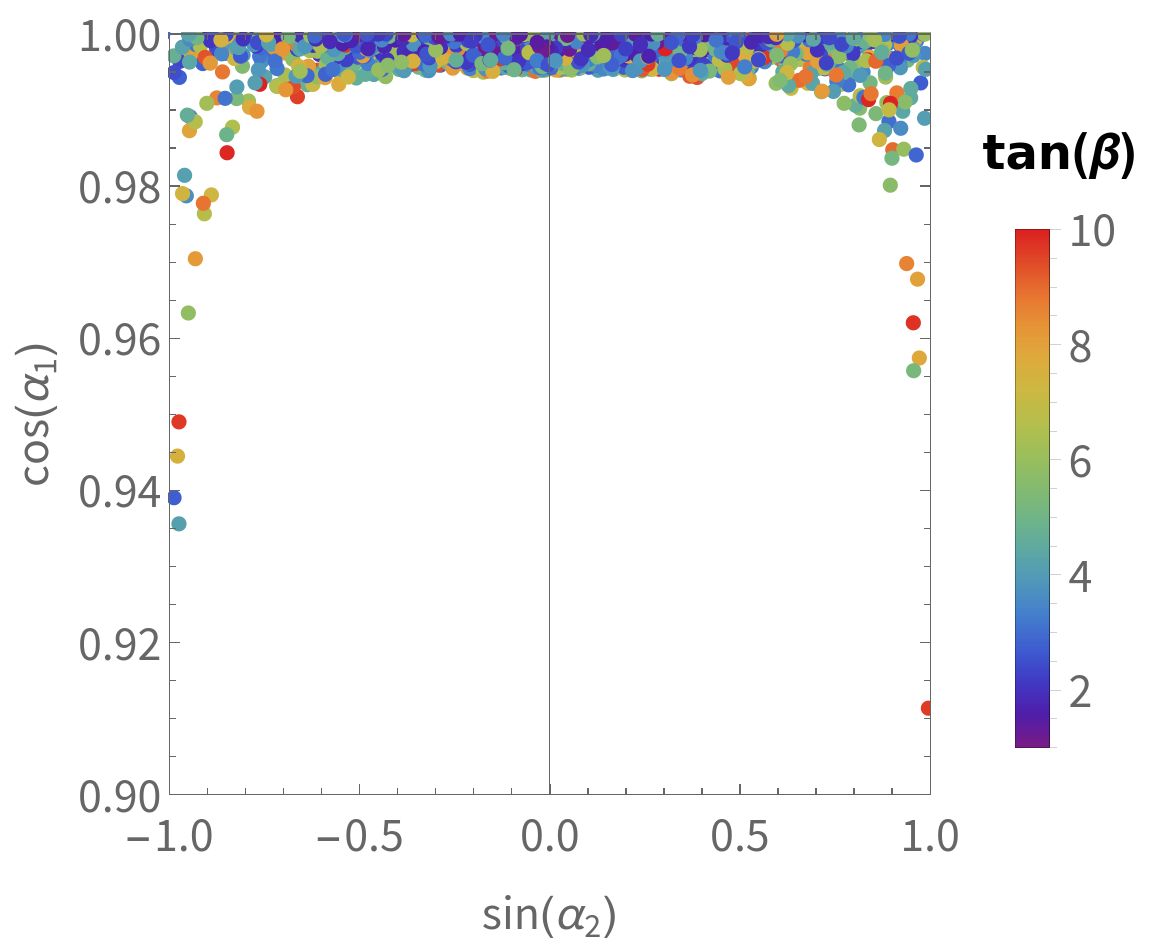}		
		\end{tabular}
		\caption{Points allowed in the plane $c_1$ vs. $s_2$ for $ |\tilde \kappa| \leq 0.5$  and $ |\kappa|  \leq 0.1$ and $1 \leq \tan \beta \leq 10$.	The scalar mass is 12 GeV and the luminosity is 3000 fb$^{-1}$. In the left plot we see the variation with $\tilde \kappa$, in the middle with $\kappa$ and on the right with $\tan \beta$.}
		\label{high}
	\end{center}
\end{figure}

In Figure~\ref{high}  we present a similar plot but now for a luminosity of 3000 fb$^{-1}$. The functional dependence of the constraint for this luminosity is the same as previously, except that now $a=0.1$ and $b=0.5$. 
Again, the left plot of Figure~\ref{high} shows the variation with $\tilde \kappa$, in the middle plot the colour bar shows the variation with $\kappa$ and on the right panel the colour code represents the variation of $\tan \beta$.
The features are similar but now  $\cos \alpha_1 >0.9$. Still, although the bound on  $\tilde \kappa$ reaches the small value of 0.5 the CP-violating angle $\alpha_2$ remains unconstrained.

%%%%%%%%%%%%%%%. Conclusions 
\section{Conclusions \label{sec:conclusions}}
\hspace{\parindent} %forca identacao

In this paper we have proposed a new reconstruction method which allowed us to search for light scalars in $t \bar t \phi$ production in dileptonic final states. While the $t\bar{t}$ system decays to two opposite charge leptons, the $\phi$ boson decays via the $\phi\rightarrow b\bar{b}$ channel. In a previous work we have discussed in detail the  $t \bar t \phi$ process for masses above 40 GeV but failed to go below this number due to degradation of the analysis. The problem was caused by the strong overlap between the jets from the hadronisation of the $b$-quarks originated in the Higgs boson decay. 
The new reconstruction method for the Higgs boson mass ($m_\phi$) overcomes this issue, and allowed to recover significantly the  analysis sensitivity to low mass Higgs bosons. The new mass reconstruction can gain, in terms of mass resolution, roughly a factor of two with respect to previous analysis methods. 
This method can be also applied to the studies of the SM Higgs boson couplings, where the same gains in mass resolution for a mass of the order of $m_H$ = 125~GeV are expected. Without loss of generality, the method can be easily extrapolated to any other two body decays of the Higgs boson 
$\phi\rightarrow \gamma\gamma$, etc., provided the decay channel is kinematically accessible.

The most sensitive CP-observables ($b_2^{t\bar{t}\phi}$ and $b_4^{t\bar{t}\phi}$) were then reconstructed and used to evaluate expected Confidence Levels (CLs) contours of exclusion limits, in the 2D ($\kappa$, $\tilde{\kappa}$) plane, for the SM with a new Higgs boson ($\phi$) against the SM hypothesis only. 
Several $\phi$ bosons, with mixed CP (both CP-even and CP-odd components) and masses that range from $m_\phi$ = 12~GeV up to 40~GeV, were considered. We have  taken the values for luminosities which typically are expected to be within reach during the RUN 3 ($\sim$300~fb$^{-1}$), up to the High Luminosity phase of the LHC (HL-LHC), with 3000~fb$^{-1}$. 
The 95\% CL exclusion limits on the ($|\kappa|$, $|\tilde{\kappa}|$) plane can be as low as, approximately, (0.10, 0.50), at the HL-LHC, for low mass Higgs bosons, 
in only the dileptonic decay channel of the $pp\rightarrow t\bar{t}\phi$ system (with $\phi\rightarrow b\bar{b}$). These results are expected to be significantly improved when the semileptonic decays of the $t\bar{t}\phi$ system are combined.
Further improvement is of course expected if other decay channels of the light Higgs boson are added. It is reasonable to expect reaching the 10$^{-2}$ level or even better both for $\kappa$ and $\tilde \kappa$ when all analyses and all Higgs decay channels are combined.
The interpretation in the framework of the C2HDM was performed assuming that the searched particle is the lightest one in the model. In that scenario, as $\kappa$ and $\tilde \kappa$ decrease $\cos \alpha_1$ gets closer to 1. If  $\tilde \kappa$ decreases even more we would start seeing regions of $\sin \alpha_2$ close to $1$ and $-1$ being excluded.

%%%%%%%%%%%%%%%%%%%%%%%%%%%%%%%%%%%%%%%%%%%%%%%%%%%%%%%
\subsubsection*{Acknowledgments}
%\hspace{\parindent} %forca identacao
%

DA, RC and RS are supported by the Portuguese Foundation for Science and Technology (FCT), Under Contracts UIDB/00618/2020, UIDP/00618/2020,  
PTDC/FIS-PAR/31000/2017,  CERN/FISPAR/0002/2017,  CERN/FIS-PAR/0014/2019, and by the HARMONIA project, contract UMO-2015/18/M/ST2/0518.
AO is partially supported by FCT, under the Contract CERN/FIS-PAR/0029/2019. DA is supported by ULisboa - BD2018.
EG is supported by FCT grant PDPD/BD/128231/2016 and project CERN/FIS-PAR/0002/2019.

%\vspace*{0.5cm}

%%%%%%%%%%%%%%%%%%%%%%%%%%%%%%%%%%%%%%%%%%%%%%%%%%%%%%%
%%%%%%%%%%%%%%%%%%%%%%%%%%%%%%%%%%%%%%%%%%%%%%%%%%%%%%%%%%%%

\vspace*{1cm}
\bibliographystyle{h-physrev}
\bibliography{papernovo.bib}
%\bibliography{direct.bib}
%\end{comment}

\end{document}